\documentclass[fleqn,usenatbib]{mnras}
\usepackage[english]{babel}
\usepackage[T1]{fontenc}
\usepackage{ae,aecompl}
\usepackage{fancyhdr}
\usepackage{amsfonts}
\usepackage{amsmath}
\usepackage{amssymb}
\usepackage{multicol}
\usepackage{layout}
\usepackage{graphicx}
\usepackage{multirow}
\usepackage{color}
\usepackage{multirow}
\usepackage{times}
\usepackage{natbib}
\def\be{\begin{equation}}
\def\ee{\end{equation}}

\usepackage{lineno}

\usepackage[utf8]{inputenc}
\usepackage{lmodern}

\definecolor{valecol}{rgb}{0,0.5, 1.}

\newcommand{\jt}[1]{{\textcolor{black}{#1}}}

\newif\ifAMStwofonts
\AMStwofontstrue

\graphicspath{{./fig/}}
\title[ $f(T)$ cosmology against the cosmographic method]{ $f(T)$ cosmology against the cosmographic method: A new study using mock and observational data}

\author[Sabiee, Malekjani and Mohammad Zadeh Jassur]{
	M. Sabiee$^{1}$, M. Malekjani \thanks{malekjani@basu.ac.ir}$^{2}$ and D. Mohammad Zadeh Jassur$^{1}$ \\ 
	$^1$ Department of Theoritical Physics and Astrophysics, Tabriz University, Tabriz 51664, Iran\\
	$^2$ Department of Physics, Bu-Ali Sina University, Hamedan 65178, Iran}
\date{Accepted ?, Received ?; in original form \today}
\pagerange{\pageref{firstpage}--\pageref{lastpage}} \pubyear{2022}

\begin{document}

\label{firstpage}
\maketitle	
\begin{abstract}
	In this paper, we study the power-law $f(T)$ model using Hubble diagrams of type Ia supernovae (SNIa), quasars (QSOs), Gamma Ray Bursts (GRBs) and the measurements from  baryonic acoustic oscillations (BAO) in the framework of the cosmographic method. Using mock data for SNIa, QSOs and GRBs generated based on the power-law $f(T)$ model, we show whether different cosmographic methods are  suitable to reconstruct the distance modulus or not. In particular, we investigate the rational PADE polynomials $(3,2)$ and $(2,2)$ in addition to the fourth- and fifth- order Taylor series. We show that PADE $(3,2)$ is the best approximation that can be used in the cosmographic method to reconstruct the distance modulus at both low and high redshifts. In the context of PADE $(3,2)$ cosmographic method, we show that the power-law $f(T)$ model is well consistent with the real observational data from the Hubble diagrams of SNIa, QSOs and GRBs. Moreover, we find that the combination of the Hubble diagram of SNIa and the BAO observation leads to better consistency between the model-independent cosmographic method and the power-law $f(T)$ model. Finally, our observational constraints on the parameter of the effective equation of state of DE, described by the power-law $f(T)$ model, show the phantom-like behavior, especially when the BAO observations are included in our analysis.  
 	
\end{abstract}


\section{Introduction}
Observational data from distant Type Ia supernovae (SNIa) revealed a hidden fact that the present Universe is expanding in an accelerated phase \citep{Riess:1998cb,Perlmutter:1998np,Kowalski:2008ez}. This important fact has been confirmed by other cosmological observations such as the cosmic microwave background (CMB) \citep{Komatsu2009,Jarosik:2010iu,Ade:2015rim}, large-scale structure (LSS) and baryonic acoustic oscillation (BAO) \citep{Tegmark:2003ud,Cole:2005sx,Eisenstein:2005su,Percival2010,Blake:2011en,Reid:2012sw}, high-redshift galaxies \citep{Alcaniz:2003qy}, high-redshift galaxy clusters \citep{Wang:1998gt,Allen2004} and weak gravitational lensing \citep{Benjamin:2007ys,Amendola:2007rr,Fu:2007qq}. The accelerated phase of the expansion can be interpreted  
by modifying the standard theory of gravity on cosmological scales or by adding an exotic cosmic fluid with negative pressure, the so-called dark energy (DE) \citep{Riess:1998cb,Perlmutter:1998np,Kowalski:2008ez}. Historically, Einstein's cosmological constant $\Lambda$ with a constant equation of state (EoS) parameter equal to $-1$ is the first and simplest DE model to interpret the current accelerated phase of the expansion. The standard model of cosmology, the $\Lambda$CDM model, which accounts for about $70\%$ of the total energy density of the Universe from $\Lambda$ and about $30\%$ from cold dark matter (CDM), is consistent with the most of the cosmological data. From a theoretical point of view, however, this model has two fundamental problems, namely the problem of fine-tuning and the problem of cosmic coincidence \citep{Weinberg:1988cp,Sahni:1999gb,Carroll:2000fy,Padmanabhan2003,Copeland:2006wr}. In addition, the standard model of cosmology suffers  from the big tension between the local measurement values of Hubble constant $H_0$ with that of the Planck CMB estimation. Moreover, there are big tensions between the Planck CMB data with weak lensing measurements and redshift surveys, concerning the value of non-relativistic matter density $\Omega_{m}$ and the amplitude of the growth of perturbations $\sigma_8$. The above observational tensions can make the standard model as an approximation of a general gravitational scenario yet to be found. For a recent review about the cosmological tensions of the standard model, we refer the reader to see \citep{Abdalla:2022yfr}.
In this concern, a large family of dynamical DE models with time-varying EoS parameters has been proposed in the literature \citep[for some earlier attemts, see][]{Caldwell2002,ArmendarizPicon:2000dh,Padmanabhan2002,Elizalde:2004mq}. Parallel to the solution of DE, the positive cosmic acceleration can be seen as the expression of a new theory of gravity on large cosmological scales. Indeed, modifying the standard Einstein-Hilbert action in the context of the Friedmann-Robertson-Walker (FRW) metric leads to the modified Friedmann equations, which can be used to justify the current accelerated expansion of the Universe without resorting to DE fluid. One of the most popular modified gravity theories is the $f(R)$ scenario, in which the Lagrangian of the modified Einstein-Hilbert action is extended to the function of the Ricci scalar $R$ \citep{Capozziello:2007ec,Capozziello:2007ms,Sotiriou:2008rp,Nojiri:2010wj}. Besides the $f(R)$ theories of gravity, the so-called $f(T)$ theory of gravity is the other solution to solve the puzzle of cosmic acceleration. 
This theory is defined on the basis of the old definition of the teleparallel equivalent of general relativity (TEGR), first introduced by Einstein \citep{Einstein29} and extended by \citep{Hayashi:1979qx,Maluf:1994ji}. A comprehensive study on the theory of teleparallel gravity (TG) can be found in the recent review by \citep{Bahamonde:2021gfp}. In general, an extended theory of gravity involves curvature, torsion and non-metricity components. If only torsion is non-vanishing, one can obtain the torsional teleparallel geometry which is the basic geometry for the most $f(T)$ theories in literature. TG theory can be used to to formulate a TEGR formalism which is dynamically equivalent to GR but may have different behaviors for other scenarios, such as quantum gravity. The Horndeski gravity can also be formulated using the teleparallel geometry as a possible revival modes for regular Hordenski gravity models \citep[see][for more detils]{Bahamonde:2021gfp}. TG as a theory built on the tangent sapce must be invariant under general coordinate transformation and local lorentz transformation. In the first formulation of teleparallel theories, it was assumed that the spin connection was always zero and then torsion tensor depends on the tetrads. This torsion tensor is a particular case which is computed in the so-called Weitzenb\"{o}ck gauge \citep[see section 2.2.3 of][]{Bahamonde:2021gfp}. By taking a local Lorentz transformation only in the tetrads, one can conclude that the torsion tensor is non-covariant quantity under the local Lorentz transformation. In this context, the action of TEGR has a total divergence term which can be removed as being a boundary term. Hence, the TEGR action is invariant under local Lorentz transformation up to a boundary term. In modified teleparallel theories of gravity like $f(T)$ gravity, we have no boundary term anymore, meaning that $f(T)$ gravity models break the local Lorentz invariance. Notice that the problem of breaking the local Lorentz invariance is related to the particular Weitzenb\"{o}ck gauge (spin connection zero) \citep{Krssak:2015oua}. If we take the simultaneous transformations in the tetrads and the spin connection, the teleparallel theories are then fully invariant (diffeomorphisms and local Lorentz). Thus, any action and consequently field equations constructed based on the torsion tensor will be fully invariant.
 In the context of cosmology, there is a large body of works that has examined the cosmological properties of various $f(T)$ models. In this framework, the dynamics and various aspects of the Universe with homogeneous and isotropic background are studied in recent works  \citep{Bengochea:2008gz,Linder:2010py,Myrzakulov:2010vz,Dent:2011zz,Zhang:2011qp,Geng:2011aj,Bamba:2010wb,Bamba:2012vg,Wu:2010av,Wu:2010xk}. In general, cosmological consequences for the various formulations of TG at both background and perturbation levels have been discussed in \citep{Bahamonde:2021gfp}. The $f(T)$ models have been studied and constrained using the various cosmological data \citep{Wu:2010mn,Capozziello:2015rda,Iorio:2015rla,Nunes:2016qyp,Saez-Gomez:2016wxb}. 
 As a more recent study, we refer the work of \citep{Briffa:2021nxg} where the various versions of $f(T)$ model have been studied observationally. Using the combinations of cosmological data including the cosmic chronometers, the SNIa observations from Pantheon catalogue, BAO observations and different model-independent values for Hubble constant $H_0$, \cite{Briffa:2021nxg} studied a detailed analysis of the impact of the $H_0$ priors on the late time properties of $f(T)$ cosmologies. They found a higher value of $H_0$ compared to equivalent analysis without considering $H_0$ priors. In addition, they showed that the $f(T)$ model produces the higher value of $H_0$ and slightly lower value of matter density $\Omega_{m0}$ as compared to standard $\Lambda$CDM model. In general, studies on the cosmological tensions and their possible alleviation in TG formalism have been addressed in section 10 of review article \citep{Bahamonde:2021gfp}. For a review of cosmological tensions $H_0$ and $\sigma_8$, we refer the reader to see the current studies in \citep{Verde:2019ivm,DiValentino:2021izs,DiValentino:2020vvd,DiValentino:2020zio,Perivolaropoulos:2021jda}.\\

On the other hand, we can study the expansion history of the Universe, without considering any particular cosmological model. These approaches are so-called model-independent methods. Some relevant model-independent methods are the Gaussian process (GP), the smoothing methods \citep{Shafieloo:2005nd} and Machine learning based on the neural network \citep{Escamilla-Rivera:2020fxq}. As popular methods, the GP methods have been extensively applied in cosmology \citep{Shafieloo:2012ht,Seikel:2013fda,Yang:2015tzc,Wang:2017jdm,Cai:2019bdh,Gomez-Valent:2018hwc,Zhang:2018gjb,Aljaf:2020eqh,Li:2019nux,Liao:2019qoc,Escamilla-Rivera:2021rbe,Dhawan:2021mel}. In particular, the $f(T)$ gravity has been investigated using the GP method in \citep{Briffa:2020qli} and references therein. Using different combinations of observational data including CC, SNIa, BAO and various $H_0$ priors, \cite{Briffa:2020qli} reconstructed the function $f(T)$ in terms of torsion $T$ with GP method. They showed that for a data sample without $H_0$, the reconstructed $f(T)$ model is well consistent wit the standard $\Lambda$CDM model, while datasets including $H_0$ priors prefer a slight change deviation from standard model.
 Another model-independent method that we use in this work is the cosmographic method. The cosmographic method has been extensively used in the context of cosmology \citep{Sahni:2002fz,Alam:2003sc,Capozziello:2009ka,Capozziello:2011tj,Capozziello:2018jya,Benetti:2019gmo,Escamilla-Rivera:2019aol,Lusso:2019akb,Capozziello:2020ctn,Rezaei:2020lfy,Bargiacchi:2021fow}. This approach was first introduced by \cite{Sahni:2002fz} and \cite{Alam:2003sc} to distinguish between different models of dark energy. \cite{Capozziello:2009ka} applied the cosmographic method to study the dynamics of galaxy clusters and compared the results with those of the $f(R)$ theory of gravity. They showed that the cosmographic method can be used to distinguish GR from alternative modified theories of gravity. \cite{Capozziello:2011tj} also used the cosmographic method to study the expansion history of the Universe and showed that results may differ from the standard model of cosmology.
 Recently, based on the cosmographic method and using Hubble diagrams of SNIa, QSOs, and GRBs, \cite{Lusso:2019akb} have shown that there is a big tension between cosmographic parameters of the standard $\Lambda$CDM model and those of the model-independent cosmographic approach. The cosmographic method presented in \citep{Lusso:2019akb}, based on the logarithmic expansion of the luminosity distance, was extended to a general case by applying the orthogonalized logarithmic polynomials of the luminosity distance \citep{Bargiacchi:2021fow}. In this context, and by using the Hubble diagrams of SNIa and QSOs, \cite{Bargiacchi:2021fow} presented a strong tension ($>4\sigma$) between the cosmographic parameters of the standard model and those of the model-independent cosmographic method. While the standard model of cosmology based on the results of \citep{Lusso:2019akb,Bargiacchi:2021fow} suffers from the strong tension, the dynamical DE models studied in \citep{Rezaei:2020lfy} have a better agreement with the cosmographic method, at least at the $2\sigma$ level. In detail, cosmographic parameters of the wCDM model, CPL-like and PADE-like  equation of state of DE were shown to agree with the $2\sigma$ confidence level of the cosmographic parameters obtained from the model-independent approach, even when we use the Hubble diagrams of QSOs and GRBs at higher redshifts \citep{Rezaei:2020lfy}. In the same study, \cite{Pourojaghi:2021den} have shown that cosmographic parameters of the holographic model of DE, constrained by the various combinations of Hubble diagrams from SNIa, QSOs and GRBs, are consistent with the same parameters in the model-independent method. As mentioned above, we encounter with the puzzling results for the standard model at higher redshifts. Another possible interpretation of the strong discrepancy between the standard model cosmographic parameters and the model-independent values is that the standard Taylor approximation in the cosmographic approach may not work optimally at the higher redshifts where QSOs, and GRBs are included in our analysis. In this line, \cite{Yang:2019vgk} pointed out that the cosmographic method cannot be extended to redshifts higher than $z\sim 2$, due to the considerable error truncation of the Taylor expansion of the Hubble parameter (or equivalently the logarithmic polynomials of the luminosity distance). We mention that this conclusion is valid only for the standard $\Lambda$CDM model. In another paper, using mock data for the Hubble diagram of QSOs, \cite{Banerjee:2020bjq} have shown that the strong discrepancy between the cosmographic parameters of the standard model and the cosmographic parameters of the model-independent cosmographic approach can be interpreted as a false tension due to an error truncation of the logarithmic polynomials imposed on the cosmographic method at redshifts higher than $\sim 2$. Since mock data for QSOs have been generated based on the standard $\Lambda$CDM model, we expect that the model and the cosmographic method are to agree with each other. Thus, any tension between the standard model and the cosmographic approach has no physical meaning and can be considered as a false tension. We mention that the result of \cite{Banerjee:2020bjq} is valid for the standard $\Lambda$CDM model. On the other hand, \cite{Rezaei:2020lfy} and \cite{Pourojaghi:2021den} have shown the consistency between dynamical DE models and the model-independent cosmographic method at higher redshifts. Due to the problems of the cosmographic method with the standard model at redshifts higher than $2$, we modify the Taylor expansion used in the cosmographic method to a more general mathematical approximation, the PADE expansion. The PADE expansion is a rational series that generally performs better than the linear Taylor expansion and avoids the convergence problem at higher redshifts \citep{Capozziello:2020ctn}. In addition to PADE approximation, the  Chebyshev approximant is the other rational polynomials that has been used in cosmographic method \citep{Capozziello:2017nbu}. It has been shown that the rational Chebyshev polynomials significantly reduces the error propagation with respect to standard Taylor series \citep{Capozziello:2017nbu}. Recently, \cite{Munoz:2020gok} examined the cosmographic polynomials using the machine learning methods. In the context of inverse cosmographic approach \citep[see][]{Escamilla-Rivera:2019aol} and using the observational SNIa sample, trained SNIa sample using deep learning architecture (DL) and the combination of SNIa + DL samples, \cite{Munoz:2020gok} tested some popular CPL, Redshift Squared (RS), PADE and Chebyshev polynomials. The inverse cosmographic method can compute a generic EoS parameter of DE without assuming directly a cosmographic series. Hence it is possible to relax the truncation issues over the series/polynomials entered in usual cosmographic methods. The advantage of observational SNIa + DL sample is that we can obtain a big data sample in a redshift range of $0.01 < z < 4$, which is more extended than the observable one. In the context of inverse cosmography, it has been shown that the PADE-like EoS parameter of DE can transit from $w_{\Lambda}=-1.0$ at low redshift to quintessence regime ($w_{DE}<-1$) at higher redshifts. Moreover, the PADE-like EoS parameter does not have any divergence at observational redshift \citep{Munoz:2020gok}. Although the Chebyshev-like EoS parameter mimics the PADE EoS at high redshifts, but it has a divergence at low redshifts \citep{Munoz:2020gok}. On the basis of the above justification in the context of inverse cosmography, we are satisfied to select the PADE approximation compared to Chebyshev approximate.\\ 
To complete this section, we give here a brief review of the previous cosmographic studies on $f(T)$ cosmology.
In \citep{Capozziello:2011hj}, the authors constrained the cosmographic parameters of the $f(T)$ model using observational data from SNIa, H(z) and BAO. They found that the best-fit values of the cosmographic parameters in the $f(T)$ model are consistently close to those of the standard model. Using observational data, including SNIa from Union 2.1, OHD, and local measurements of $H_0$, \cite{Aviles:2013nga} reconstructed the $f(T)$ model in the context of the  cosmographic method based on the linear Taylor series up to the foruth order of approximation. \cite{Capozziello:2015rda} studied the transition redshift from the early matter dominated phase to the current accelerated phase within $f(T)$ model using the cosmographic method and Union 2.1 supernova compilation. In the context of cosmographic method, \cite{Cai:2019bdh} investigated the $f(T)$ cosmology using the measurements of the $H(z)$ data set, observations of the BAO and local measurements of $H_0$ and showed that the $H_0$ tension can be alleviated. In addition, \cite{Farias:2021jdz} examined the generalized $f(R,T)$ model in the context of the cosmographic method.\\
In the line of above studies, here we study the viable $f(T)$ teleparallel cosmological model, namely the power-law model, in the framework of the cosmographic approach based on the PADE expansion. Using the Hubble diagrams of SNIa, QSOs and GRBs, we investigate how the power-law $f(T)$ model performs at higher redshifts in the context of PADE cosmographic method. To ensure that the PADE approximation works properly at redshifts higher than $z\sim 2$, we create mock data for the Hubble diagrams of SNIa, QSOs and GRBs based on the power-law $f(T)$ cosmology. Since mock data are generated based on the model, any discrepancy between model and cosmographic method is related to the error truncation of the mathematical approximation used in cosmographic method. So, if the PADE cosmographic approach works correctly, we expect the cosmographic parameters from the $f(T)$ model to match those from the PADE cosmographic approach.\\
The layout of the paper is as follows: In Sect .\ref{sect:sectionII} we give a brief explanation of the power-law $f(T)$ model in  homogeneous and isotropic FRW Universe. In Sect. \ref{sect:sectionIII}, we investigate the power-law $f(T)$ model in the framework of cosmographic approach and also extend the cosmographic parameters to a more general case based on the PADE approximation. We report our numerical analysis and observational constraints using observational and mock data for Hubble diagrams of SNIa, QSOs and GRBs in Sect. \ref{sect:result}. Finally, we summarize this work in Sect. \ref{sect:conlusion}. 
\section{The power-law $f(T)$ model}\label{sect:sectionII}
In this section, we briefly present the main equations for the power-law version of the $f(T)$ cosmology, where $f(T)=-T+\mathcal{F}$ and $\mathcal{F}\propto (-T)^b$. We ignore here the main derivations of the equations and refer the reader to see \citep{Briffa:2021nxg}, for more details. In the context of a flat homogeneous and isotropic FRW metric $ds^2=dt^2-a(t)\delta_{ij}dx^idx^j$ and tetrad $e_{\mu}^A={\rm diag}(1,a,a,a)$ \citep{Krssak:2015oua,Tamanini:2012hg}, the modified Freidmann equations are written as
\begin{eqnarray}\label{eq:eq2}
&&H^2= \frac{8\pi G_N}{3}(\rho_m+\rho_r)
+\frac{\mathcal{F}}{6}-\frac{T}{3}\mathcal{F}_T,\\\label{eq:eq3}
&&\dot{H}=-\frac{4\pi G_N(\rho_m+\rho_r+P_r)}{\left(1 - \mathcal{F}_T - 2T\mathcal{F}_{TT}\right)},
\end{eqnarray}
where $\mathcal{F}_{T}=\partial \mathcal{F}/\partial T$, $\mathcal{F}_{TT}=\partial^{2} \mathcal{F}/\partial
T^{2}$.
 The additional terms in Eqs. (\ref{eq:eq2} \& \ref{eq:eq3}) represent the modification of gravity beyond the standard form of Freidmann equations. We can easily reconstruct the energy density of the equivalent DE component as follows \cite[see also][]{Linder:2010py}:
\begin{eqnarray}\label{eq:eq4}
&&\rho_{\rm de}\equiv\frac{3}{8\pi
	G_N}\left[\frac{\mathcal{F}}{6}-\frac{T\mathcal{F}_T}{3}\right]. \label{rhoDDE}
\end{eqnarray}
Hence, the corresponding effective equation of state (EoS) parameter
is obtained as 
\begin{eqnarray}\label{eq:eq5p}
\label{wfT}
w_{\rm de}=
-1+\frac{1}{3}\frac{d{\rm ln}T}{d{\rm ln}a}
\frac{\mathcal{F}_{T}+2T\mathcal{F}_{TT}}{[(\mathcal{F}/T)-2\mathcal{F}_{T}]}.
\end{eqnarray}
The dimensionless Hubble parameter, from the modified Freidmann equation (\ref{eq:eq3}), reads
\begin{eqnarray}\label{eq:eq6}
E^2(a)=\Omega_{m0}a^{-3}+\Omega_{r0}a^{-4}+\Omega_{F0} X(a),
\end{eqnarray}
where $\Omega_{F0}=1-\Omega_{m0}-\Omega_{r0}$ and $X(a)$ is given by
\begin{eqnarray}
\label{distortparam}
X(a)=\frac{1}{T_0\Omega_{F0}}\left(-\mathcal{F}+2T\mathcal{F}_T\right).
\end{eqnarray}
Using $T=-6H^2$ ($T_0=-6H_0^2$) and functional form of the power-law $\mathcal{F}=\alpha (-T)^b$ pattern \citep{Bengochea:2008gz,Briffa:2021nxg} where $\alpha=(6H_0^2)^{(1-2b)}\frac{\Omega_{F0}}{1-2b}$, we can rewrite the dimensionless Hubble parameter as \citep[see also][]{Briffa:2021nxg}:
\begin{eqnarray}
\label{eq:eq7} E^2(a,b)=\Omega_{\rm m0}a^{-3}+\Omega_{\rm
	r0}a^{-4}+\Omega_{\rm F0} E^{2b}(a,b) \;.
\end{eqnarray}	
We explicitly see that by putting $b=0$, the standard flat-$\Lambda$CDM model is recovered. For small values of $b$ , we can perform the Taylor expansion of $E^2(a,b)$ around $b=0$ as 
\begin{eqnarray}\label{eq:eq8}
E^2(a,b)=E^2(a,0)+\left.\frac{dE^2(a,b)}{db}\right|_{b=0} b+...
\end{eqnarray}
where $E^2(a,0)=E^2_{\Lambda}(a)$, represents the dimensionless parameter of standard $\Lambda$CDM model. For $b<<1$, we can truncate the Taylor expansion to the first exponent of $b$ and obtain the following relation for the Hubble parameter of the power-law $f(T)$ cosmology in a flat geometry \citep{Basilakos:2016xob}:
\begin{eqnarray}
\label{eq:eq9}
E^2(a,b)\simeq
E^2_\Lambda(a)+\Omega_{F0}\ln\left[E^2_\Lambda(a)\right]b \;,
\end{eqnarray}
where $E^2_{\rm \Lambda}(a)=\Omega_{\rm m0}a^{-3}+\Omega_{\rm r0}a^{-4}+\Omega_{\rm F0}$. Thus, the evolution of the Hubble parameter in power-law $f(T)$ model depends on the values of the model parameter $b$ and the cosmological parameter $\Omega_{\rm m0}$. Note that in the late times of the history of the Universe, we neglect the energy density of radiation, which means that $\Omega_{F0}=1-\Omega_{\rm m0}$. Finally, the EoS parameter of equivalent DE in the power-law $f(T)$ model can be obtained from Eq. (\ref{eq:eq5p}) as follows
 \begin{eqnarray}
 \label{eq:eq5pp}
w_{\rm de}(a) = -1-\frac{2b}{3}\frac{d\ln{E}}{d\ln{a}}\;,
 \end{eqnarray}
where $E$ is given by Eq. (\ref{eq:eq9}). In the next section, we define the cosmographic parameters as the time derivative of the Hubble parameter and obtain specific forms of these parameters for the power-law $f(T)$ cosmology, by performing various time derivatives of Eq.(\ref{eq:eq9}).
\section{The Cosmographic approach}\label{sect:sectionIII}
Dynamics of the scale factor $a(t)$ as a function of cosmic time is one of the most important quantities in cosmology. Based on the Freidmann equations, the functional form of the scale factor depends on the energy densities of the Universe. Therefor, we cannot obtain a specific cosmic scale factor when  dealing with cosmological models. In parallel, the cosmographic method as a method which is independent of cosmological models, can propose a mathematical form of the scale factor as a function of cosmic time using the Taylor series or other extended polynomials beyond the Taylor expansion. Indeed, we can mathematically expand the unknown function $a(t)$ around its present value $a(t_0)=1$, without presupposing any specific cosmological model. In this section, we first introduce the cosmographic method based on the linear Taylor approximation and then present the extended cosmographic method, based on the rational PADE approximation.

\subsection{Cosmography based on the Taylor series}

 The reconstructed scale factor in the context of cosmographic method defined on the basis of the Taylor approximation reads 
 
\begin{align}
\label{eq:eq10}
&a(t)\simeq 1+ H_0 (t-t_0) - \frac{q_0}{2} H_0^2 (t-t_0)^2 + 
&\frac{j_0}{6} H_0^3 (t-t_0)^3 +\nonumber \\ 
&\frac{s_0}{24} H_0^4 (t-t_0)^4 + \frac{l_0}{120} H_0^5 (t-t_0)^5\;,
\end{align}

where we truncate the Taylor expansion at the fifth exponent of $(t-t_0)$. The coefficients of the Taylor series are the time derivatives of the scale factor at the present time. Note that the truncation of Taylor series can cause a large error and hence make a significant deviation between the reconstructed scale factor and real scale factor that we are exploring. In this paper we will discuss in detail the error truncation of the Taylor approximation used in the cosmographic method. The coefficients of Eq. (\ref{eq:eq10}) are obtained based on the present-time values of the following quantities namely the cosmographic parameters \citep{Visser:2003vq}: $H(t)=\frac{\dot{a}}{a}$ (Hubble parameter), $q(t)=-\frac{\ddot{a}}{aH^2}$ (deceleration parameter), $j(t)=\frac{\dddot{a}}{aH^3}$ (jerk parameter), $s(t)=\frac{\ddddot{a}}{aH^4}$ (snap parameter) and $l(t)=\frac{\ddddot{\dot a}}{aH^5}$ (lerk parameter). In this framework, we can equate various time derivatives of the Hubble parameter with cosmographic parameters as follows:

\begin{align}\label{eq:eq11}
\frac{\dot{H}}{H^2} &= -q - 1\;,\nonumber\\
\frac{\ddot{H}}{H^3} &= j + 3 q + 2\;,\nonumber\\
\frac{\dddot{H}}{H^4} &= - 4 j - 3 q^{2} - 12 q + s - 6\;,\nonumber\\
\frac{\ddot{\ddot{H}}}{H^5} &= 10 j q + 20 j + l + 30 q^{2} + 60 q - 5 s + 24\;,\nonumber\\
\frac{\ddot{\dddot{H}}}{H^6} &= - 10 j^{2} - 120 j q - 120 j - 6 l + m - 30 q^{3} - 270 q^{2} + 15 q s\nonumber\\ &\quad - 360 q + 30 s - 120\;,
\end{align}

Instead of the scale factor, we can expand the Hubble parameter around its present-time value $H_0$ using the Taylor series to reconstruct the expansion of the Universe at low redshifts, as follows:

\begin{align}\label{eq:eq13}
H(z) \simeq H_0 + \frac{dH}{dz}\vert_{z=0} z + \frac{d^2H}{dz^2}\vert_{z=0} \dfrac{z^2}{2} + 
\frac{d^3H}{dz^3}\vert_{z=0} \dfrac{z^3}{6}\nonumber\\
 + \frac{d^4H}{dz^4}\vert_{z=0} \dfrac{z^4}{24} 
+ \frac{d^5H}{dz^5}\vert_{z=0} \dfrac{z^5}{120}\;,
\end{align}
where the coefficients $\dfrac{d^{i}H}{dz^{i}}\vert_{z=0}$ ($i$ counts 1 to 5) are related to the present-time values of the cosmographic parameters through Eq.(\ref{eq:eq11}). Notice that here we truncated the Taylor series at $z^5$. It is easy to understand that the reconstructed Hubble parameter based on Eq. (\ref{eq:eq13}) is not valid beyond $z=1$ due to the divergence problem of the Taylor expansion. To avoid the divergence problem one proposed way is changing  $z$ variable to $y=z/(1+z)$ variable \citep{Vitagliano_2010,Capozziello:2011tj,Rezaei:2020lfy}. The reconstructed Hubble parameter based on the $y$ variable can be written as \citep{Pourojaghi:2021den}:
\begin{eqnarray}\label{eq:eq14}
&&E(y)=\frac{H(y)}{H_0}\simeq 1+C_1 y + \dfrac{C_2 y^2}{2}+\dfrac{C_3 y^3}{6}+\dfrac{C_4 y^4}{24}+\nonumber\\
&&\dfrac{C_5 y^5}{120}\;,
\end{eqnarray}
where coefficients $C_i$ can be obtained in terms of present-time values of the cosmographic parameters as follows:

\begin{align}\label{eq:eq15}
C_1=& (q_{0} + 1)\;\nonumber\\
C_2=& (j_{0} - q_{0}^{2} + 2 q_{0} + 2)\;,\nonumber\\
C_3=& (- 4 j_{0} q_{0} + 3 j_{0} + 3 q_{0}^{3} - 3 q_{0}^{2} + 6 q_{0} - s_{0} + 6)\;,\nonumber\\
C_4=& (- 4 j_{0}^{2} + 25 j_{0} q_{0}^{2} - 16 j_{0} q_{0} + 12 j_{0} + l_{0} - 15 q_{0}^{4} +\nonumber\\ &\quad 12 q_{0}^{3} - 12 q_{0}^{2} + 7 q_{0} s_{0} + 24 q_{0} - 4 s_{0} + 24)\;,\nonumber\\
C_5=& (70 j_{0}^{2} q_{0} - 20 j_{0}^{2} - 210 j_{0} q_{0}^{3} + 125 j_{0} q_{0}^{2} - 80 j_{0} q_{0} + \nonumber\\ &\quad 15 j_{0} s_{0} + 60 j_{0} - 11 l_{0} q_{0} + 5 l_{0} - m_{0}+ 105 q_{0}^{5} - 75 q_{0}^{4} +\nonumber\\ &\quad 60 q_{0}^{3} - 60 q_{0}^{2} s_{0} - 60 q_{0}^{2} + 35 q_{0} s_{0} + 120 q_{0} - 20 s_{0} + 120)\;.
\end{align}

In section \ref{sect:result}, by using the generated mock data for the Hubble diagrams of SNIa and QSOs, we examine the cosmographic method defined on the basis of the $y$-variable. To do this, we compute the error truncation of the cosmographic method at redshifts higher than $z\sim 1$. Let us now introduce the other approximation namely the rational PADE series to avoid the convergence problem of the $z$-variable Taylor expansion.

\subsection{Cosmography based on the PADE approximation}

The PADE approximation is a particular and classical type of the rational approximations. The main idea of the PADE approximation is to expand a given function as a ratio of two power series and determining both the numerator and denominator coefficients by using the coefficients of the Taylor series of given function. Thus a given function $f(z)=\sum\limits_{i=0}^{\infty}c_iz^i$, where $c_i$ is the set of coefficients in Taylor series, can be approximated by means of PADE polynomials as follows

\begin{eqnarray}
\label{eq:eq16}
P_{n,m}(z)=\frac{\sum\limits_{i=0}^{n}P_iz^i}{1+\sum\limits_{j=0}^{m}Q_jz^j}\;.
\end{eqnarray}  

In general, we can equate the coefficients of PADE appropriation of the Hubble function $H(z)$ with those of the linear Taylor expansion of $H(z)$\citep{Capozziello:2020ctn}. Notice that to find the coefficients of the $P_{n,m}(z)$ in Eq.(\ref{eq:eq16}), we should use the Taylor expansion $\sum\limits_{i=0}^{n+m}c_i z^i$. For example $P_{2,2}(z)$ ($P_{3,2}(z)$) approximation of $H(z)$ is equivalent with the linear Taylor approximation $\sum\limits_{i=0}^{4}c_iz^i$ ($\sum\limits_{i=0}^{5}c_iz^i$). Hence the PADE approximations $P_{2,2}(z)$ and $P_{3,2}(z)$ for dimensionless Hubble parameter $E(z)=H(z)/H_0$ can be written as follows, respectively \citep{Capozziello:2018jya}:

\begin{align}\label{eq:eq17}
P_{2,2}=& \frac{P_0 + P_1 z + P_2 z^2}{1 + Q_1 z + Q_2 z^2}\;,\nonumber\\
P_0=& 1\;,\nonumber\\
P_1=& H_1 + Q_1\;,\nonumber\\
P_2=& H_1Q_1 + \frac{H_2}{2} + Q_2\;,\nonumber\\
Q_1=& \frac{-H_1H_4 + 2H_2H_3}{4H_1H_3 - 6H_2^2}\;,\nonumber\\
Q_2=& \frac{3H_2H_4 - 4H_3^2}{24H_1H_3 - 36H_2^2}\;,
\end{align}
and

\begin{align}\label{eq:eq18}
P_{3,2}=& \frac{P_0 + P_1 z + P_2 z^2 + P_3 z^3}{1 + Q_1 z + Q_2 z^2}\;,\nonumber\\
P_0=& 1\;,\nonumber\\
P_1=& H_1 + Q_1\;,\nonumber\\
P_2=&  H_1Q_1 + \frac{H_2}{2} + Q_2\;,\nonumber\\
P_3=& H_1Q_2 + \frac{H_2Q_1}{2} + \frac{H_3}{6}\;,\nonumber\\
Q_1=& \frac{-3H_2H_5 + 5H_3H_4}{15H_2H_4 - 20H_3^2}\;,\nonumber\\
Q_2=& \frac{4H_3H_5 - 5H_4^2}{60H_2H_4 - 80H_3^2}\;.
\end{align}
Note that in Eqs. (\ref{eq:eq17} \& \ref{eq:eq18}), $H_i=\dfrac{1}{H_0}\frac{d^iH}{dz^i} \vert_{z=0}$ where $\frac{d^iH}{dz^i} \vert_{z=0}$ is defined based on Eq. (\ref{eq:eq13}).
In Sect. \ref{sect:result}, we will show that the PADE approximation reconstructs the Hubble parameter better than the Taylor expansion defined based on $y=z/(1+z)$ variable.

\subsection{Cosmographic parameters for $f(T)$ model}

We now introduce the cosmographic parameters of the power-law $f(T)$ cosmology. By calculating the time derivatives of Eq. (\ref{eq:eq9}), we obtain the following relations for the first through fifth time derivatives of the Hubble parameter in power-law $f(T)$ cosmology:

\begin{align}\label{eq:eq19}
\left(\frac{\dot{H}}{H^2}\right)_{t=t_0} =& (\dfrac{-3}{2})\Omega_{m0}\Big(1+b(1-\Omega_{m0})) \Big) \;,\nonumber\\
\left(\frac{\ddot{H}}{H^3}\right)_{t=t_0} =& (\dfrac{9}{2})\Omega_{m0}\Big(1+b(1-\Omega_{m0})^2) \Big)\;,\nonumber\\
\left(\frac{\dddot{H}}{H^4}\right)_{t=t_0} =& \Big(\dfrac{-27}{4}\Omega_{m0}\Big[2+\Omega_{m0}+b(1-\Omega_{m0})\nonumber\\
&(2+\Omega_{m0}(-4+b(1-\Omega_{m0})^2+3\Omega_{m0}))\Big]\Big)\;,\nonumber\\
\left(\frac{\ddddot{H}}{H^4}\right)_{t=t_0} =& \dfrac{81}{4}\Omega_{m0}\Big[2+4\Omega_{m0}-b(1-\Omega_{m0})\Big(-2+\Omega_{m0}\nonumber\\
&(6+\Omega_{m0}(-13+6\Omega_{m0})+b(1-\Omega_{m0})^2\nonumber\\
&(-4+7\Omega_{m0}))\Big)\Big]\;,\nonumber\\
\left(\frac{\ddot{\dddot{H}}}{H^5}\right)_{t=t_0} =& \dfrac{243}{8}\Omega_{m0}\Bigg[b^3(1-\Omega_{m0})^4\Omega_{\rm m0}^2(-4+7\Omega_{m0})\nonumber\\
&-2[2+\Omega_{m0}(11+2\Omega_{m0})]+b^2(1-\Omega_{m0})^2\nonumber\\
&\Omega_{m0}\Big(-22+\Omega_{m0}[112+\Omega_{m0}(-172+79\Omega_{m0})]\Big)\nonumber\\
&-b(1-\Omega_{m0})\Big(4+\Omega_{m0}[-16+\Omega_{m0}(88\nonumber\\
&-87\Omega_{m0}+30\Omega_{m0}^2)]\Big)
\Bigg]\;.
\end{align}

Inserting relations of Eq. (\ref{eq:eq19}) in the left hand side of Eq. (\ref{eq:eq11}), we can obtain the present-time values of the cosmographic parameters in the power-law $f(T)$ cosmology as follows:

\begin{align}\label{eq:eq20}
q_0 =& \dfrac{3\Omega_{m0}}{2}\Big[1+b(1-\Omega_{m0})\Big]-1 \;,\nonumber\\
j_0 =& 1-\dfrac{9}{2}b(1-\Omega_{m0})\Omega_{m0}^2\;,\nonumber\\
s_0 =& 1+\dfrac{9}{4}\Omega_{m0}\Big[-2 -b(1-\Omega_{m0})\Big(2+\Omega_{m0}\nonumber\\
&(-10+3(3-b(1-\Omega_{m0}))\Omega_{m0})\Big)\Big]\;,\nonumber\\
l_0 =& 1+\dfrac{3}{4}\Omega_{m0}\Big[4+18\Omega_{m0}-b(1-\Omega_{m0})\nonumber\\
&\Big(-4+3\Omega_{m0}(44-102\Omega_{m0}+54\Omega_{m0}^2-\nonumber\\
&3b(1-\Omega_{m0})(2+3\Omega_{m0}(-6+7\Omega_{m0})))\Big)\Big]\;,\nonumber\\
m_0 =& 1+\dfrac{9}{8}\Omega_{m0}\Big[-6(2+3\Omega_{m0}(4+\Omega_{m0}))+\nonumber\\
&9b^3(-1+\Omega_{m0})^3\Omega_{m0}^2(2+3\Omega_{m0}(-6+7\Omega_{m0}))\nonumber\\
&+9b^2(1-\Omega_{m0})^2\Omega_{m0}\Big(-8+\Omega_{m0}(154+\Omega_{m0}\nonumber\\
&(-382+237\Omega_{m0}))\Big)-2b(1-\Omega_{m0})\Big(6+\Omega_{m0}\nonumber\\
&(-212+81\Omega_{m0}(11+\Omega_{m0}(-13+5\Omega_{m0})))\Big)\Big].
\end{align}

Inserting $b=0$ in the above equation, we can recovery the cosmographic relations for the standard $\Lambda$CDM model as:

\begin{align}\label{eq:eq21}
q_0 =& \dfrac{3\Omega_{m0}}{2}-1 \;,\nonumber\\
j_0 =& 1\;,\nonumber\\
s_0 =& 1-\dfrac{9}{2}\Omega_{m0}\;,\nonumber\\
l_0 =& 1+ 3\Omega_{m0} +\frac{27}{2}\Omega_{m0}^2\;,\nonumber\\
m_0 =& 1-\dfrac{27}{2}\Omega_{m0}-81\Omega_{m0}^2-\dfrac{81}{4}\Omega_{m0}^3.
\end{align}
If we also substitute $b=0$ and $\Omega_{m0}=0.3$, we get $q_0=-0.55$, $j_0=1$, $s_0=-0.35$, $l_0=3.11$ and $m_0=-18.17$, for the 
values of the cosmographic parameters for the standard flat-$\Lambda$CDM model, in agreement with the previous studies \citep{Rezaei:2020lfy,Lusso:2019akb}.

\begin{figure*} 
	\centering
	\includegraphics[width=8.0cm]{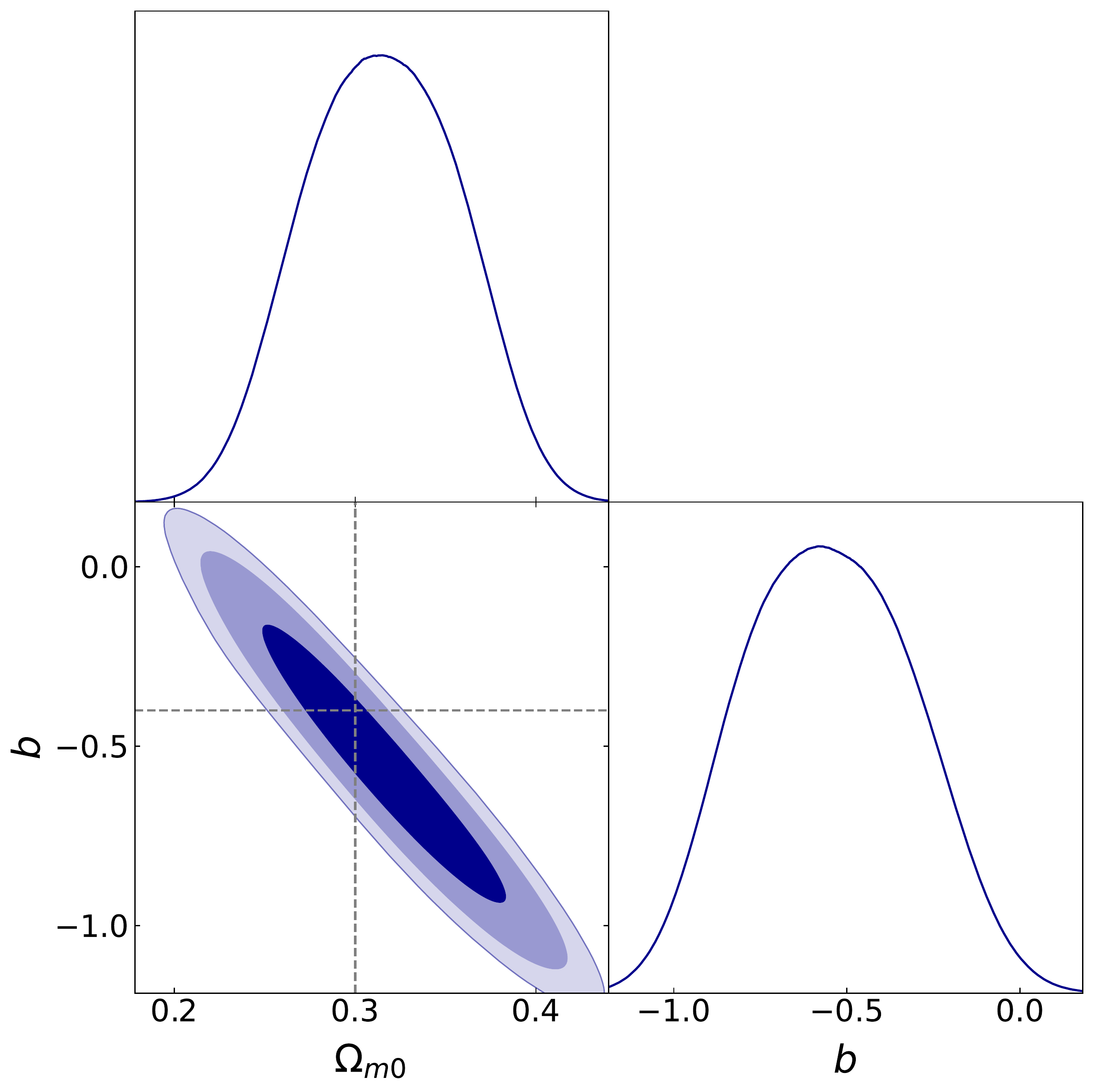}
	\includegraphics[width=8.0cm]{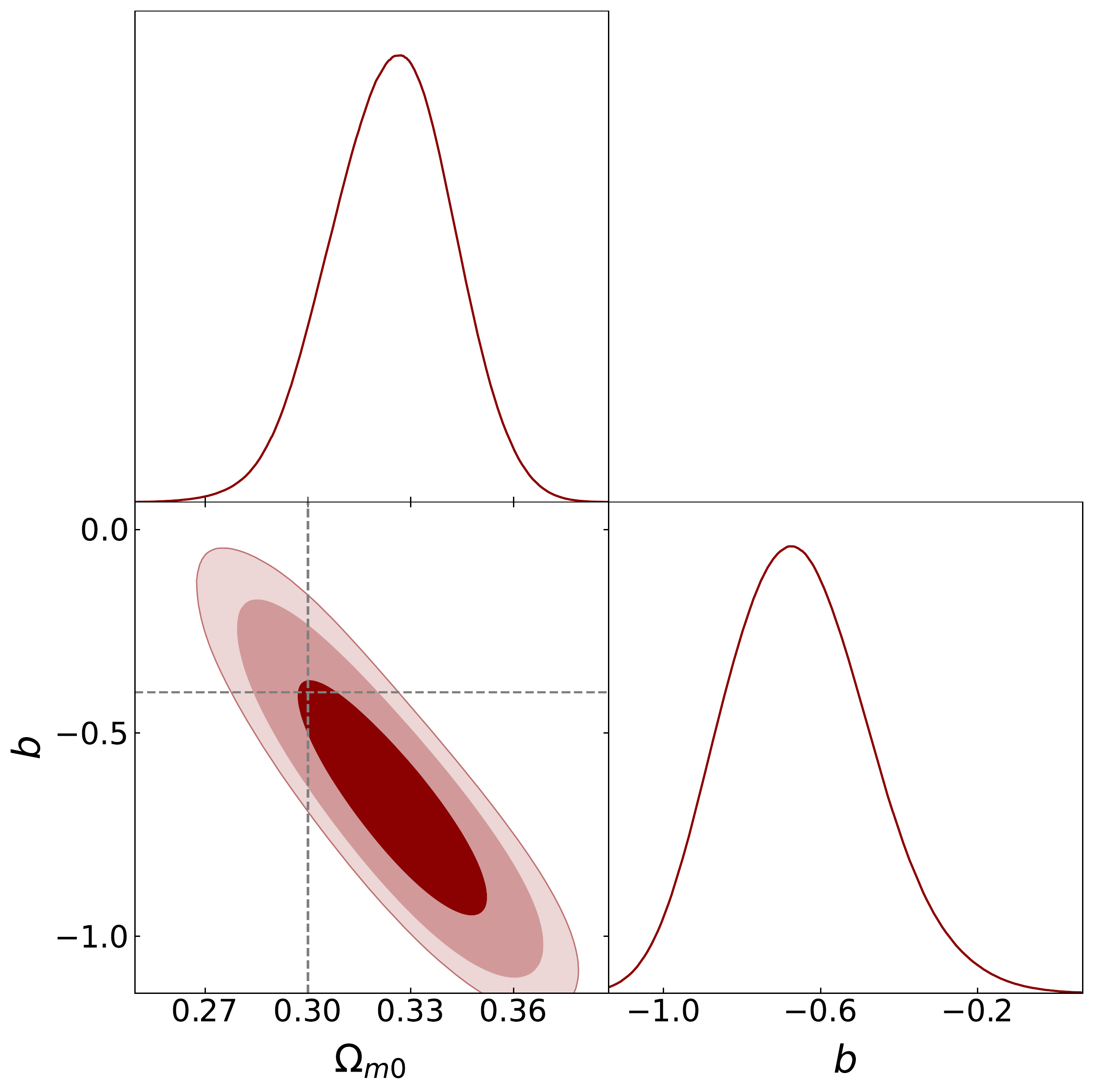}
	\includegraphics[width=8.0cm]{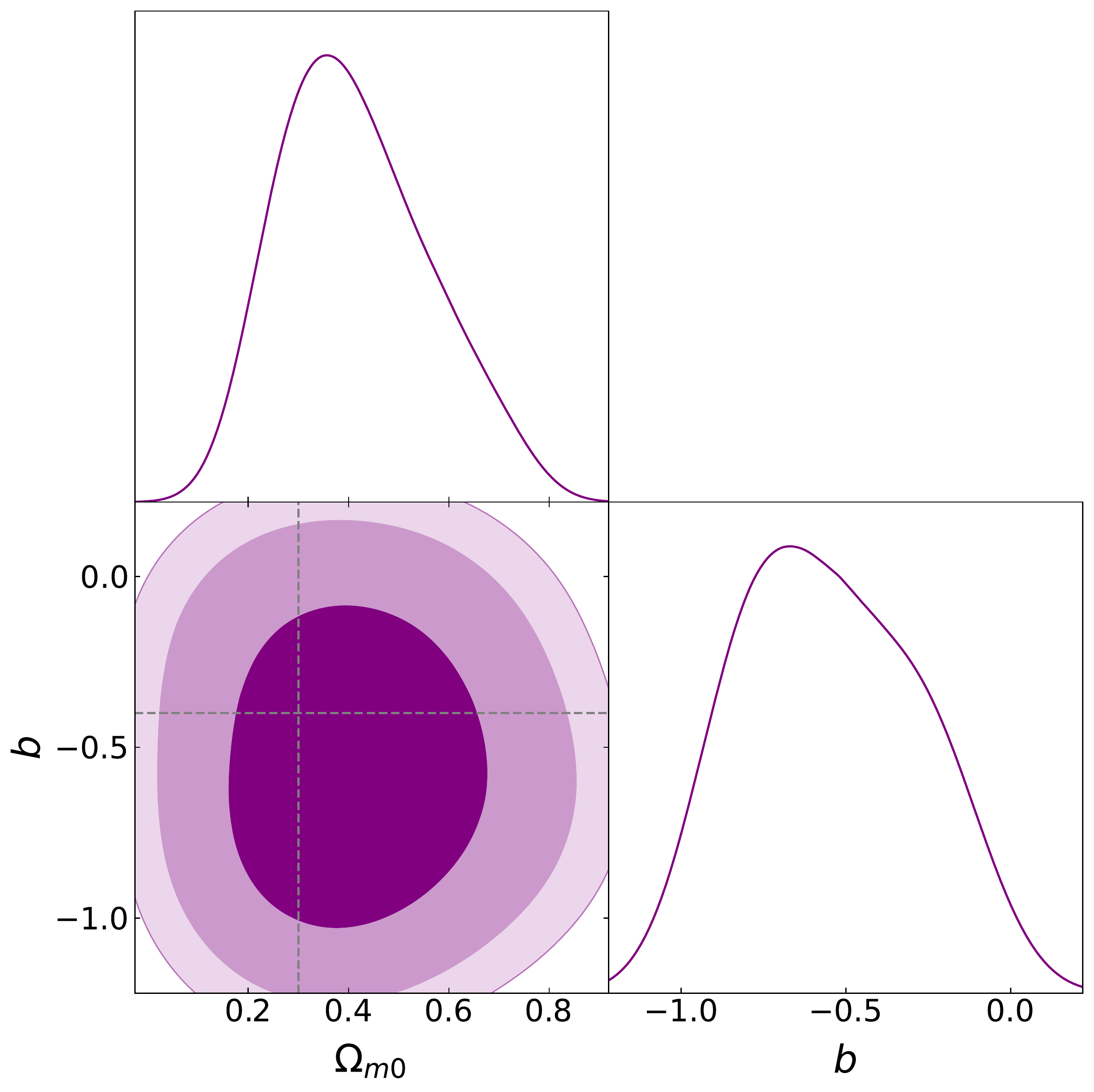}
	\caption{ Up-left panel: Constraints with $1\sigma$, $2\sigma$ and $3\sigma$ confidence regions to the cosmological parameters of the power-law $f(T)$ model, obtained from mock Hubble diagram of SNIa. The horizontal and vertical lines represent the canonical values of $b$ and $\Omega_{m0}$, respectively. Up-right panel: Same as up-left panel, but for mock Hubble diagram of QSOs. Down panel: Same as up-left panel, but for mock Hubble diagram of GRBs.}
	\label{fig:fig1}
\end{figure*}

\begin{table*}
	\centering
	\caption{Best-fit with $1\sigma$, $2\sigma$ and $3\sigma$ confidence intervals for the current values of the cosmographic parameters obtained in the power-law $f(T)$ model, using the mock data for Hubble diagrams of SNIa, QSOs and GRBs.
	}
\fontsize{5pt}{5pt}
	\begin{tabular}{c  c  c c c c }
		\hline \hline
		& $q_0$ & $j_0$ & $s_0$& $l_0$&$m_0$ \\
		\hline
		mock SNIa:&&&&\\
		& $-0.713^{+0.032,+0.060,+0.073}_{-0.032,-0.063,-0.076}$ & $1.20^{+0.10,+0.20,+0.25}_{-0.12,-0.19,-0.23}$ & $-0.49^{+0.23,+0.40,+0.49}_{-0.23,-0.41,-0.47}$ &  $4.27^{+0.93,+1.3,+1.5}_{-0.63,-1.5,-1.8}$ &
		$-10.41^{+0.50,+2.4,+3.8}_{-1.2,-1.7,-2.0}$
		\\\hline
		mock QSOs:&&&&\\
		&
		$-0.734^{+0.041,+0.081,+0.11}_{-0.041,-0.079,-0.097}$ & $1.22^{+0.071,+0.13,+0.17}_{-0.071,-0.13,-0.17}$ & $-0.517^{+0.098,+0.19,+0.25}_{-0.098,-0.19,-0.24}$ &  $4.48^{+0.46,+0.71,+0.86}_{-0.33,-0.81,-1.1}$ &
		$-10.45^{+0.44,+1.3,+1.6}_{-0.7,-0.99,-1.3}$\\
		\hline 
			mock GRBs:&&&&\\
		&
		$-0.48^{+0.22,+0.48,+0.59}_{-0.30,-0.42,-0.50}$ & $1.26^{+0.14,+0.29,+0.37}_{-0.18,-0.27,-0.32}$& $1.19^{+0.73,+1.4,+1.7}_{-0.73,-1.3,-1.7}$ &  $6.0^{+1.5,+4.5,+5.1}_{-2.4,-3.4,-4.0}$&
		$-23.6^{+18,+20,+22}_{-6.8,-33,-40}$\\
		\hline 	
	\end{tabular}\label{tab:tab1}
\end{table*}
\begin{figure*} 
	\centering
	\includegraphics[width=8.5cm]{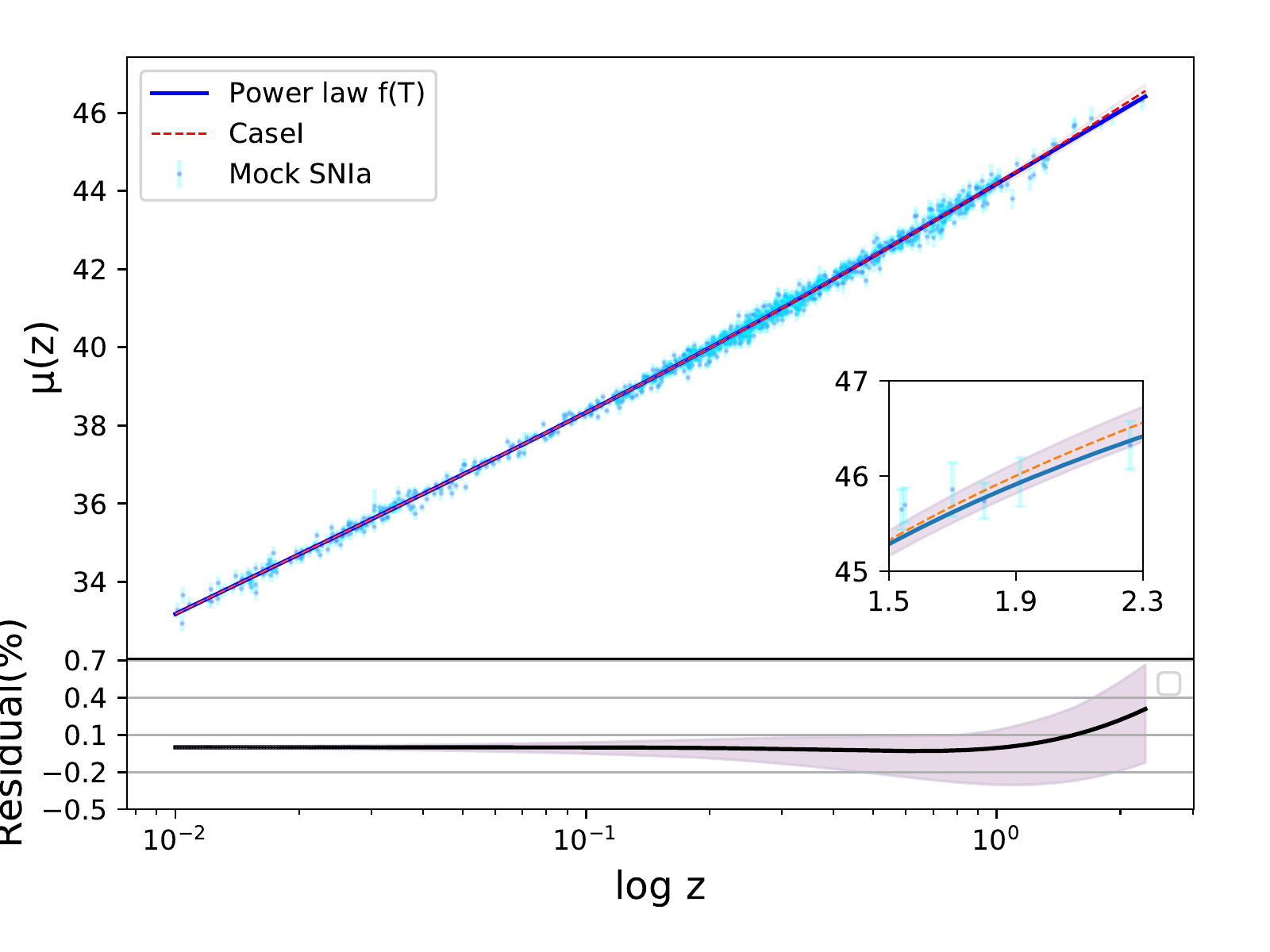}
	\includegraphics[width=8.5cm]{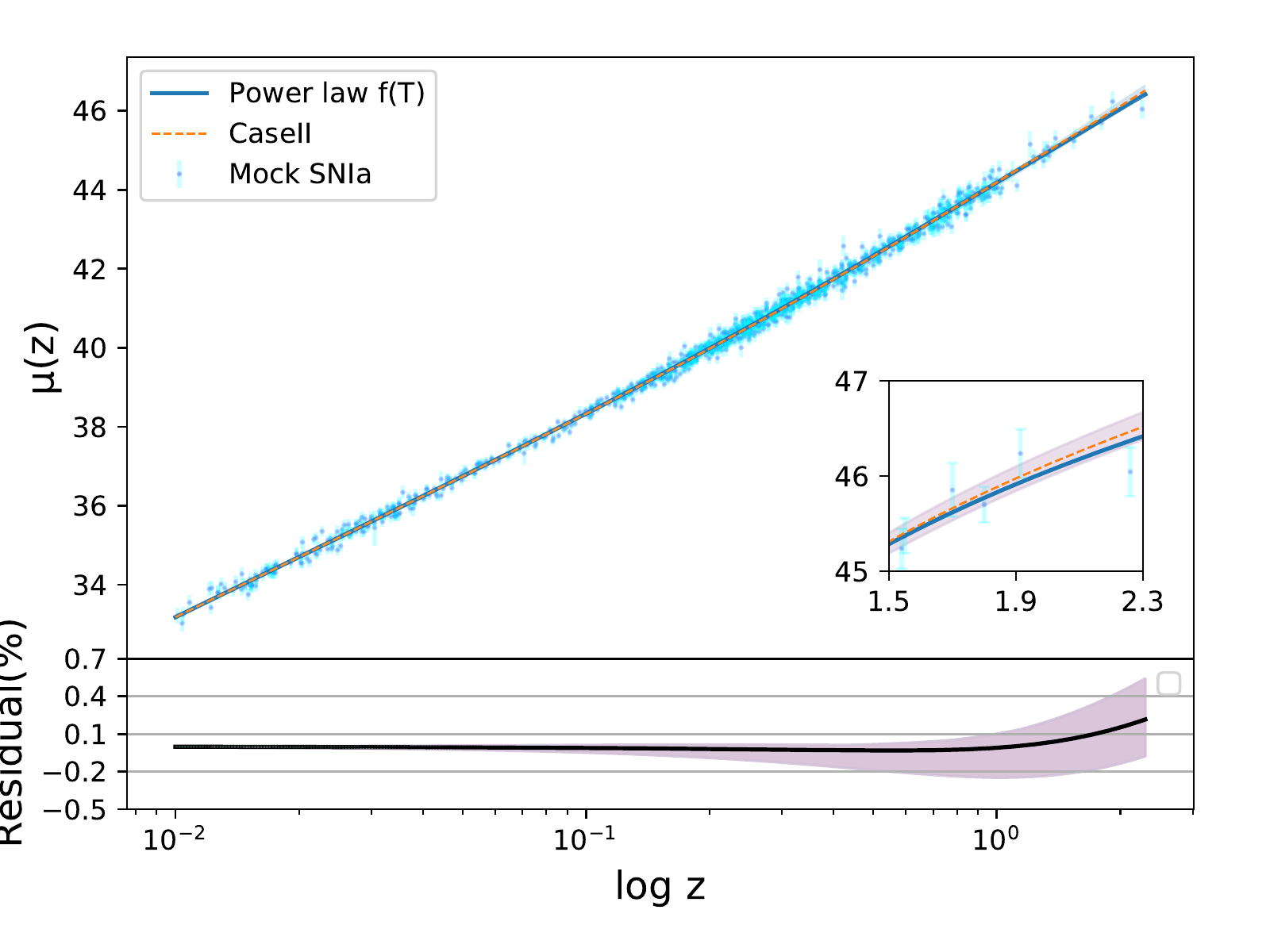}
	\includegraphics[width=8.5cm]{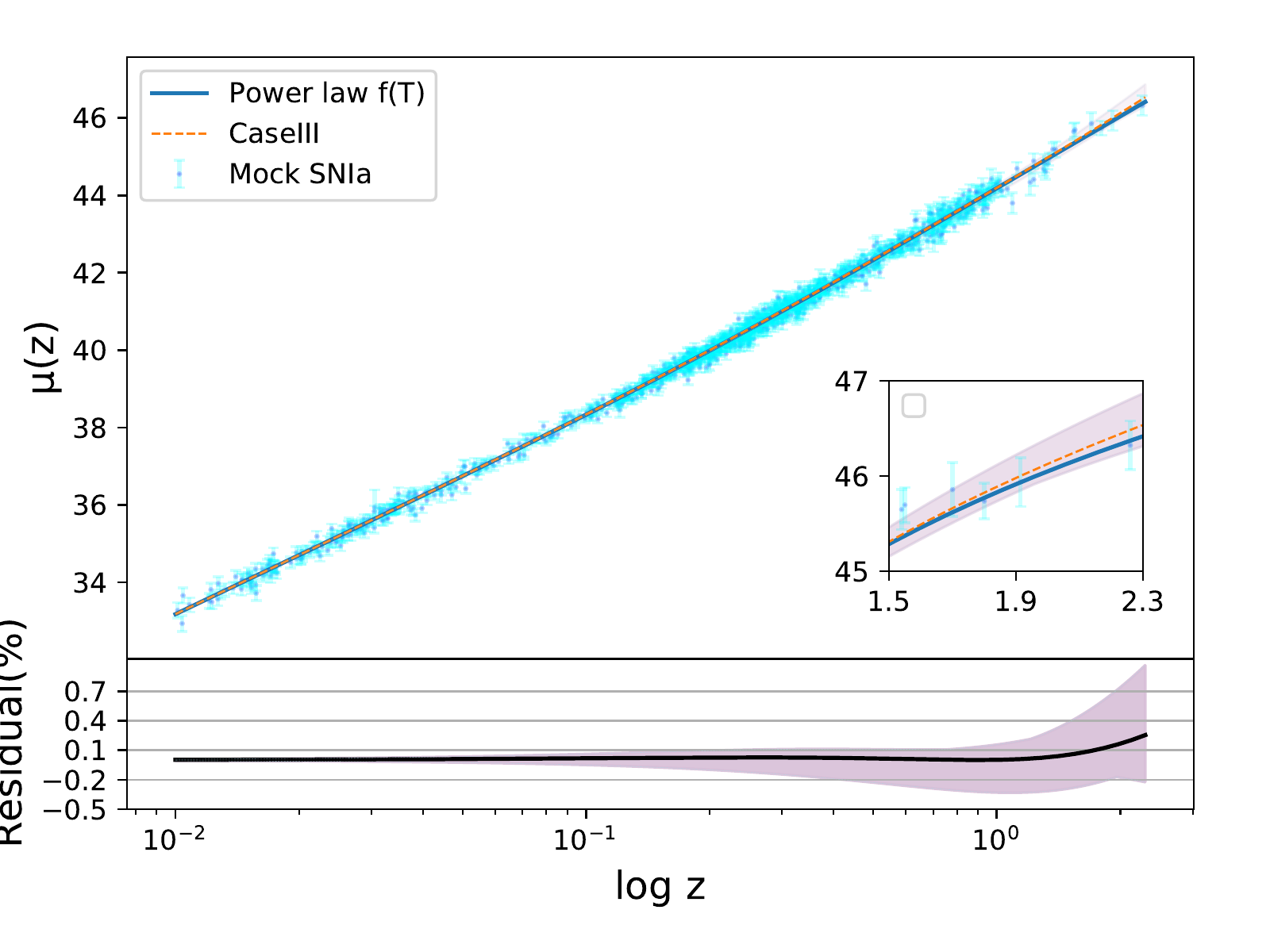}
	\includegraphics[width=8.5cm]{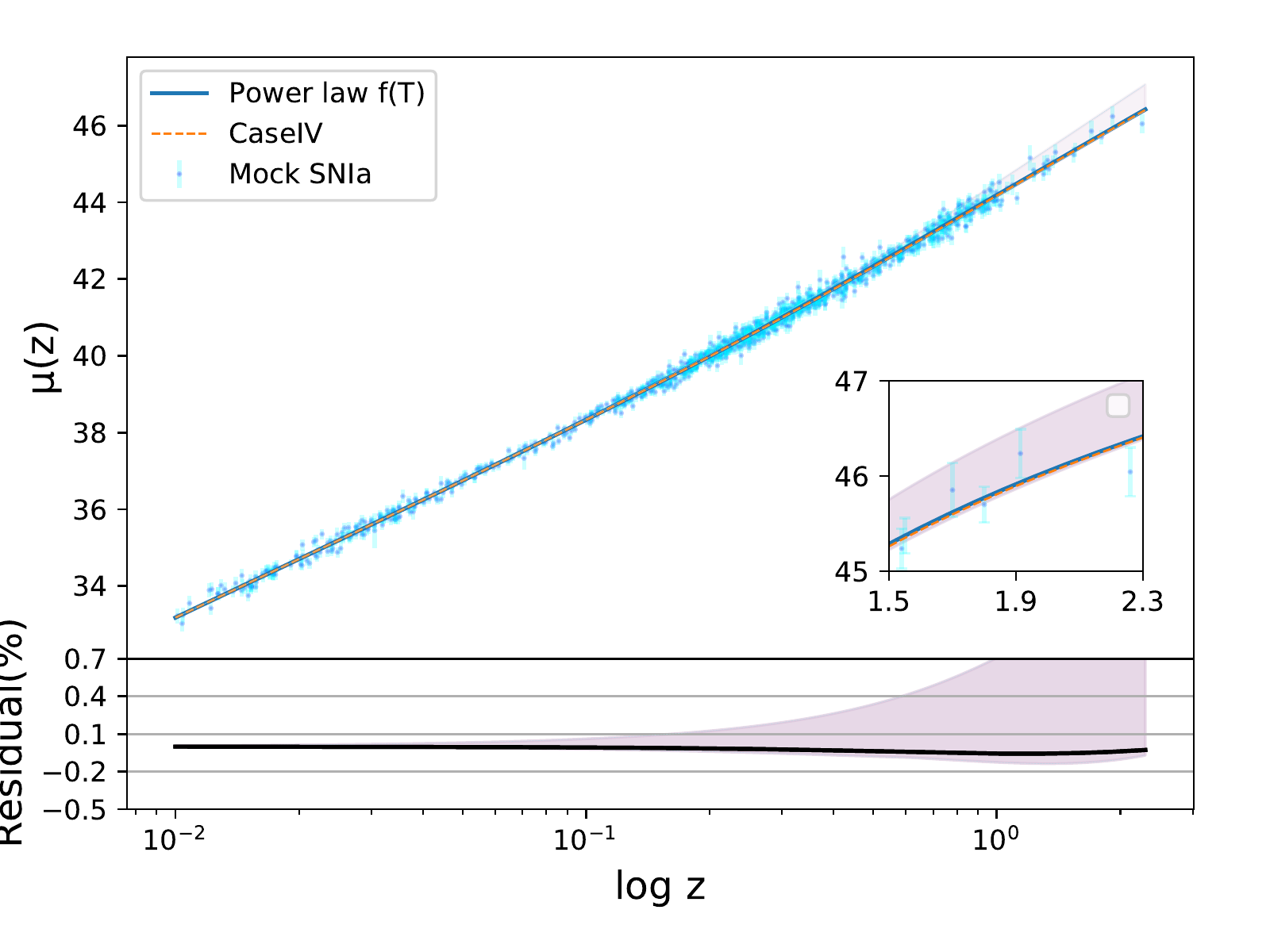}
	\caption{ Upper-left panel: The reconstructed distance modulus in the context of cosmographic method (case I), using mock data for Hubble diagram of SNIa. The distance modulus for the power-law $f(T)$ model is shown by solid-blue line for comparison. Mock data for Hubble diagram of SNIa are shown by colored data points. The percentage difference between the cosmographic method and the $f(T)$ model is shown in residual plane. Upper-right panel: Same as upper-left panel, but for case II of cosmographic method. Lower-left panel: Same as upper-left panel, but for case III of cosmographic method. Lower-right panel: Same as upper-left panel, but for case IV of cosmographic method.}
	\label{fig:fig2}
\end{figure*}

\begin{figure*} 
	\centering
	\includegraphics[width=8cm]{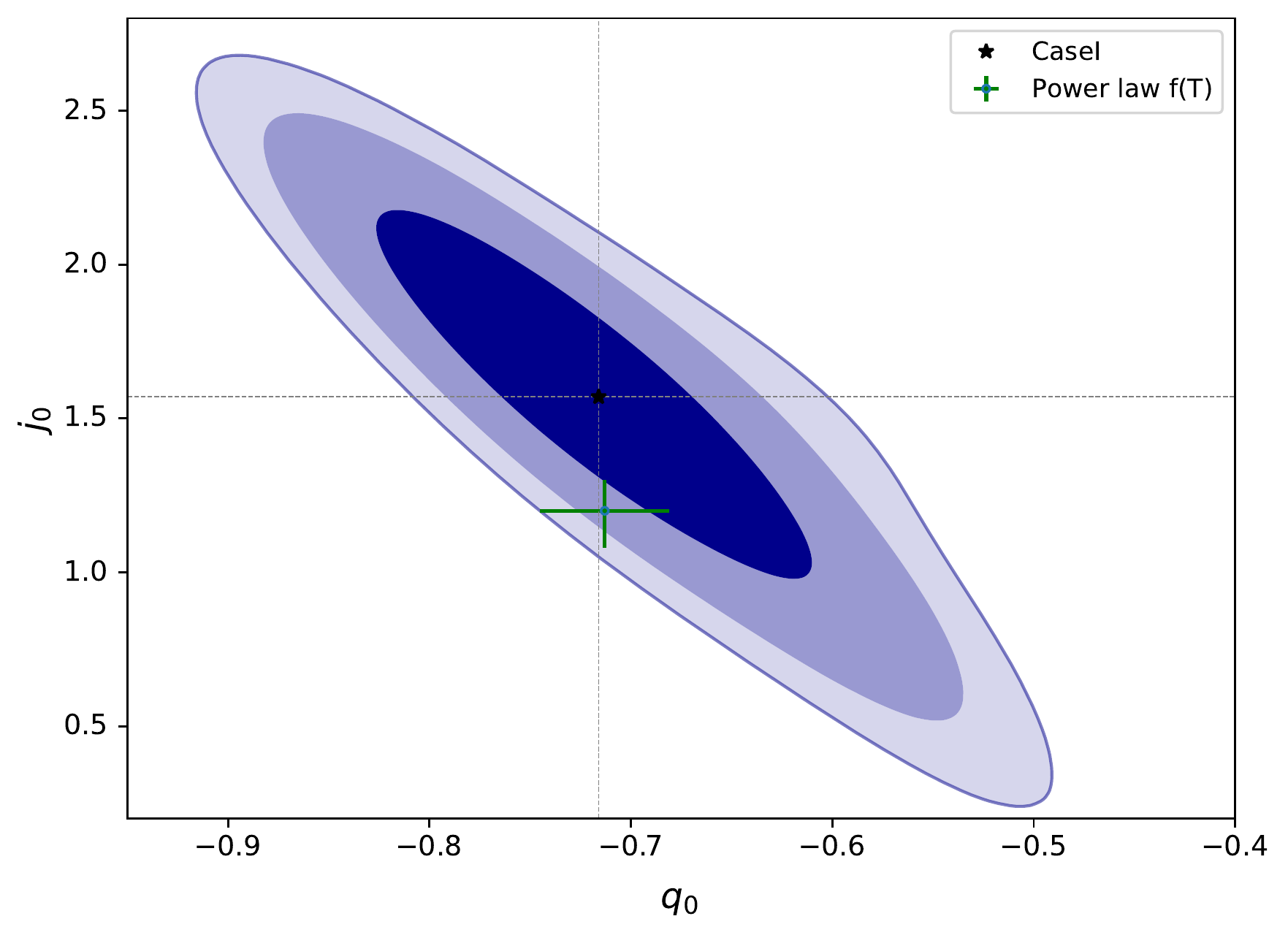}
	\includegraphics[width=8cm]{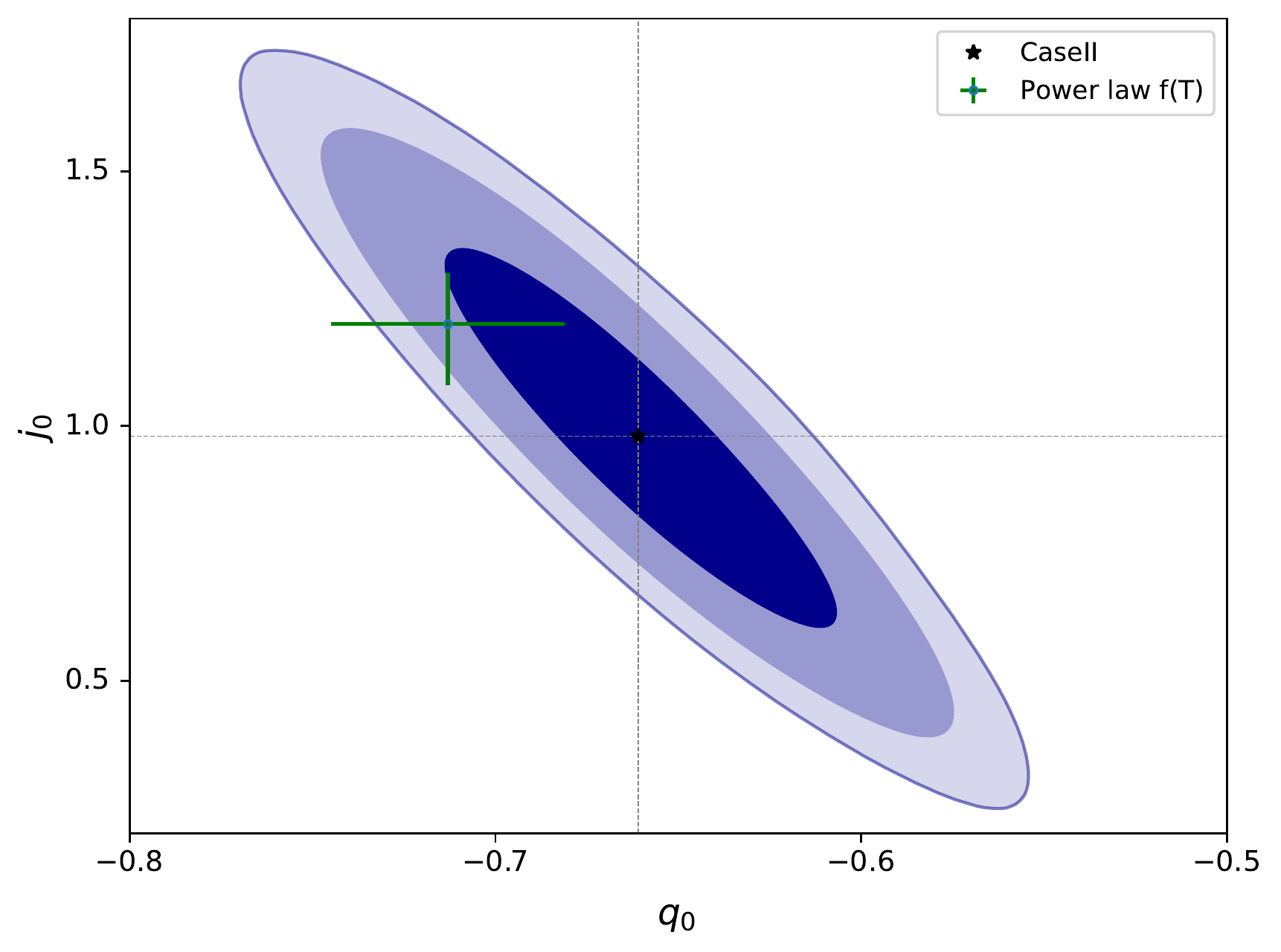}
	\includegraphics[width=8cm]{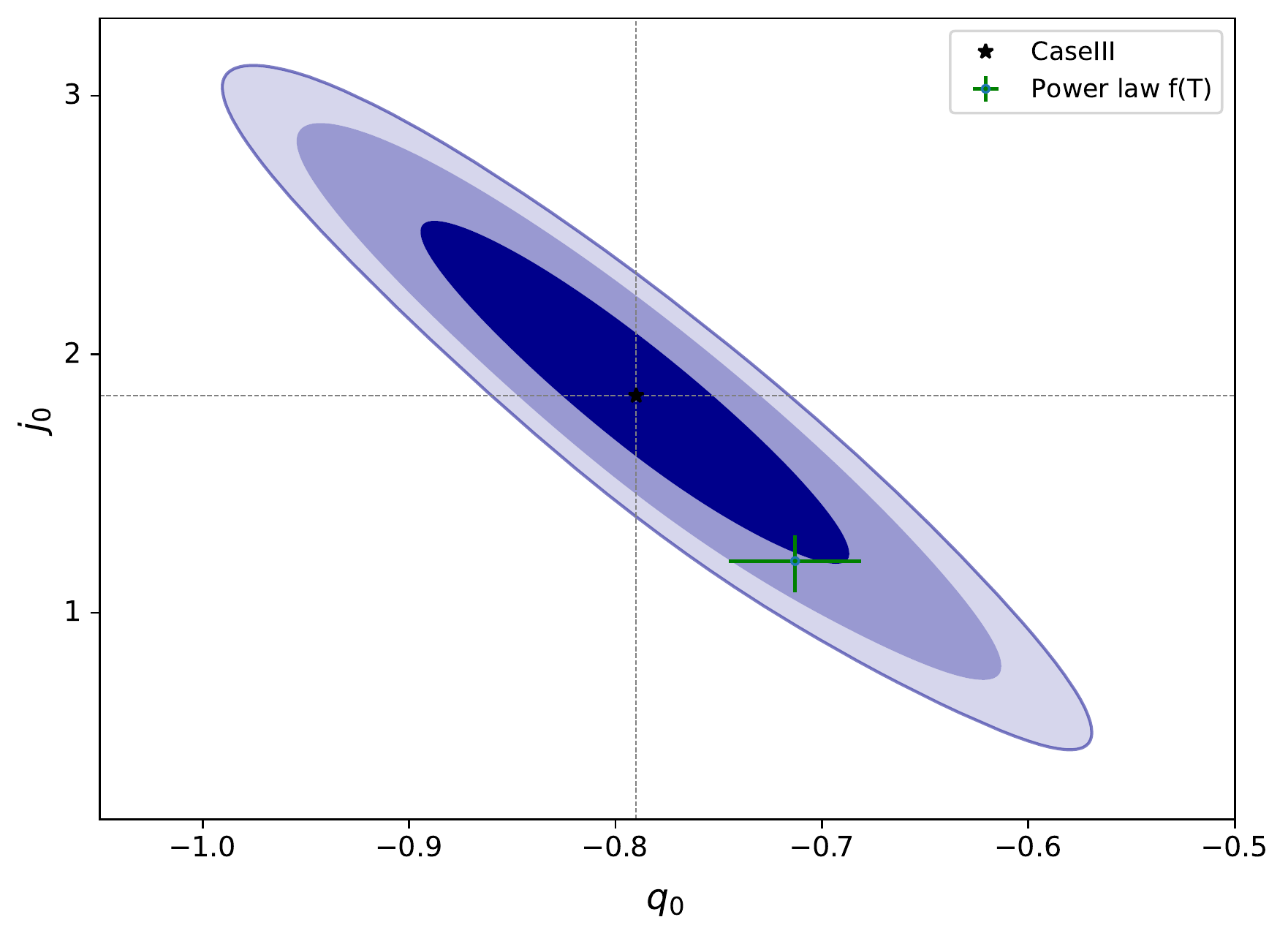}
	\includegraphics[width=8cm]{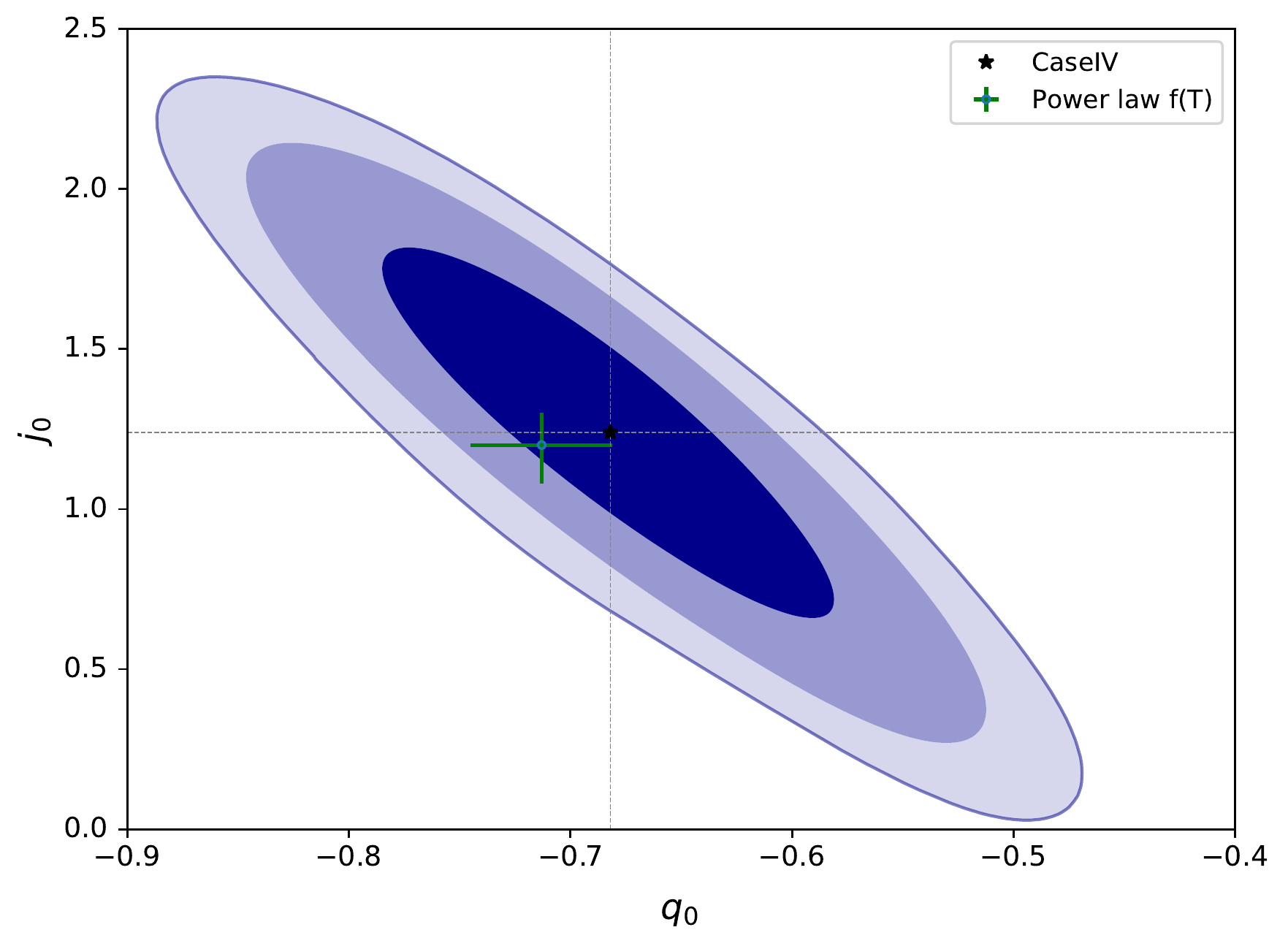}
	\caption{ Upper-Left panel: $1\sigma$, $2\sigma$ and $3\sigma$ Constraints to the cosmographic parameters $q_0$ and $j_0$ in the context of cosmographic method (case I), using mock data for Hubble diagram of SNIa. The vertical and horizontal lines indicate the best-fit of $q_0$ and $j_0$, respectively. The location of the power-law $f(T)$ model is also shown as described in the legend.
		Upper-right panel: Same as upper-left panel, but for case II of cosmographic method. Lower-left panel: Same as upper-left panel, but for case III of cosmographic method. Lower-right panel: Same as upper-left panel, but for case IV of cosmographic method.}
	\label{fig:fig3}
\end{figure*}

\begin{table*}
	\centering
	\caption{Best-fit and  $1\sigma$, $2\sigma$, $3\sigma$ confidence regions for the current values of the cosmographic parameters obtained in the context of cosmographic methods, using mock data for Hubble diagram of SNIa .}
	\begin{tabular}{c c  c  c c c c }
		\hline \hline
		Model & $q_0$ & $j_0$ & $s_0$& $l_0$&$m_0$ \\
		\hline
		
		Case I& $-0.716^{+0.060,+0.14,+0.18}_{-0.075,-0.12,-0.15}$ & $1.57^{+0.40,+0.70,0.87}_{-0.33,-0.77,-1.00}$ & $0.3^{+1.1,+2.2,+2.3}_{-1.1,-2.3,-2.3}$ &  $0.73^{+1.4, +2.1,+2.5}_{-0.91,-2.5,-2.9}$ & --
		\\\hline
		Case II	&$-0.661^{+0.033,+0.066,+0.078}_{-0.033,-0.063,-0.086}$ & $0.98^{+0.23,+0.46,+0.59}_{-0.23,-0.44,-0.50}$ & $-1.23^{+0.36,+1.1,+1.3}_{-0.43,-0.78,-0.81}$ &  $2.7^{+2.0,+3.8,+4.7}_{-2.0.-3.8,-4.7}$ &
		$-11.5^{+4.2,+11.0,+15.0}_{-7.9,-8.6,-9.1}$\\
		\hline \hline
		Case III	&$-0.790^{+0.066,+0.13,+0.18}_{-0.066,-0.13,-0.16}$ & $1.84^{+0.42,+0.83,+0.96}_{-0.42,-0.90,-1.1}$ & $0.71^{+0.67,+0.93,+1.1}_{-0.42,-1.1,-1.5}$ &  $4.1^{+1.0,+1.8,+1.9}_{-1.0.-1.8,-2.0}$ &
		$-$\\
		\hline 
		Case IV&$-0.682^{+0.059,+0.13,+0.16}_{-0.066,-0.12,-0.15}$ & $1.24^{0.38,+0.62,+0.82}_{-0.32,-0.71,-0.93}$ & $-0.44^{+0.61,+1.3,+1.7}_{-0.61,-1.3,-1.6}$ &  $2.3^{+1.8,+3.3,+4.0}_{-2.0.-3.2,-3.8}$ &
		$-10.1^{+5.5,+9.7,+11}_{-5.5,-9.7,-10}$\\
		\hline \hline
	\end{tabular}\label{tab:tab2}
\end{table*}

\section{Observational constraints on cosmographic parameters}\label{sect:result}
In this section, we use Hubble diagrams for the low and high redshift SNIa, QSOs and GRBs, to constrain the cosmographic parameters in the context of model-independent cosmographic method. We also perform the same analysis and obtain the observational constraints on the cosmographic parameters of the $f(T)$ model. Before using the observational data, we should check the validity of the cosmographic method at higher redshifts. Using the generated mock data, we first calculate the percentage difference between the reconstructed distance modulus in model-independent cosmographic method and one that from $f(T)$ model.
We test both cosmographic approaches based on the linear Taylor approximation and the PADE polynomials presented in the previous section. Let us start with mock data generated for Hubble diagrams of SNIa, QSOs \jt{and GRBs}, and show how we can test the validation of the cosmographic methods.
\subsection{Mock data and cosmographic methods}
Here we generate mock data for Hubble diagrams of SNIa, QSOs \jt{and GRBs} based on the power-law $f(T)$ cosmology. For this purpose, we set the cosmological parameters  of the model as $\Omega_{m0}=0.3$ and $b=-0.4$. Using these canonical values, 
we obtain the current value of the effective EoS parameter of the $f(T)$ model as $w_{DE,0}=-1.086$, which means that the $f(T)$ model is currently varying in the phantom regime and deviates sufficiently from $w_{\Lambda}=-1.0$. We note that recent observational studies prefer the phantom regime for the effective EoS parameter of the power-law $f(T)$ model \citep{Basilakos:2016xob,Nesseris:2013jea}. 
Using the canonical values $\Omega_{m0}=0.3$ and $b=-0.4$ and Eq. (\ref{eq:eq9}), we calculate the distance modulus $\mu_{T}$ in the context of power-law $f(T)$ model based on the following relation:
\begin{eqnarray}\label{eq:muth}
\mu_{T}(z)=5 \log_{10}[(1+z)\int_{0}^{z}\frac{dz}{E(z)}]+\mu_{0}, 
\end{eqnarray}
where $\mu_{0}=42.384-5 \log_{10}(h)$ is the current value of the distance modulus and $h=H_0/100$. Then, using the function $\mu_{T}(z)$ at specific redshift $z_i$, we generate mock  data for distance modulus ($\mu(z_i)$,$\Delta\mu(z_i)$). Here $\Delta\mu(z_i)$ is the error bar of the distance modulus $\mu(z_i)$. The distance modulus $\mu(z_i)$ is chosen using the normal distribution with the mean value $\mu_T(z_i)$ \citep[see also][]{Yang:2019vgk,Banerjee:2020bjq}. We set redshifts $z_i$ to the observational redshifts in which the SNIa, QSOs \jt{and GRBs} have been observed. We also set $\Delta\mu(z_i)$ to the observational errors for the Hubble diagrams for SNIa, QSOs \jt{and GRBs} data. In the case of SNIa, we use the Pantheon catalogue which contains 1048 data points in the redshift range of $0.03 < z < 2.3$ \citep{Scolnic:2017caz}. 
In the case of QSOs, we use the main quasar sample composed of 1598 data points in the redshift range of $0.04 < z < 5.1$ \citep{Lusso:2017hgz}. \jt{Finally for GRBs, we use the collected data in the redshift range $0.03<z<9.3$ form \citep{Escamilla-Rivera:2021vyw}.}
 Having mock data, we first perform the minimization of the $\chi^2$ function based on the MCMC algorithm for the power-law $f(T)$ model to obtain the best fit values of the free parameters $\Omega_{m0}$ and $b$. If the process of the generation of mock data is correct, we should expect that the best-fit values of the free parameters $\Omega_{m0}$ and $b$  consistently closed to their canonical values. In other words, since we are generating mock data using the canonical values of $\Omega_{m0}$ and $b$, the best-fit values of these parameters should not significantly deviate from the canonical values. Our results are presented in Fig. (\ref{fig:fig1}). The up-left (up-right) panel of Fig. (\ref{fig:fig1}), shows constraints on the parameters $\Omega_{m0}$ and $b$ within $3\sigma$ confidence levels obtained from  mock data for SNIa (QSOs).\jt{ The down panel shows the confidence regions obtained from mock GRBs data.} For all panels, we explicitly observe that the canonical values $\Omega_{m0}=0.3$ and $b=-0.4$ are within in $1\sigma$ confidence interval. We mention that in the case of mock SNIa sample, the best-fit values of $\Omega_{m0}$ and $b$ with $3\sigma$ uncertainties are obtained as $\Omega_{m0}=0.315^{+0.042,+0.075,+0.098}_{-0.042,-0.076,-0.099}$ and $b=-0.55^{+0.24,+0.46,+0.58}_{-0.24,-0.45,-0.55}$. In the case of mock QSOs sample, we obtain $\Omega_{m0}=0.324^{+0.017,+0.033,+0.040}_{-0.017,-0.035,-0.047}$ and $b=-0.66^{+0.18,+0.37,+0.50}_{-0.18,-0.35,-0.43}$. \jt{ Finally in the case of mock GRBs, the best-fit values are $\Omega_{m0}= 0.41^{+0.13,+0.31,+0.37}_{-0.18,-0.27,-0.29}$ and $-0.55^{+0.27,+0.54,+0.67}_{-0.32,-0.50,-0.62}$.}\\
We now calculate the best-fit values of the cosmographic parameters for the power-law $f(T)$ model, using the MCMC chain for $\Omega_{m0}$ and $b$ in Eq.(\ref{eq:eq20}). Our numerical results for mock SNIa, QSOs \jt{ and GRBs} samples are presented in Tab. (\ref{tab:tab1}). In the next step, we perform the MCMC analysis to find the best- fit values of cosmographic parameters in the context of model-independent cosmographic approach by using  mock data for SNIa, QSOs \jt{ and GRBs} discussed above. In the case of model-independent cosmographic method based on the Taylor approximation, we substitute the reconstructed Hubble function from Eq.(\ref{eq:eq14}) into Eq.(\ref{eq:muth}), where the coefficients of $E(z)$ in Eq.(\ref{eq:eq14}) are related to the cosmographic parameters (as free parameters) via Eq.(\ref{eq:eq15}). Notice that here we consider two cases of Taylor expansions up to the orders of $y^4$ (Case I) and $y^5$ (Case II). Obviously in Case II, we have one more cosmographic parameter $m_0$ and one more algebraic term compared to Case I. In the case of cosmographic approach based on the PADE approximation of Hubble parameter, we use Eqs. (\ref{eq:eq17} \& \ref{eq:eq18}) and insert them in the Eq.(\ref{eq:muth}) to reconstruct the distance modulus respectively for $P_{2,2}$ (Case III) and $P_{3,2}$ (Case IV). On the other hand, inserting the exact Hubble parameter of the power-law $f(T)$ model (Eq.\ref{eq:eq9}) into Eq. (\ref{eq:muth}), we can directly obtain the distance modulus of SNIa, QSOs \jt{ and GRBs} in the context of the power-law $f(T)$model. It is mentioned that the free parameters $\Omega_{m0}$ and $b$ in the Hubble parameter (Eq.\ref{eq:eq9}) are fixed to their best-fit values obtained from MCMC analysis obtained from mock SNIa, QSOs \jt{ and GRBs} data.

\begin{figure*} 
	\centering
	\includegraphics[width=8.5cm]{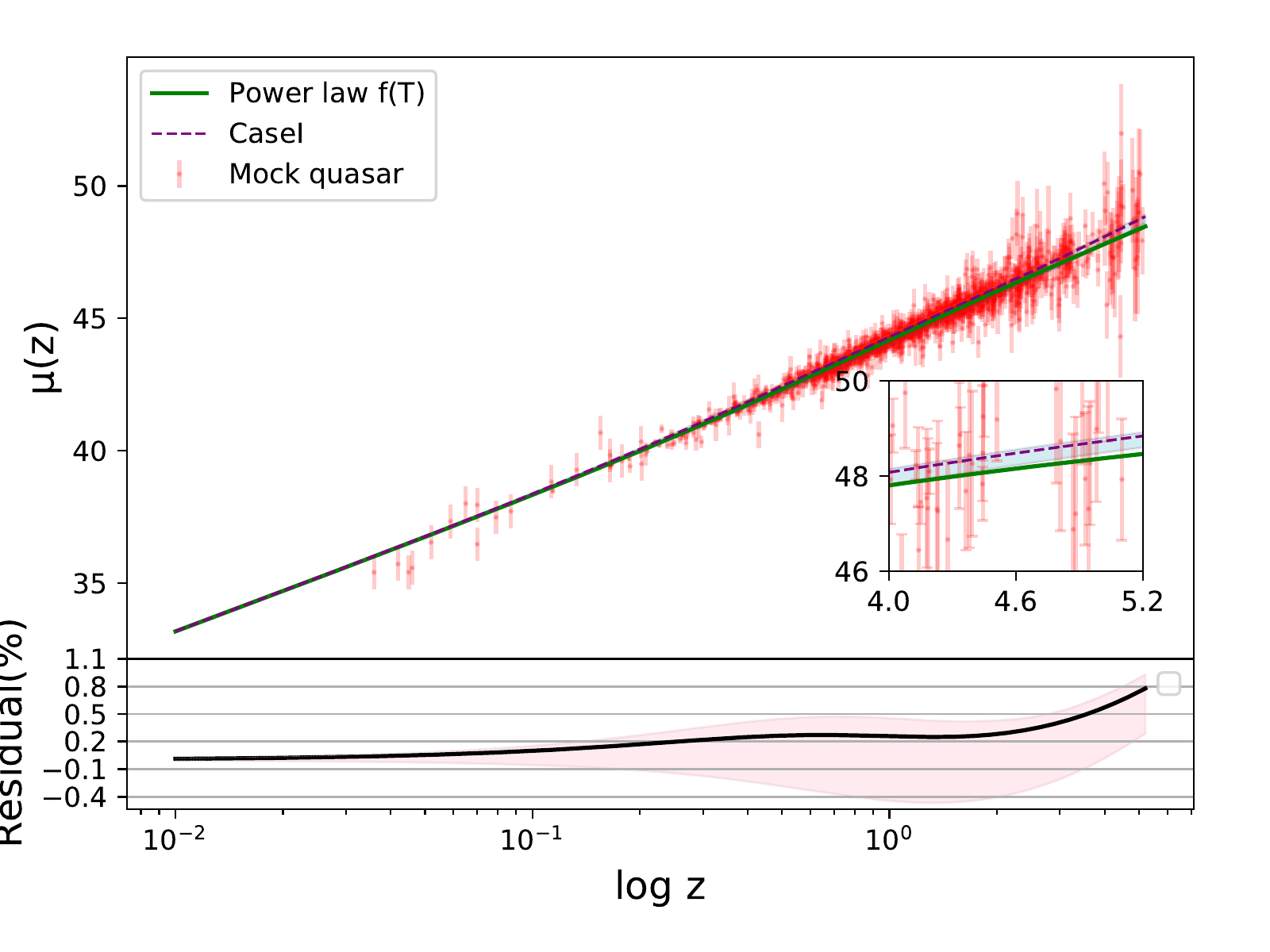}
	\includegraphics[width=8.5cm]{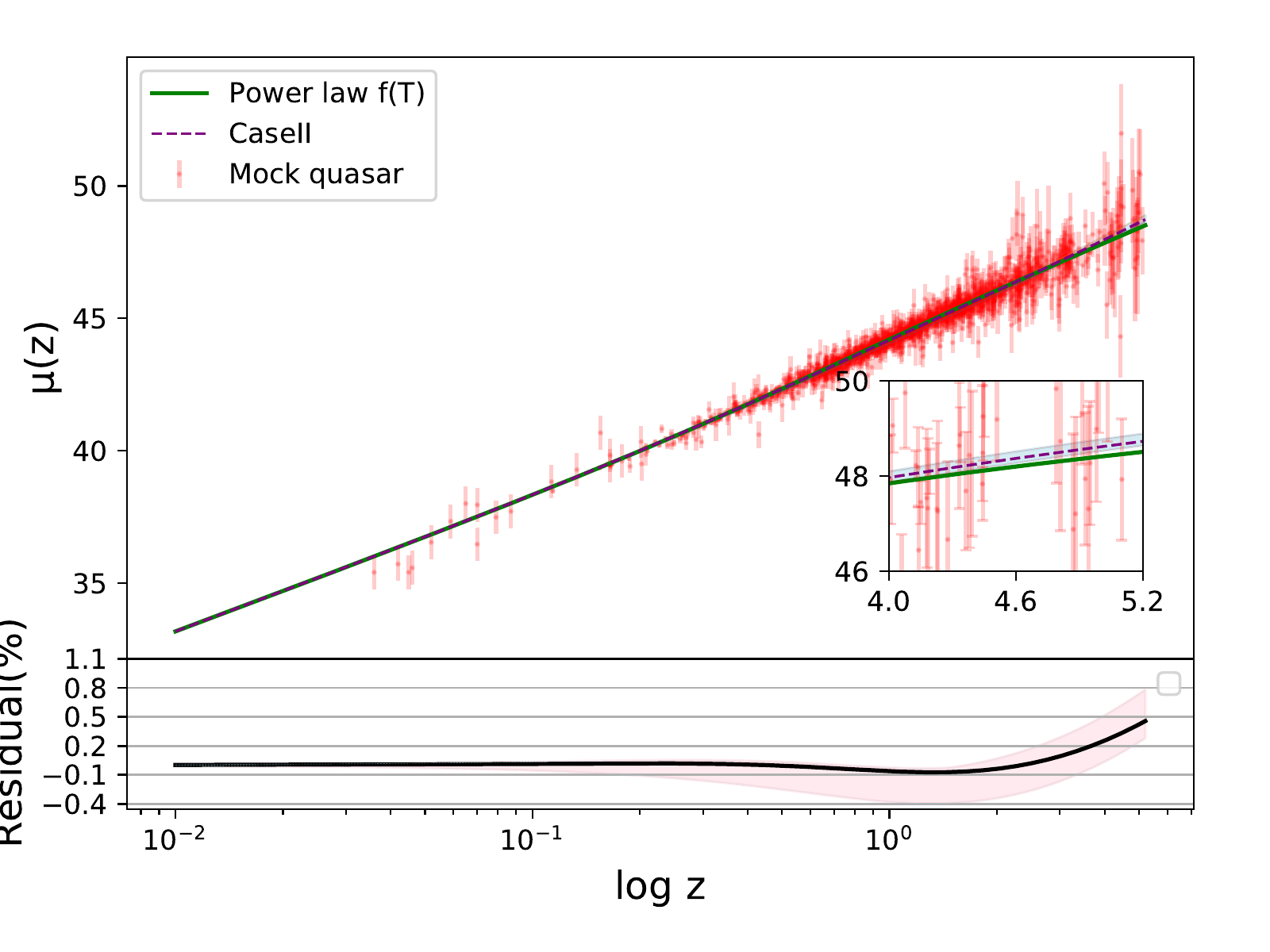}
	\includegraphics[width=8.5cm]{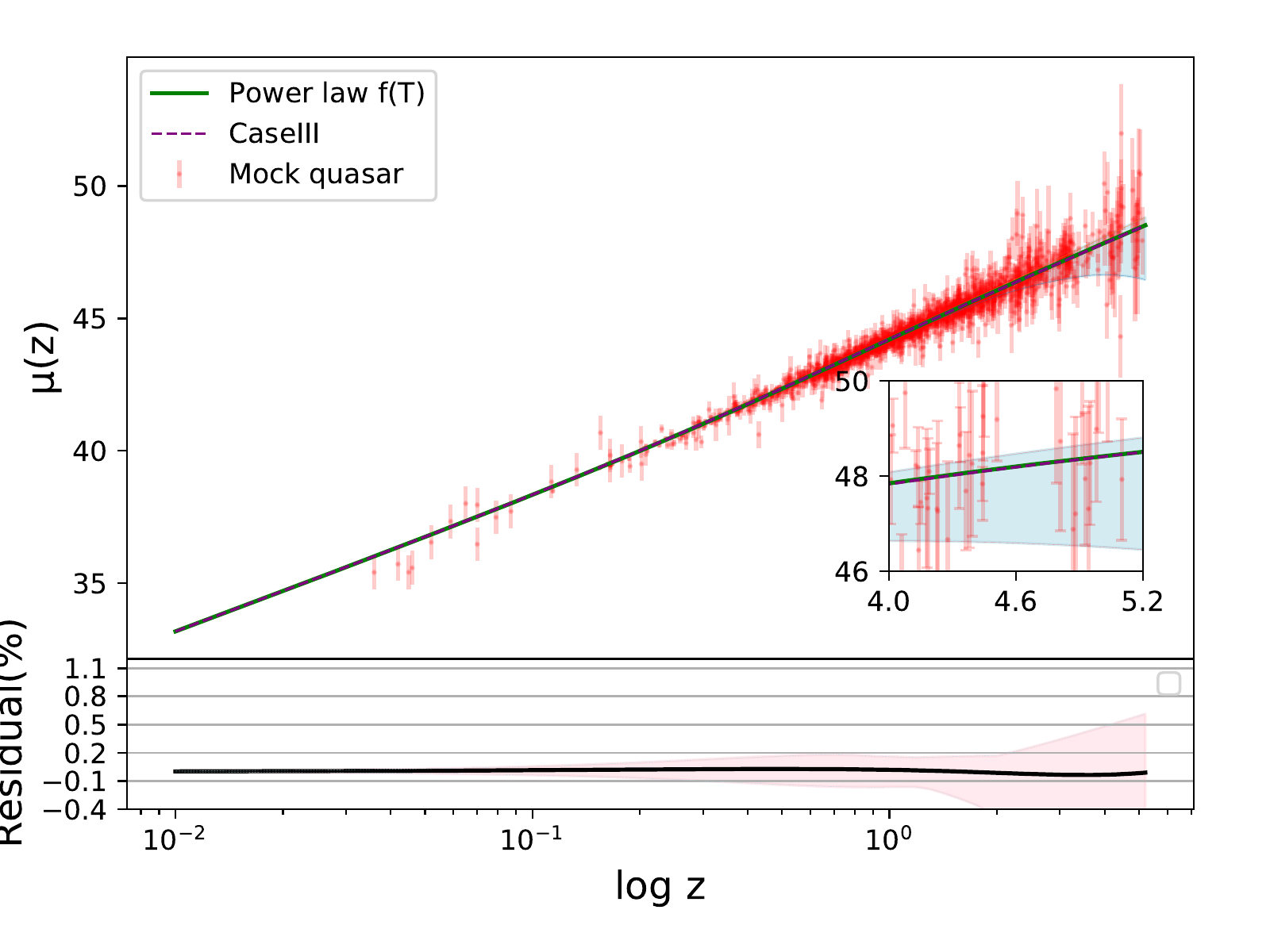}
	\includegraphics[width=8.5cm]{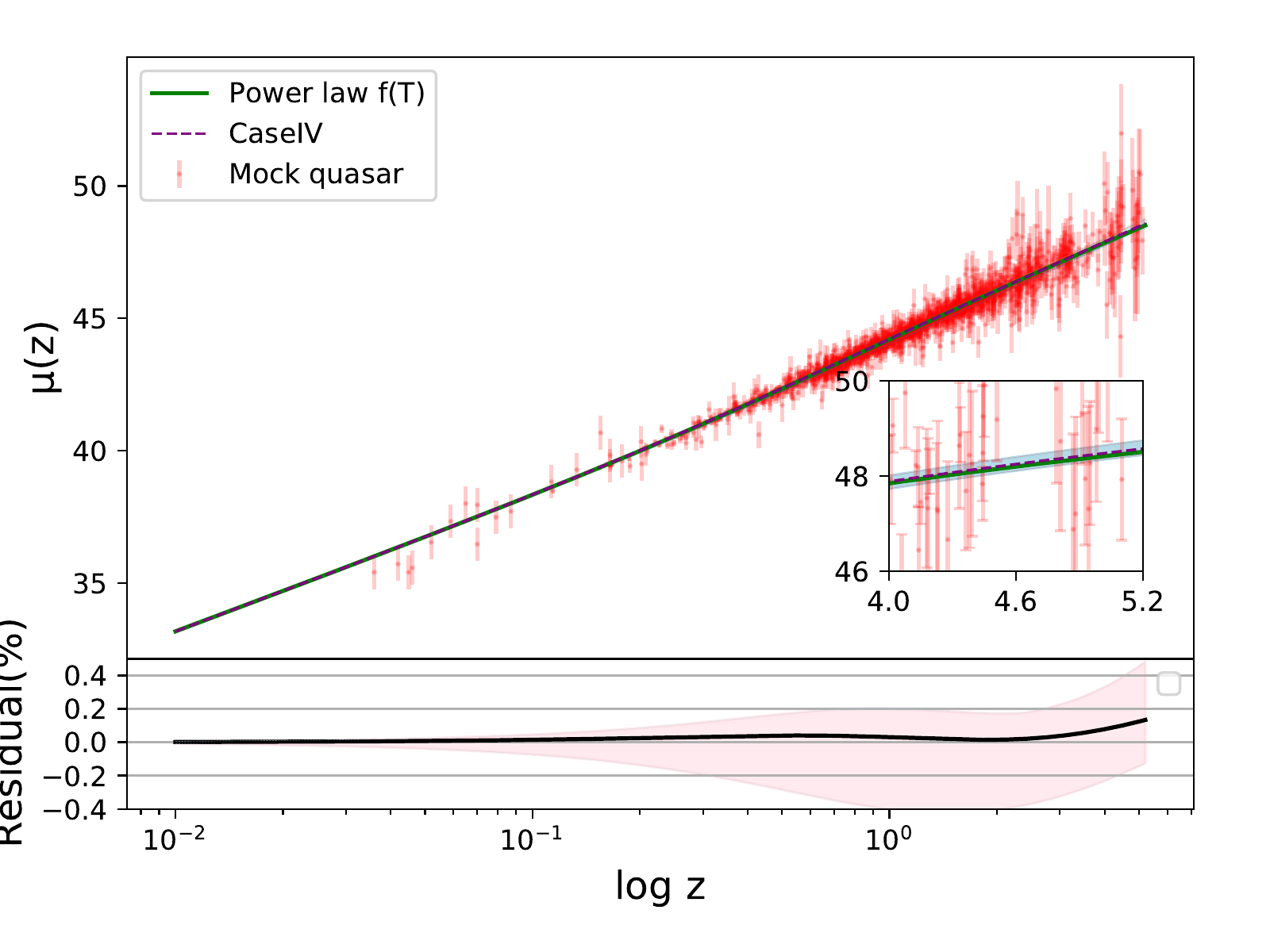}
	\caption{Same as Fig. (\ref{fig:fig2}), but by using mock data for the Hubble diagram of QSOs.}
	\label{fig:fig4}
\end{figure*}

\begin{figure*} 
	\centering
	\includegraphics[width=8.5cm]{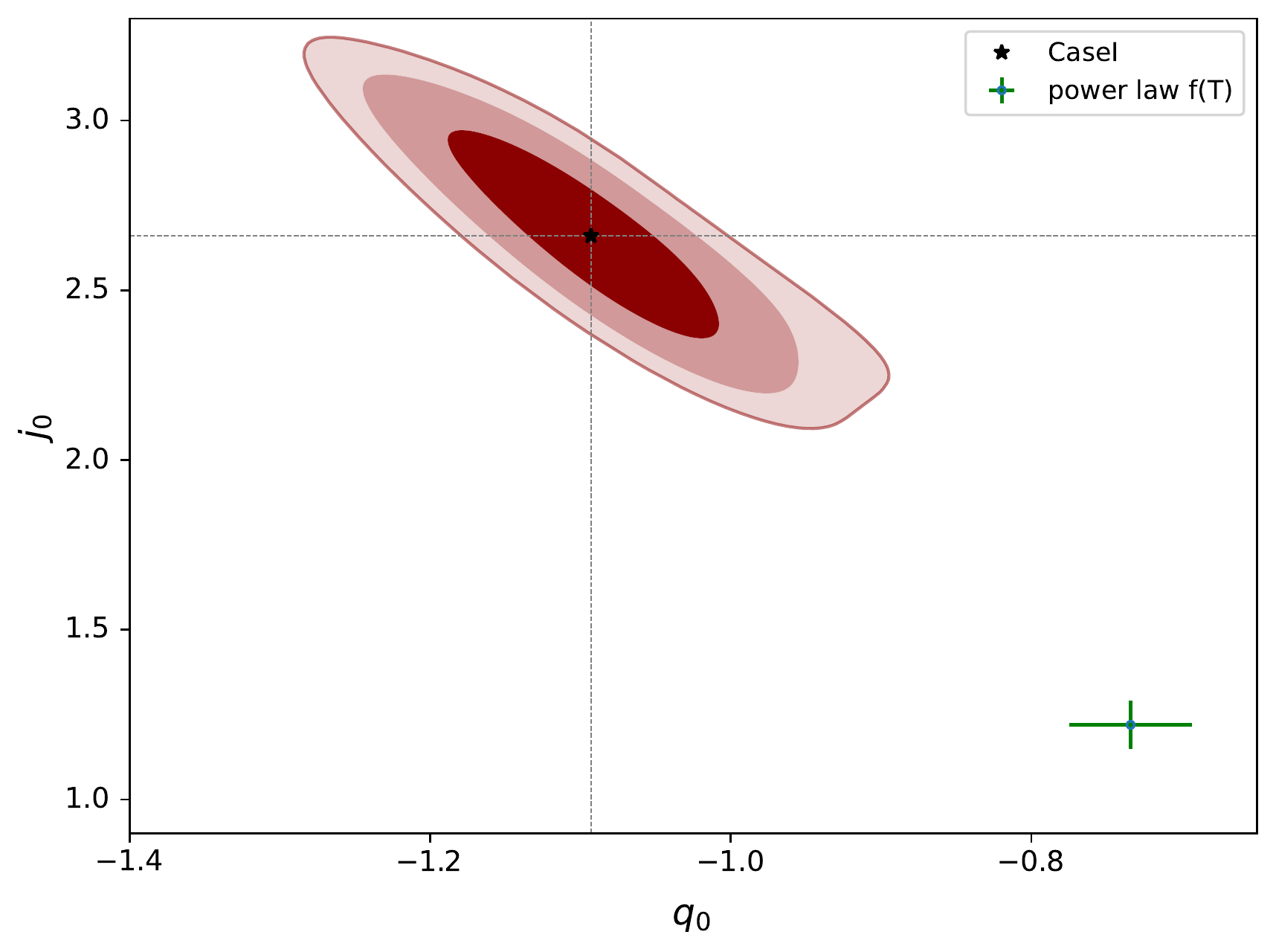}
	\includegraphics[width=8.5cm]{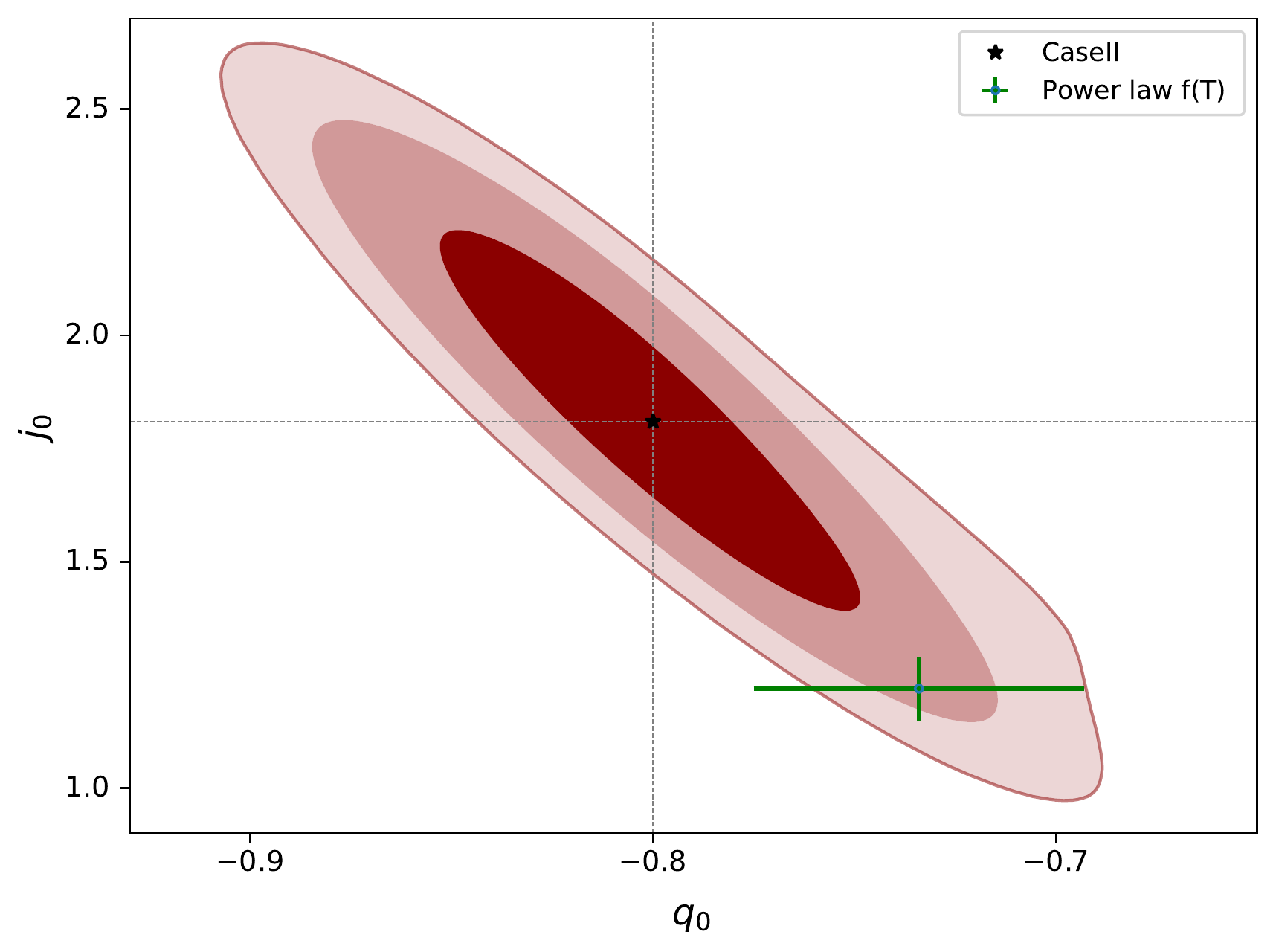}
	\includegraphics[width=8.5cm]{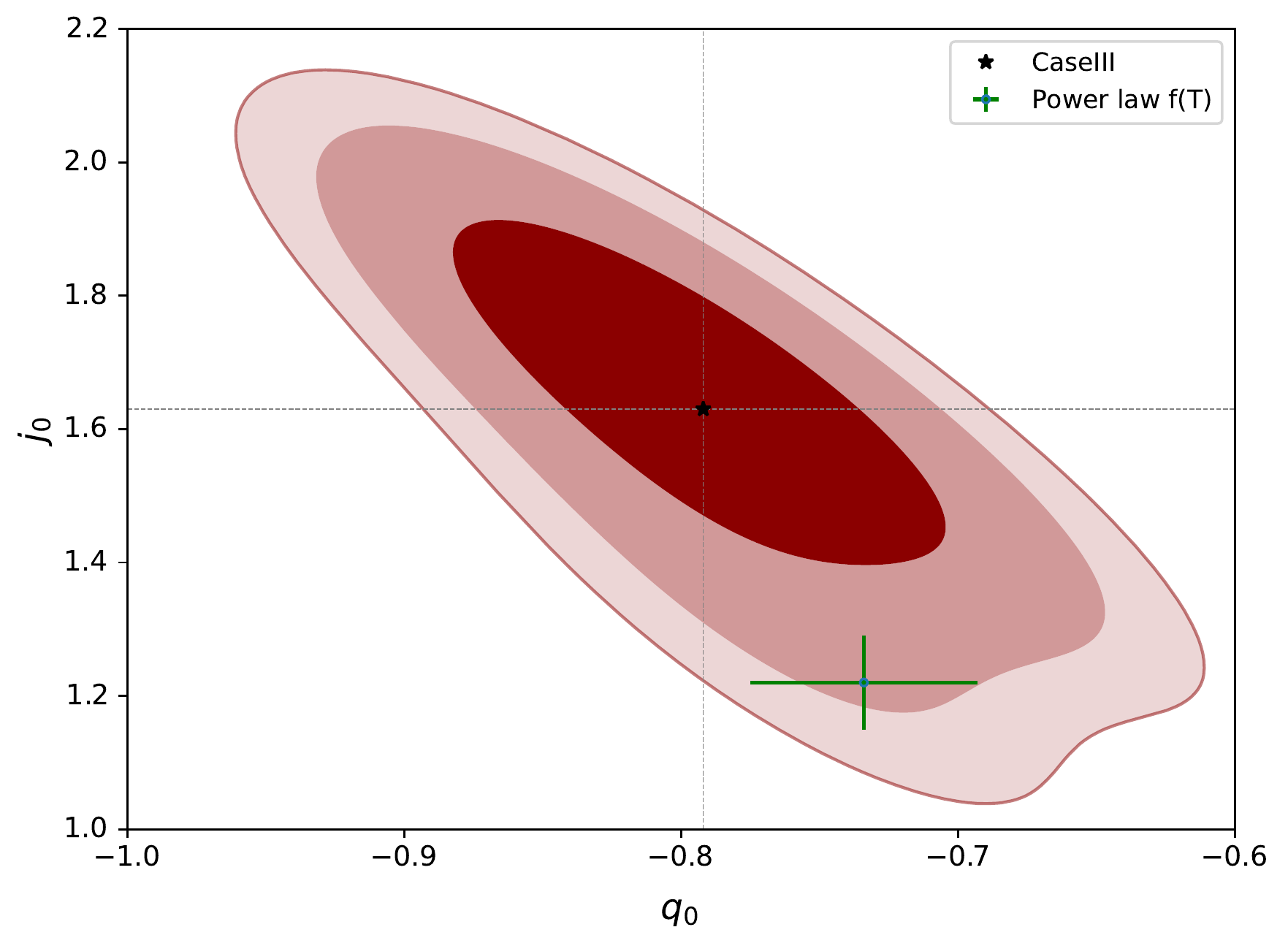}
	\includegraphics[width=8.5cm]{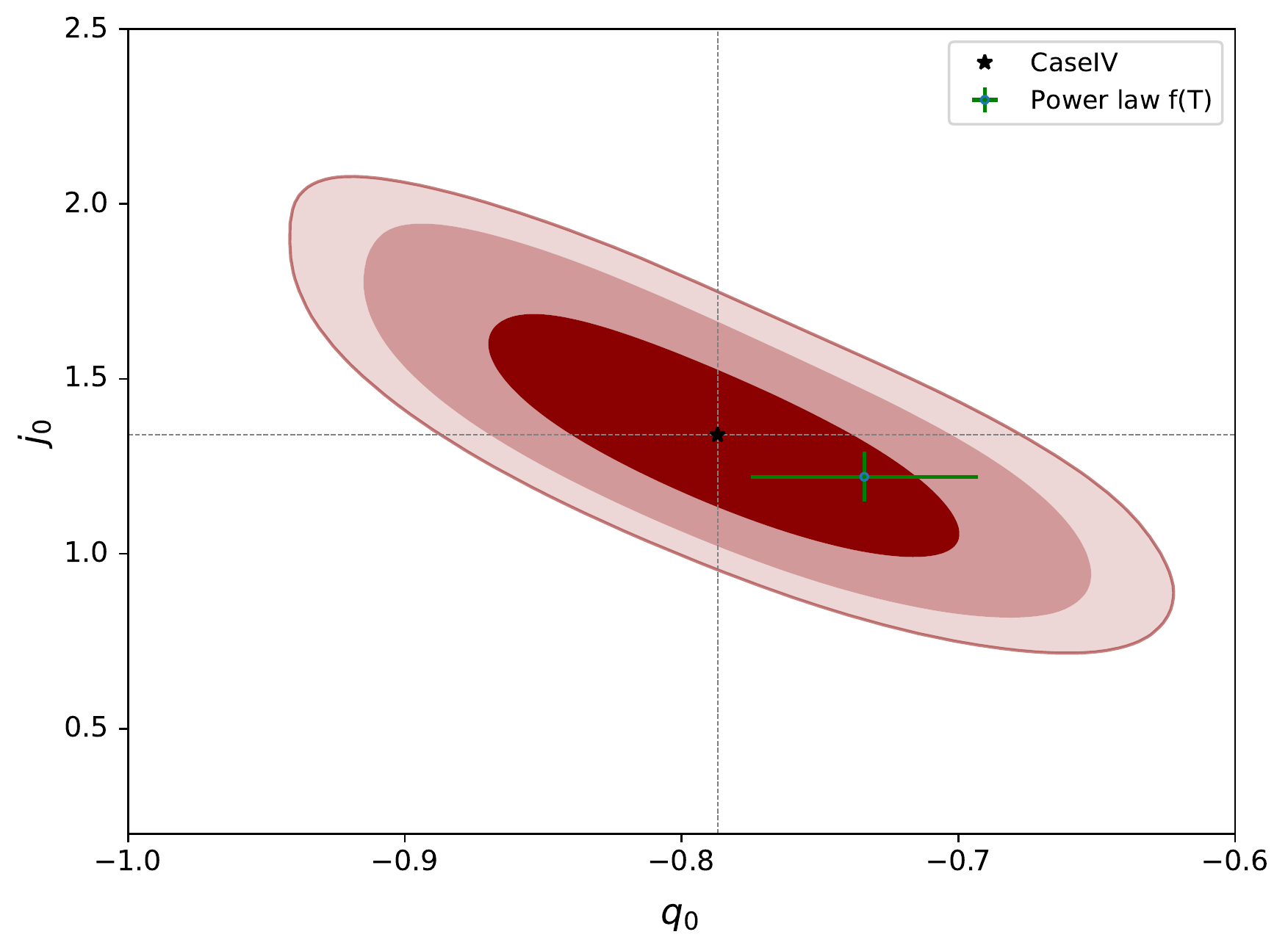}
	\caption{ Same as Fig. (\ref{fig:fig3}), but by using mock data for the Hubble diagram of QSOs.}
	\label{fig:fig6}
\end{figure*}

\begin{table*}
	\centering
	\caption{Same as Table \ref{tab:tab2}, but by using mock data for the Hubble diagram of QSOs.}
	\begin{tabular}{c c  c  c c c c }
		\hline \hline
		Model & $q_0$ & $j_0$ & $s_0$& $l_0$&$m_0$ \\
		\hline
		
		Case I& $-1.093^{+0.062,+0.10,+0.19}_{ -0.062,-0.11,-0.15}$ & $ 2.66^{+ 0.19,+0.35,+0.44}_{-0.19,-0.36,-0.47}$ & $-1.31^{+0.60,+0.94,+1.2}_{-0.47,-1.1,-1.3}$ &  $4.67^{+0.98,+2.3,+2.4}_{-1.4,-1.8,-2.3}$ & --
		\\\hline
		Case II	&$-0.800^{+0.033,+0.065,++0.095}_{ -0.033,-0.064,-0.085}$ & $1.81^{+ 0.25,+0.52,+0.66}_{-0.25,-0.52,-0.68}$ & $-0.38^{+0.86,+1.6,+2.3}_{-0.86,-1.5,-2.3}$ &  $4.9^{+1.1,+1.3,+1.6}_{-0.55,-1.8,-2.3}$ &
		$-10.1^{+1.9,+2.5,+2.9}_{-1.6,-2.2,-2.5}$\\
		\hline \hline
		Case III	&$-0.792^{0.051,+0.10,+0.15}_{0.051,-0.098,-0.11}$ & $1.63^{+0.17,+0.31,+0.39}_{-0.15,-0.34,-0.47}$ & $0.89^{+0.49,+0.66,+0.75}_{-0.25,-0.91,-1.2}$ &  $5.82^{+0.51,+1.0,+1.3}_{-0.59,-1.0,-1.3}$ &
		$-$\\
		\hline 
		Case IV&$-0.787^{ +0.054,+0.10,+0.13}_{-0.054,-0.099,-0.13}$ & $1.34^{+0.20,+0.44,+0.58}_{-0.24,-0.41,-0.49}$ & $-1.29^{ +0.47,+0.92,+1.0}_{ -0.47,-0.88,-1.1}$ &  $1.9^{+1.7,+3.5,+4.3}_{-2.1,-3.2,-3.9}$ &
		$-6.1^{+4.4,+7.3,+8.3}_{-4.4,-8.3,-9.8}$\\
		\hline \hline
	\end{tabular}\label{tab:tab3}
\end{table*}

\begin{itemize}
	\item Our results for the reconstructed distance modulus of SNIa are shown in Fig. (\ref{fig:fig2}). We observe that in all cases of cosmographic method, the percentage difference between the reconstructed distance modulus of SNIa and that of the exact function in the residual plane is less than $1\%$. In cosmographic method based on the foruth-order Taylor expansion (case I), the percentage difference at redshifts lower than one is negligible and then reaches to $0.5\%$ at redshift $2.3$. In cosmographic method based on the fifth-order Taylor expansion (case II), discrepancy is vanishing at $z<1$ and reaching to $0.25\%$ at $z=2.3$, which means that we can significantly reduce the error truncation of the Taylor expansion compared to case I. In the case of cosmographic method based on the $P_{2,2}(z)$ polynomials (case III), the percentage difference in residual plane gets to the value $0.25\%$ at redshift $z=2.3$. Finally, when we use $P_{3,2}(z)$ approximation (case IV), the percentage difference is less than $0.05\%$. 
	Obviously, any difference (even small) between the reconstructed distance modulus in the context of cosmographic method and the exact distance modulus from the $f(T)$ model can be interpreted as the approximation of cosmographic method, due to the error truncation (even small) of the Taylor or PADE approximations. We emphasize that the mock data for the distance modulus of SNIa are generated based on the $f(T)$ model. Hence, we expect to see that $f(T)$ model fits on the data. We find that the $P_{3,2}$ approximation performs better than the $P_{2,2}$ approximation and both perform better than the Taylor series. 
	In particular, $P_{3,2}(z)$ is the best approximation to reconstruct the distance modulus of SNIa among all the cases considered in the cosmographic method . However, this result does not suggest that the other cases do not perform well. In Fig. (\ref{fig:fig3}), we compare the best-fit values of the cosmographic parameters $q_0$ and $j_0$ obtained from power-law $f(T)$ model with the confidence regions of these parameters obtained from the various cases of cosmographic methods discussed above. We observe the full consistency of all cases of cosmographic method with the power-law $f(T)$ model in the $q_0-j_0$ plane. This consistency means that the errors due to the truncation of the Taylor expansion and the PADE polynomials are small enough, at least when we discuss the $q_0$ and $j_0$ parameters in the redshift range $z<2.3$. Our numerical results for the best-fit values of  all cosmographic parameters together with their $3\sigma$ confidence level obtained by cosmographic methods are shown in Table (\ref{tab:tab2}). We can extend our discussion of the consistency between the cosmographic method and the power-law $f(T)$ model to the higher cosmographic parameters instead of $q_0$ and $j_0$, by comparing the values of the first row of Table (\ref{tab:tab1}) with those of Table (\ref{tab:tab2}).
	In Case I of the cosmographic approach, we observe that the cosmographic parameter $s_0$ of the model is consistent with that of the cosmographic approach. However, we see a strong deviation (larger than $3\sigma$) between the cosmographic parameter $l_0$ of $f(T)$ model with that of the cosmographic method. This discrepancy indicates that the cosmographic method based on the fourth-order Taylor series of $y=z/(1+z)$ cannot well fitted to the mock SNIa data, due to the error truncation of Taylor expansion. We mention that this inconsistency occurs when we look at the higher cosmographic parameters and as shown above, we observed the agreement between Case I and $f(T)$ model in $q_0-j_0$ plane. Note that the above comparison between Case I and $f(T)$ model was performed in the redshift range of SNIa ($z<2.3$). At lower resolution, two opposing groups \citep{Lusso:2019akb} and \citep{Yang:2019vgk} agreed that the cosmographic method performs at least up to $z\sim2$. Comparing the cosmographic parameters of Case II and $f(T)$ model, we find the full consistency between the cosmographic method and the power-law $f(T)$ model. This interesting result is due to the added term proportional to $y^5$ in cosmographic method compared to Case I. With one more free parameter $m_0$, we establish a better consistency between the cosmographic method based on the Taylor series and the power-law $f(T)$ model. This result obtained in Case II is not a wonderful thing, because we added one more free parameter compared to Case I. In the case III, all cosmographic parameters of cosmographic method are consistent with those of the power-law $f(T)$ model which means that the $P_{2,2}$ approximation is the useful tool to reconstruct the distance modulus up to $z=2.3$. Moreover, we find the complete agreement between the case IV and the power-law $f(T)$ model for all cosmographic parameters within $1\sigma-3\sigma$ confidence regions. Assuming deviations of less than $3\sigma$ as statistical error (not mathematical error), we convince ourselves that the difference between $P_{3,2}$ and $P_{2,2}$ for reconstructing the distance modulus up to redshift $z=2.3$ is negligible because the cosmographic parameters of the power-law $f(T)$ model match those of both methods with an accuracy of at least $3\sigma$ uncertainties. We emphasize that the use of $P_{3,2}$ causes the statistical errors to be smaller compared to $P_{2,2}$. We observe this point in Fig. (\ref{fig:fig3}), where the distance between the power-law $f(T)$ model and Case IV in $q_0-j_0$ plane is much smaller than the same distance between the power-law $f(T)$ model and Case III. Based on the above arguments, we adopt the $P_{3,2}$ polynomial in the context of the PADE approximation as a useful tool for reconstructing the distance modulus up to $z=2.3$ without a need for higher order of PADE polynomials. 
	
\begin{figure*} 
	\centering
	\includegraphics[width=8.5cm]{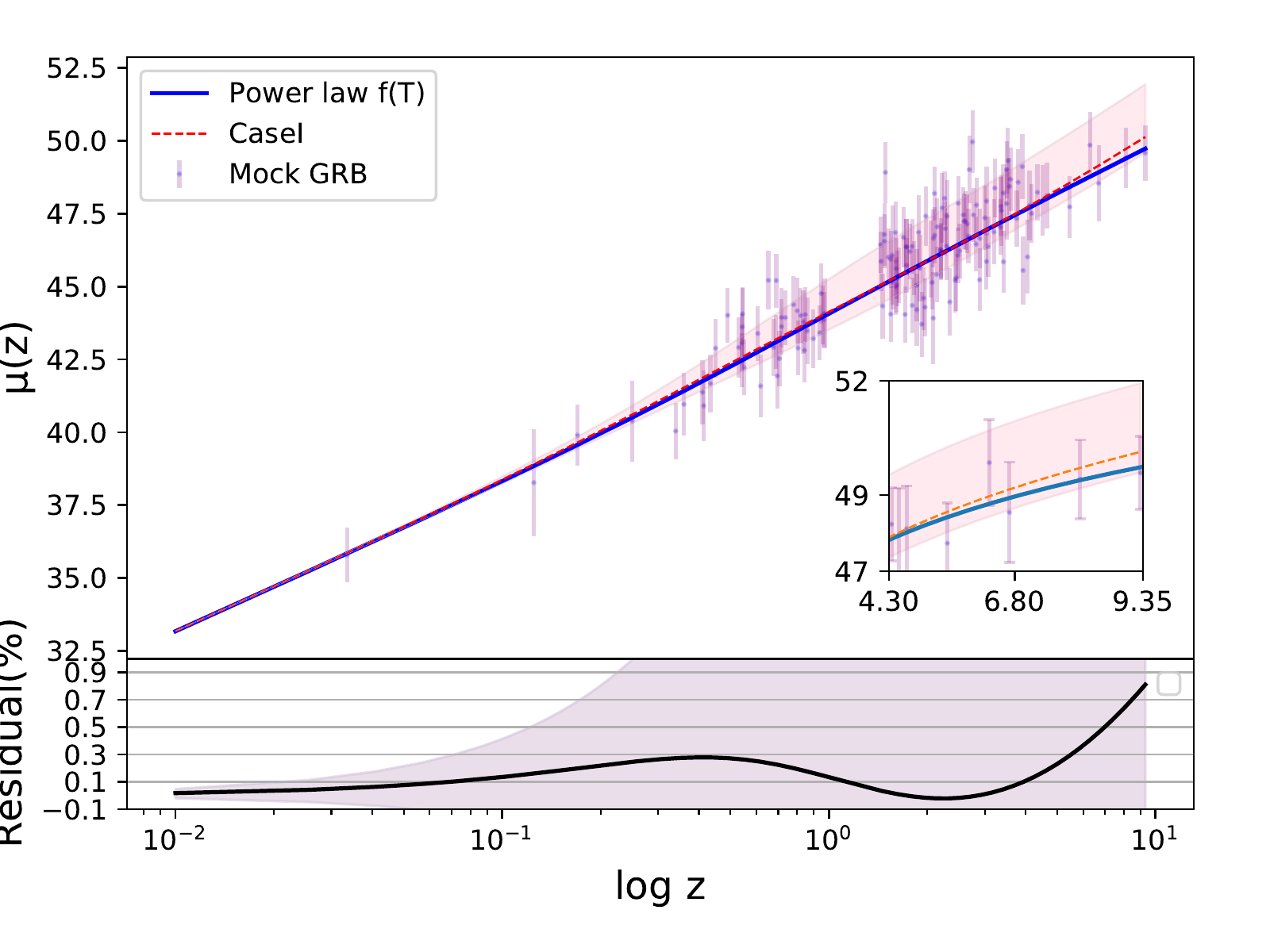}
	\includegraphics[width=8.5cm]{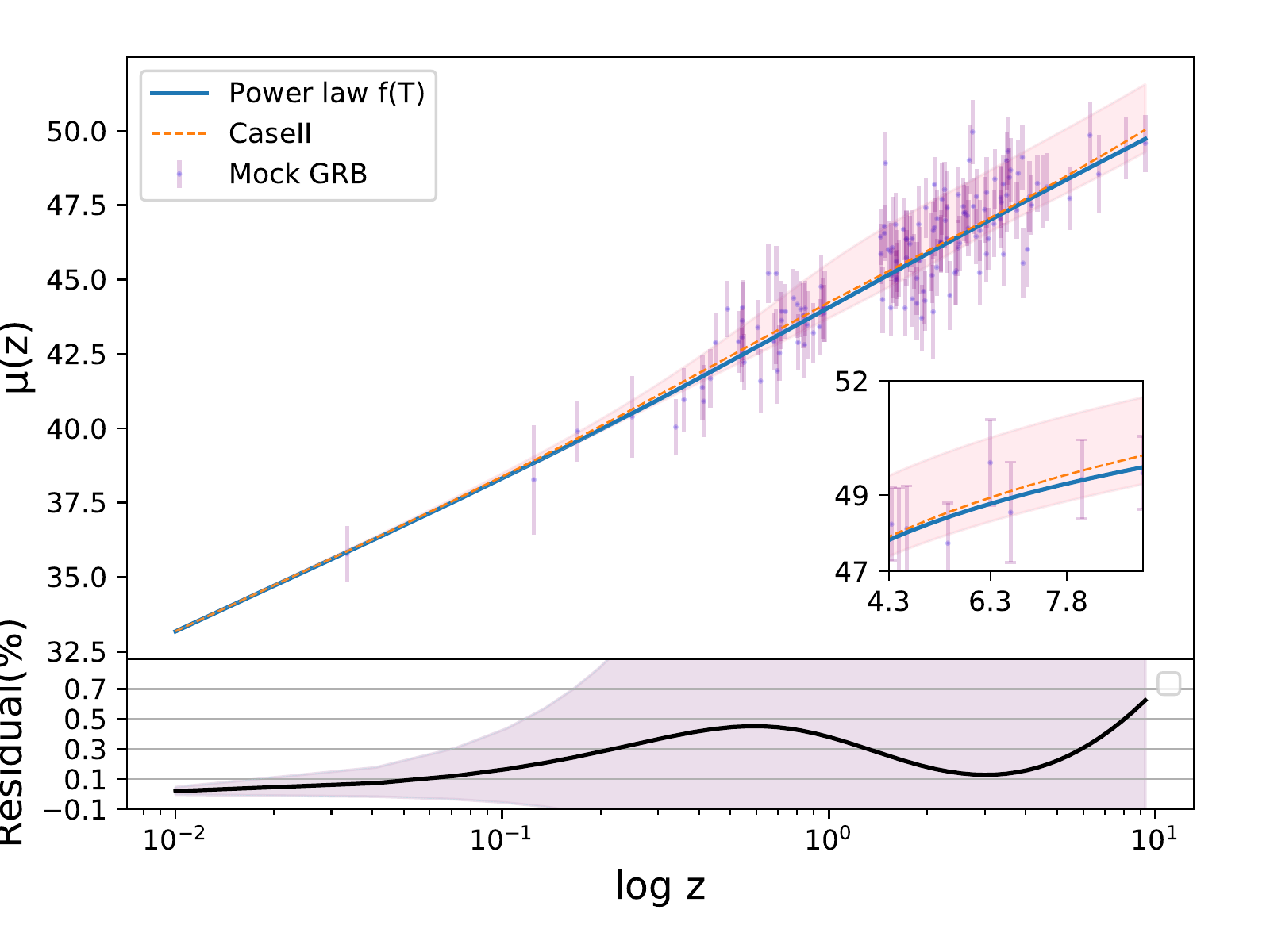}
	\includegraphics[width=8.5cm]{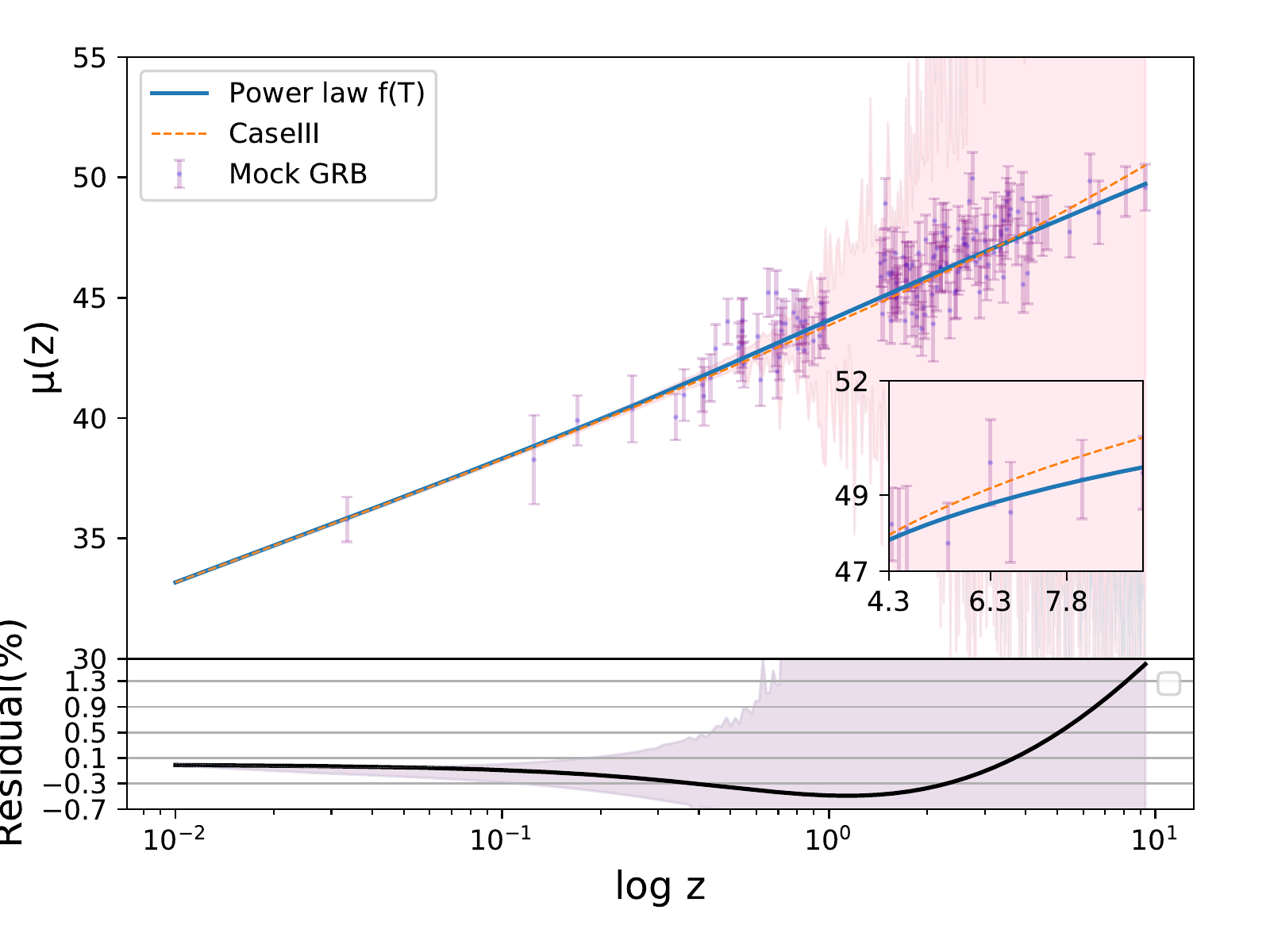}
	\includegraphics[width=8.5cm]{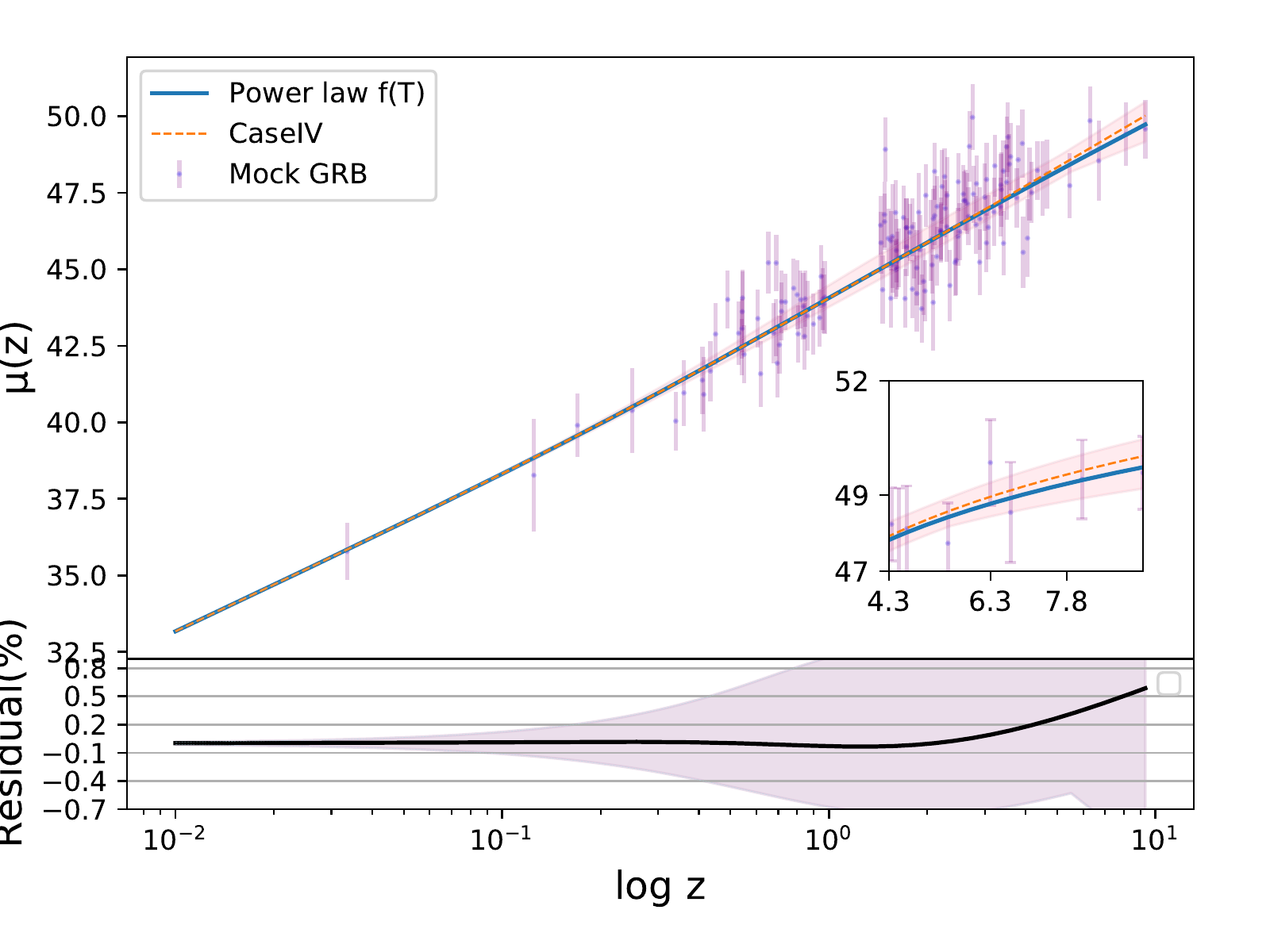}
	\caption{ \jt{Same as Fig. (\ref{fig:fig2}), but by using mock data for the Hubble diagram of GRBs.}}
	\label{fig:fig7a}
\end{figure*}
\begin{figure*} 
	\centering
	\includegraphics[width=8.5cm]{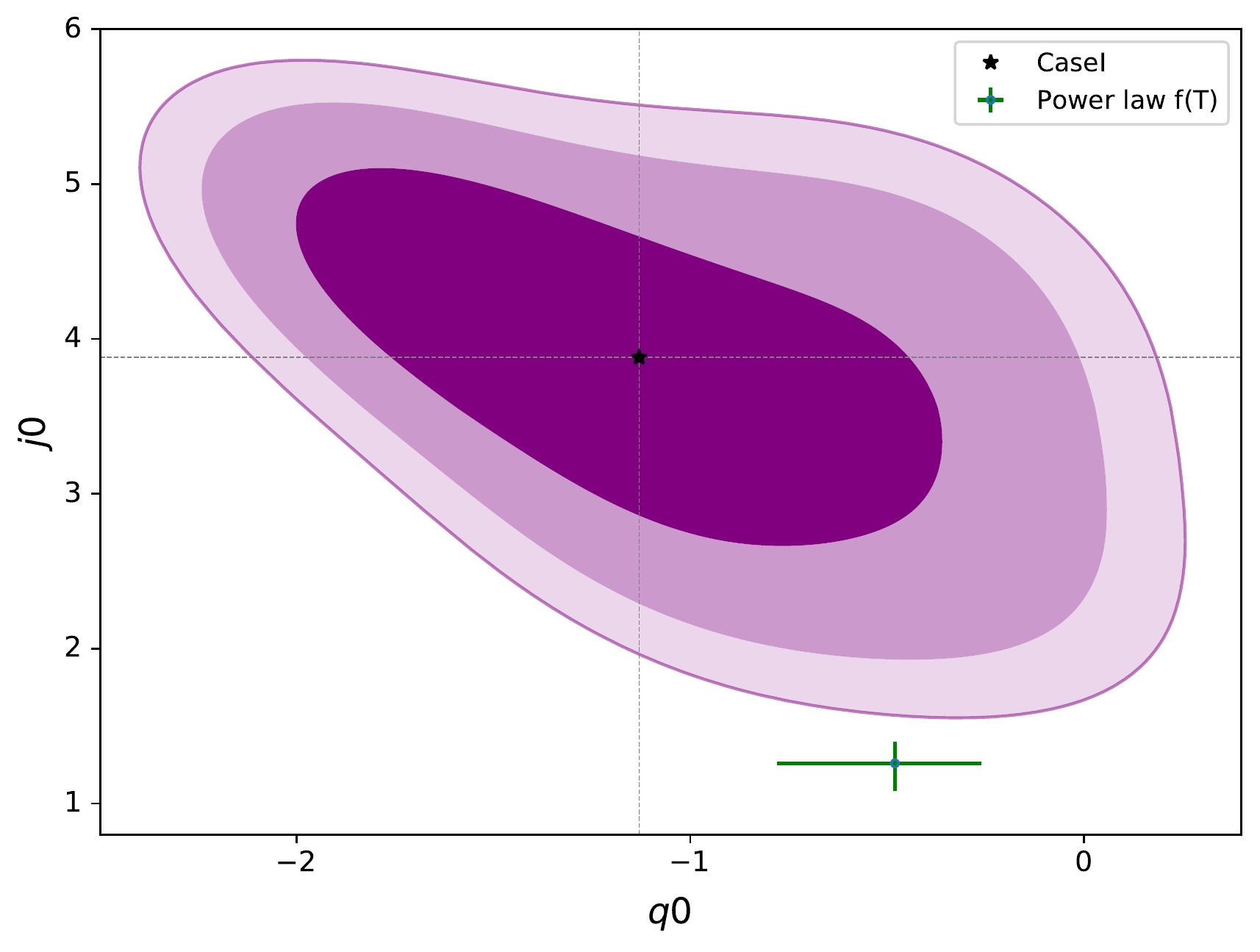}
	\includegraphics[width=8.5cm]{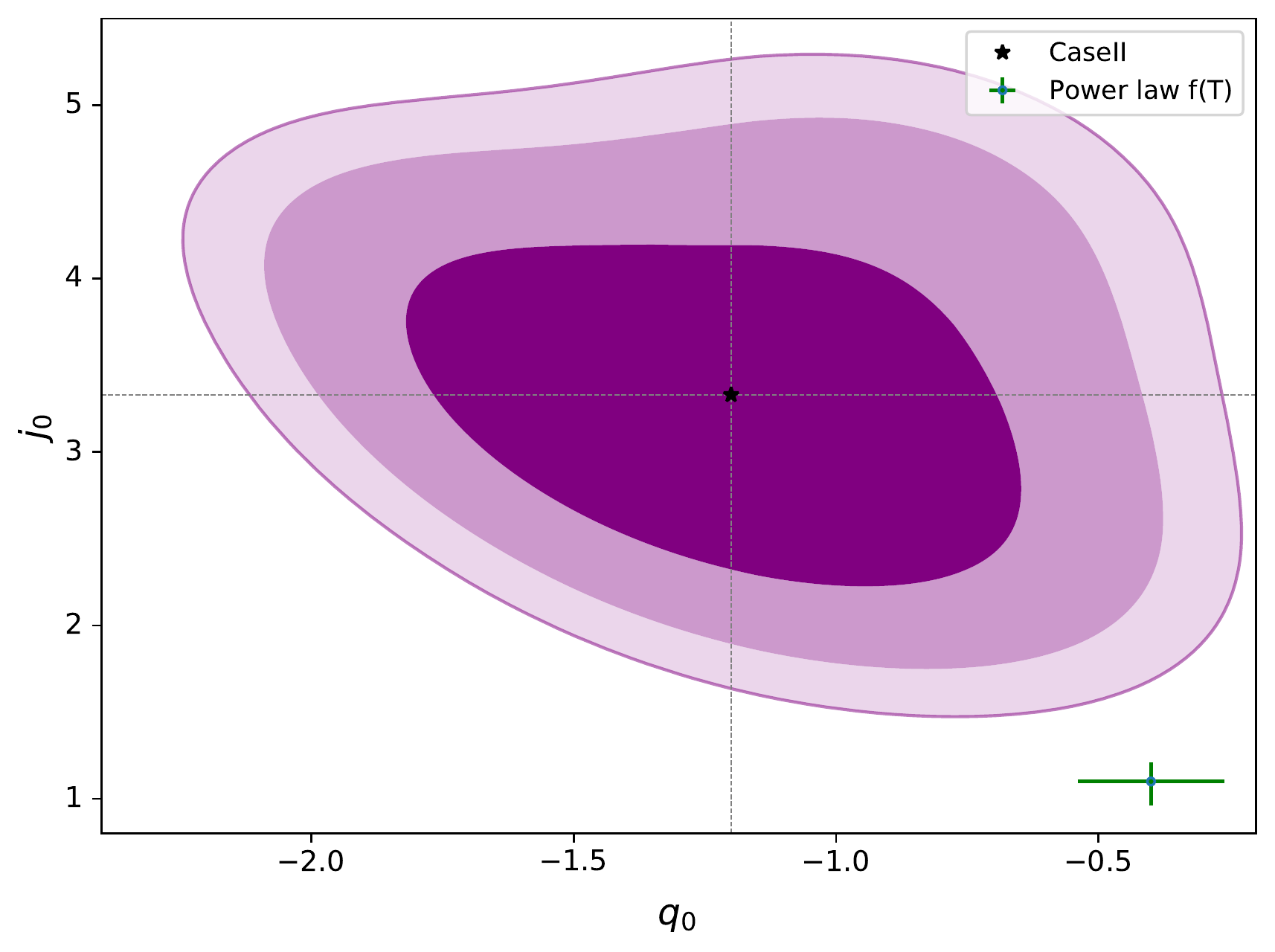}
	\includegraphics[width=8.5cm]{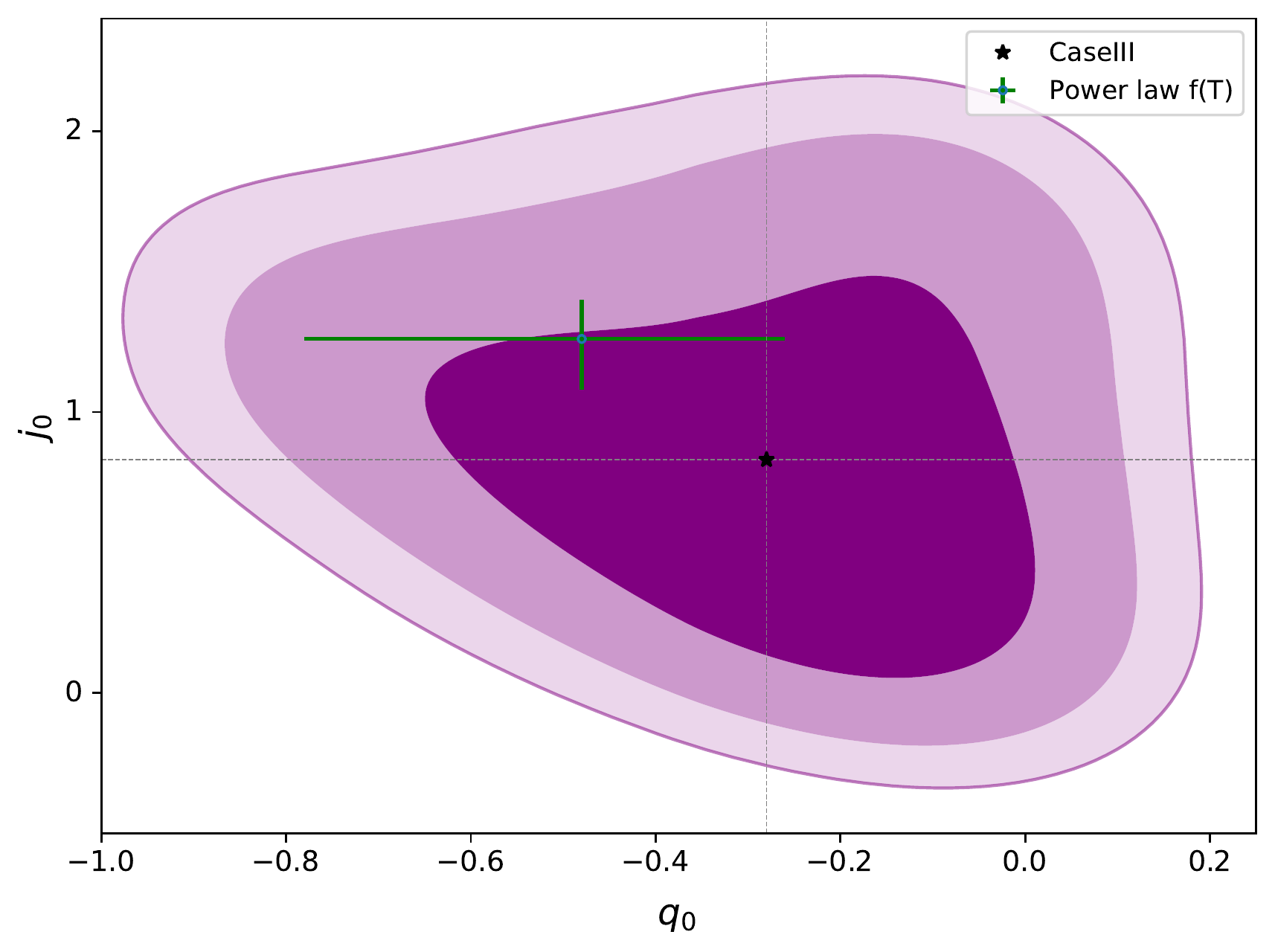}
	\includegraphics[width=8.5cm]{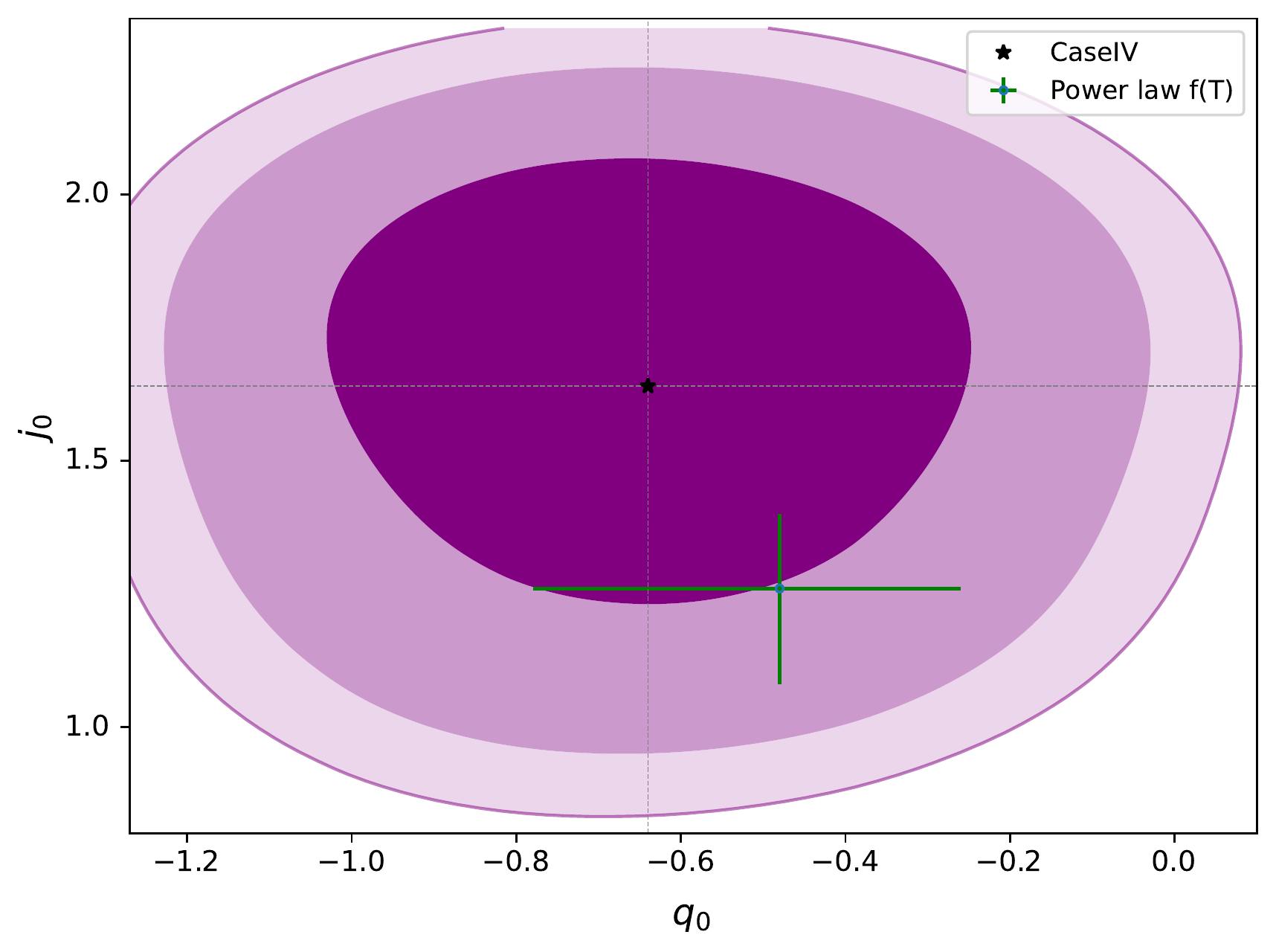}
	\caption{ \jt{Same as Fig. (\ref{fig:fig3}), but by using mock data for the Hubble diagram of GRBs.}}
	\label{fig:fig7b}
\end{figure*}

	\item Let us continue our study to the higher redshifts, where we use mock data for the Hubble diagram of QSO and reconstruct the distance modulus up to redshift $z\sim 5$ in the context of cosmographic methods. In Fig. (\ref{fig:fig4}), the reconstructed distance modulus up to redshift $z\sim 5$ has been shown for the various cases of cosmographic method discussed in this work. For each panel, we have also plotted the distance modulus in the context of the power-law $f(T)$ model obtained by using of the best-fit values of the cosmological parameters $\Omega_{m0}$ and $b$ from the right panel of Fig. (\ref{fig:fig1}). We observe that in the case I, the percentage difference between the cosmographic method and $f(T)$ model reaches to $0.8\%$ at $z\sim 5$. In the case II, this value is roughly $0.45\%$ at $z\sim 5$ which means that adding one more algebraic term proportional to $y^5$ in the Taylor series causes the decrement of the error truncation compared to the foruth-order Taylor expansion. In the case III, we observe that the percentage difference between the cosmographic method and the $f(T)$ model is less than $0.25\%$ which indicates that the $P_{2,2}$ approximation can make a better fit compared to the Taylor series with one more free parameter. Finally by using the $P_{3,2}$ polynomial, we decrease the difference between the cosmographic method and the $f(T)$ model to less than $0.15\%$ at $z\sim 5$. This value is sufficiently small to assume that the PADE approximation can be used for reconstructing the distance modulus at the higher redshifts in the context of cosmographic method. Moreover, we plot the confidence regions of the cosmographic parameters $q_0$ and $j_0$ obtained from the MCMC analysis using mock QSOs in Fig. (\ref{fig:fig6}). In the upper-left panel, we observe a tension (a deviation bigger than $3\sigma$) between the power-law $f(T)$ model and the case I of cosmographic method. Obviously, this tension is appeared due to the error truncation of the Taylor series in Case I. In particular, this tension is related to the large percentage difference between the distance modulus obtained from the case I of cosmographic method and distance modulus obtained from the power-law $f(T)$ model (see upper-left panel of Fig. \ref{fig:fig4}). Thus, we conclude that the fourth-order Taylor expansion is not suitable approximation for reconstructing the distance modulus in the context of cosmographic approach. We emphasize that this conclusion is obtained when we study the power-law $f(T)$ model and cannot be simply extended to other DE models. For example in the previous works \citep{Rezaei:2020lfy,Pourojaghi:2021den}, the authors used the cosmographic method based on fourth-order Taylor expansion in terms of $y=z/(1+z)$ variable and showed that some DE models and DE parameterizations are well consistent with the observations of QSOs and GRBs even at high redshifts around $z\sim 5$. While this result is not valid for the standard $\Lambda$CDM model \citep{Yang:2019vgk}. In the upper-right panel, we find an agreement (deviation is lower than $3\sigma$ and can be considered as a statistical error) between the case II and the power-law $f(T)$ model in $q_0-j_0$ plane. We conclude that the fifth-order Taylor expansion can remove the tension occurred in the fourth-order Taylor expansion. Moreover, we observe that the case II and the power-law $f(T)$ model are consistent to each other, when we compare the higher cosmographic parameters $s_0$, $l_0$ and $m_0$ (see Table \ref{tab:tab3} for case II and Table \ref{tab:tab1} for QSOs). Thus, the use of the fifth-order Taylor expansion can be considered as a suitable tool for reconstructing the distance modulus at relatively high redshifts in the context of cosmographic method. In the lower-left panel, we show the complete agreement between case III and the power-law $f(T)$ model within $2\sigma$ confidence region and better than what happened for Case II. We also find no tension for the higher cosmographic parameters $s_0$ and $l_0$ obtained in the case III with those of the power-law $f(T)$ model (see Table \ref{tab:tab3} for case III and Table \ref{tab:tab1} for QSOs). Finally, in the lower-right panel, we show the complete agreement within $1\sigma$ error between the power-law $f(T)$ model and the case IV in $q_0-j_0$ plane. We also find the complete agreement between the power-law $f(T)$ model and case IV, when we compare their higher cosmographic parameters $s_0$, $l_0$ and $m_0$. Hence, both of the $P_{2,2}$ and $P_{3,2}$ approximations can be employed for reconstructing the distance modulus at higher redshifts. In particular, we can choose the $P_{3,2}$ approximation, because this case has the better fit to the mock data compared to other cases discussed above. 
		\begin{table*}
		\centering
		\caption{\jt{Same as Table \ref{tab:tab2}, but by using mock data for the Hubble diagram of GRBs.}}
		\begin{tabular}{c c  c  c c c c }
			\hline \hline
			\jt{Model} & \jt{$q_0$} & \jt{$j_0$} & \jt{$s_0$}& \jt{$l_0$}&\jt{$m_0$} \\
			\hline
			\jt{Case I}&\jt{ $-1.13^{+0.47,+1.0,+1.3}_{-0.62,-0.91,-1.1}$} &\jt{ $3.88^{+0.92,+1.2,+1.4}_{-0.55,-1.5,-1.9}$} & \jt{$0.2^{+1.5,+2.4,+2.9}_{-1.3,-2.5,-3.1}$} & \jt{ $2.1^{+1.3,+1.8,+2.2}_{-1.0,-2.0,-2.3}$} & \jt{--}
			\\\hline
			\jt{Case II}	&\jt{$-1.20^{+ 0.37,+0.68,+0.83}_{-0.37,-0.69,-0.86}$} &\jt{ $3.33^{+0.65,+1.3,+1.6}_{-0.73,-1.2,-1.5}$} & \jt{$0.0^{+ 1.1,+1.9,+2.3}_{-1.1,-1.9,-2.3}$} &  \jt{$2.0^{+ 1.2,+1.9,+2.5}_{-1.2,-1.9,-2.4}$} &
			\jt{$-11.6^{+2.0,+3.3,+3.5}_{-2.0,-3.3,-3.4}$}\\
			\hline \hline
			\jt{Case III}	&\jt{$-0.28^{+0.26,+0.31,+0.33}_{-0.11,-0.41,-0.56}$} & \jt{$0.83^{+0.40,+0.91,+1.2}_{-0.53,-0.81,-0.98}$} & \jt{$0.28^{+0.80,+1.5,+1.9}_{-0.80,-1.6,-2.0}$} & \jt{ $3.02^{+1.7,+2.1,+2.3}_{-0.98,-2.4,-3.0}$} &
			\jt{$--$}\\
			\hline 
			\jt{Case IV}&\jt{$-0.64^{+0.21,+0.42,+0.53}_{-0.25,-0.40,0.49}$} & \jt{$1.64^{+0.32,+0.43,+0.52}_{-0.19,-0.54,-0.68}$} & \jt{$-1.53^{+0.27,+0.50,+0.52}_{-0.27,-0.43,-0.46}$} & \jt{ $2.78^{+1.2,+1.2,+1.3}_{-0.45,-1.8,-2.4}$} &
			\jt{$-5.0^{+5.6,+9.3,+9.8}_{-5.6,-9.3,-9.8}$}\\
			\hline \hline
		\end{tabular}\label{tab:grb}
	\end{table*}
	\item \jt{Finally, we extend our analysis to redshifts $z\sim 9$, by using the mock data for Hubble diagrams of GRBs. In Fig. (\ref{fig:fig7a}), we show the reconstructed distance modulus in the context of different cosmographic methods and also $f(T)$ model. We also show the difference between cosmographic methods and $f(T)$ model in residual plane. We observe that the percentage difference between $P_{3,2}$ and $f(T)$ model is smallest. In addition in Fig. (\ref{fig:fig7b}), we plot the confidence regions of cosmographic parameters $q_0$ and $j_0$ obtained from cosmographic methods and compare them with those of obtained from $f(T)$ model. We explicitly observe that the best fit-value of $j_0$ obtained from $f(T)$ model is out of the confidence regions obtained from fourth- and fifth-order Taylor series. This result shows that both Taylor approximations are inadequate for reconstructing the distance modulus at high redshifts $z\sim 9$. Notice that using mock QSOs, we already showed the fourth-order Taylor series is not suitable approximation even at $z\sim 5$, while the fifth-order works properly at redshifts $z\leq 5$. Here using mock GRBs extended to higher redshifts than the redshifts of QSOs, we see that the cosmographic method defined based on the fourth- and fifth-order Taylor series does not work accurately. In contrast to Taylor series, the $f(T)$ model and PADE cosmographic methods are agreeing with each other in $q_0-j_0$ plane (see lower panels of Fig.\ref{fig:fig7b}). Our numerical results for higher cosmographic parameters for different cosmographic methods obtained from mock GRBs data are reported in Tab. (\ref{tab:grb}). Comparing the values of all cosmographic parameters in Tab. (\ref{tab:grb}) with those of mock GRBs in Tab.\ref{tab:tab1}, we observe that $P_{2,2}$ (case II) and $P_{3,2}$ (case IV) are consistent with $f(T)$ model. Nevertheless, $P_{3,2}$ gives better consistency than $P_{2,2}$. Similar to what was concluded from mock Hubble diagrams of QSOs up to $z \sim 5$, our analysis using mock GRBs indicates that the cosmographic method based on the $P_{3,2}$ is an appropriate method to reconstruct the distance modulus up to high redshift $z\sim 9$. We should mention that our conclusion based on the mock analysis, is independent of the canonical values assumed for $f(T)$ model parameters $\Omega_{m0}$ and $b$. In fact, we repeated our analysis for different canonical values as $\Omega_{m0}=0.2$ and $b=-0.1$ which gives the present EoS parameter of $f(T)$ model as $w_{DE,0}=-1.02$. Creating the mock QSOs data for these new canonical values, we found that the $P_{3,2}$ cosmographic approach is the best method while the fourth-order Taylor series is insufficient at high redshifts.}
	
\end{itemize}
	\begin{figure*} 
	\centering
	\includegraphics[width=7.5cm]{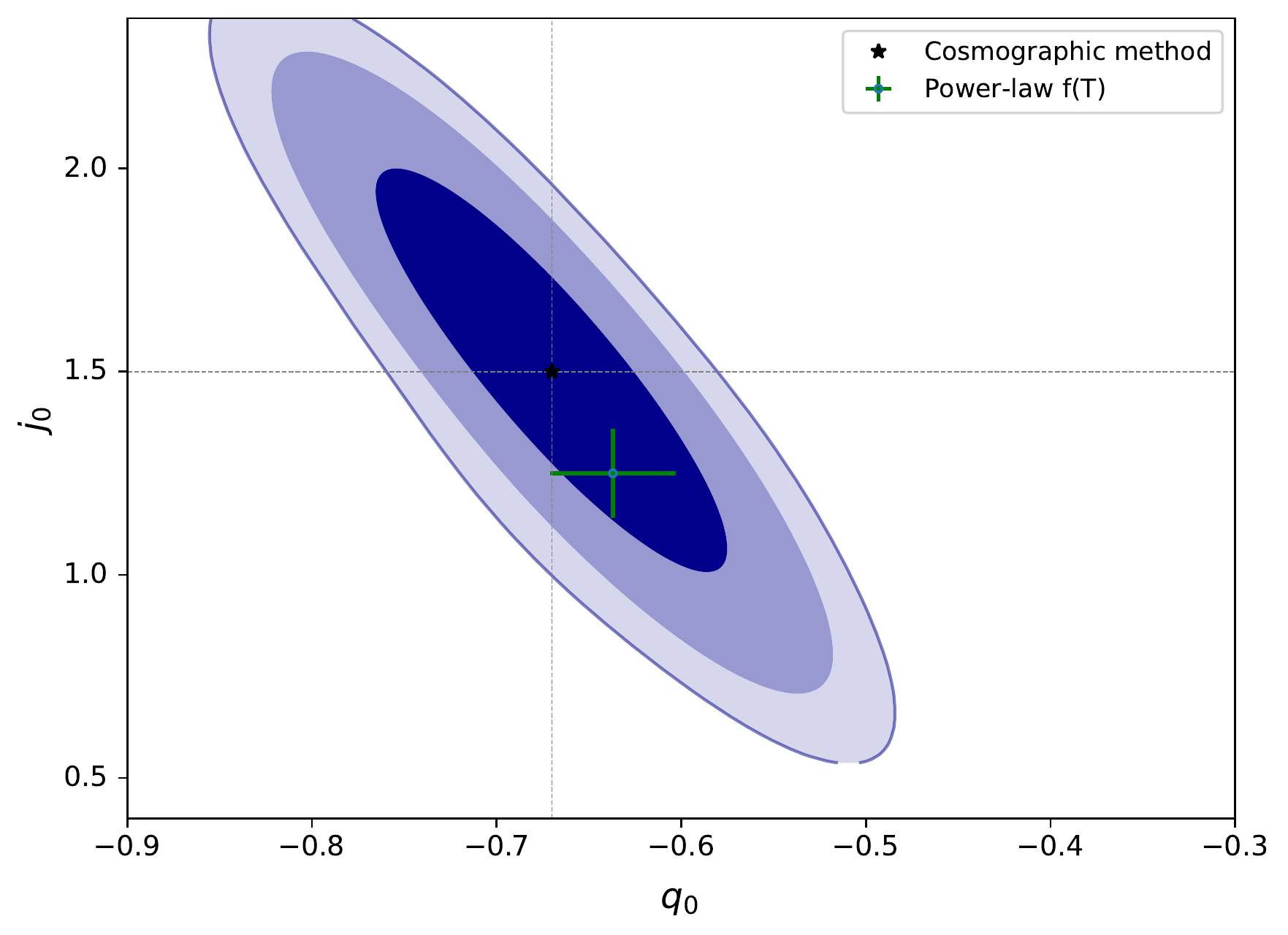}
	\includegraphics[width=7.5cm]{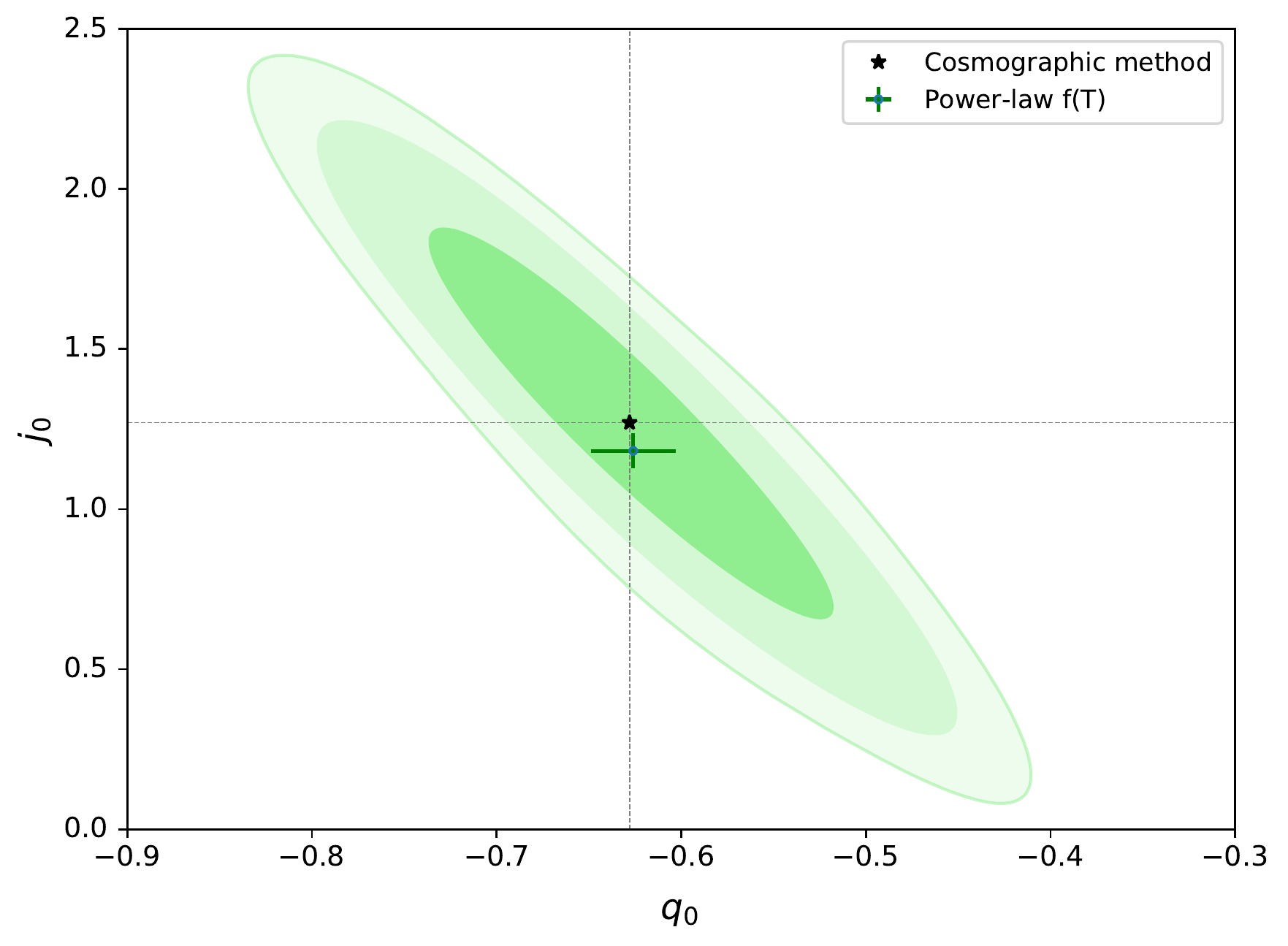}
	\includegraphics[width=7.5cm]{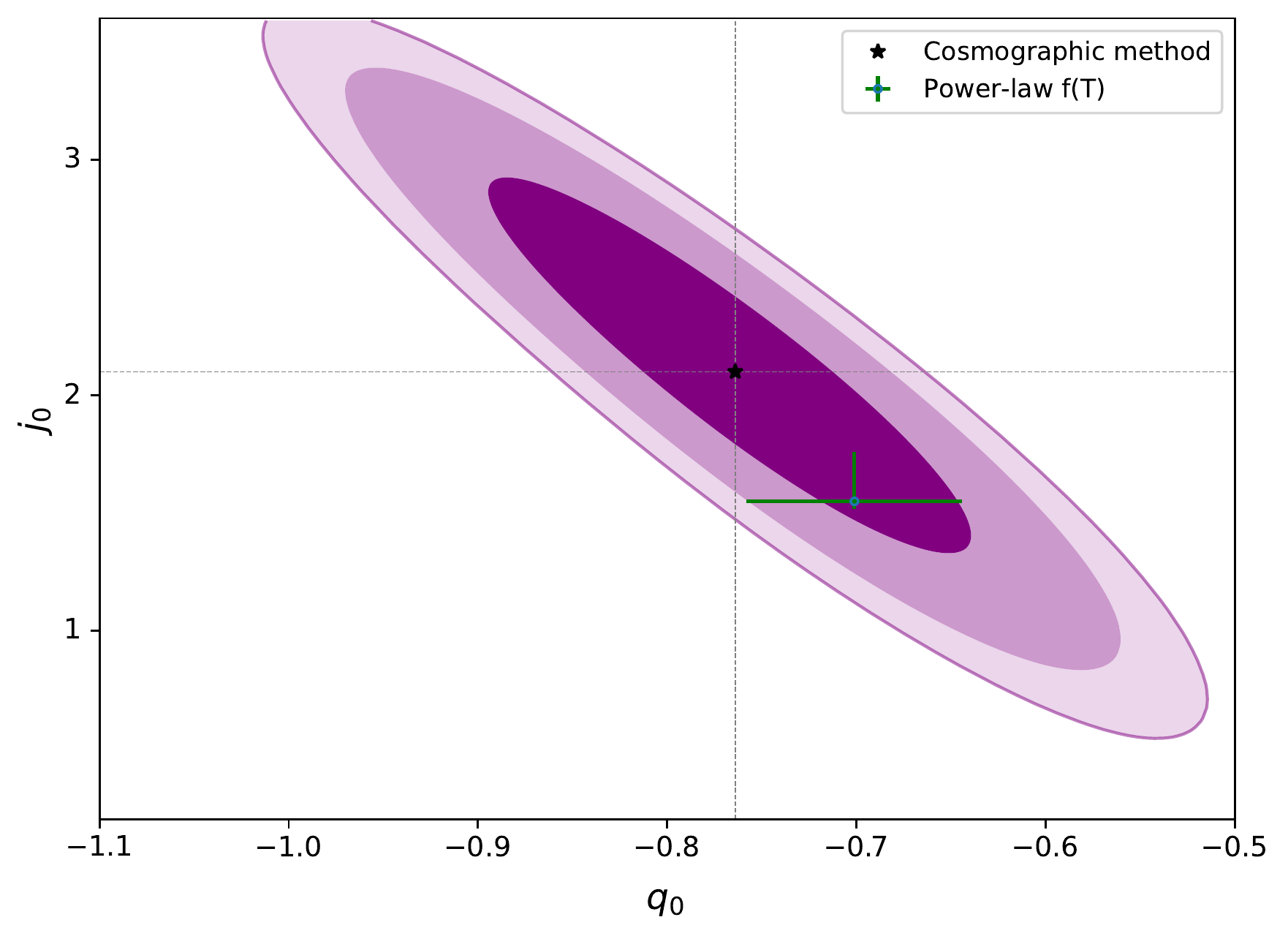}
	\includegraphics[width=7.5cm]{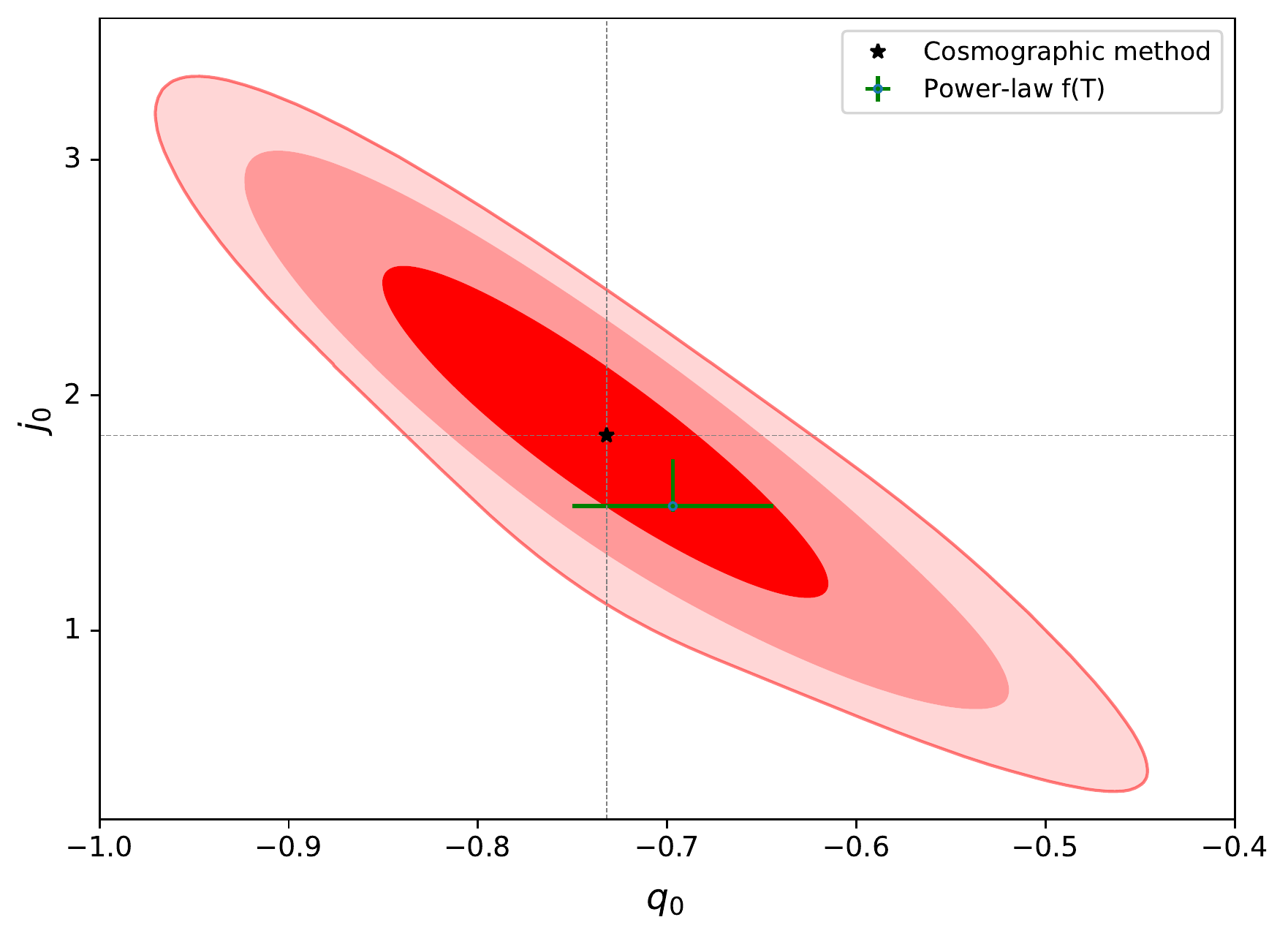}
	\caption{ Upper-left panel: Constraints with $1\sigma$, $2\sigma$ and $3\sigma$ confidence regions on the cosmographic parameters $q_0$ and $j_0$ obtained using the real observational data form the Hubble diagram of SNIa (sample {\it i}). The vertical and horizontal lines show the best-fit values of the $q_0$ and $j_0$, respectively. The location of the power-law $f(T)$ model in $q_0-j_0$ plane is shown in the contours as described in the legend. Upper-right panel: Same as upper-left panel, but for sample ({\it ii}), consisting of the observational Hubble diagram for SNIa + BAO measurements. The lower-left panel: Same as upper-left panel, but for sample ({\it iii}), consisting of the Hubble diagrams of SNIa and QSOs. The lower-right panel: Same as upper-left panel, but for sample ({\it iv}), combining of the Hubble diagrams from SNIa, QSOs and GRBs.}
	\label{fig:fig8}
\end{figure*}

	\begin{figure*} 
	\centering
	\includegraphics[width=8.5cm]{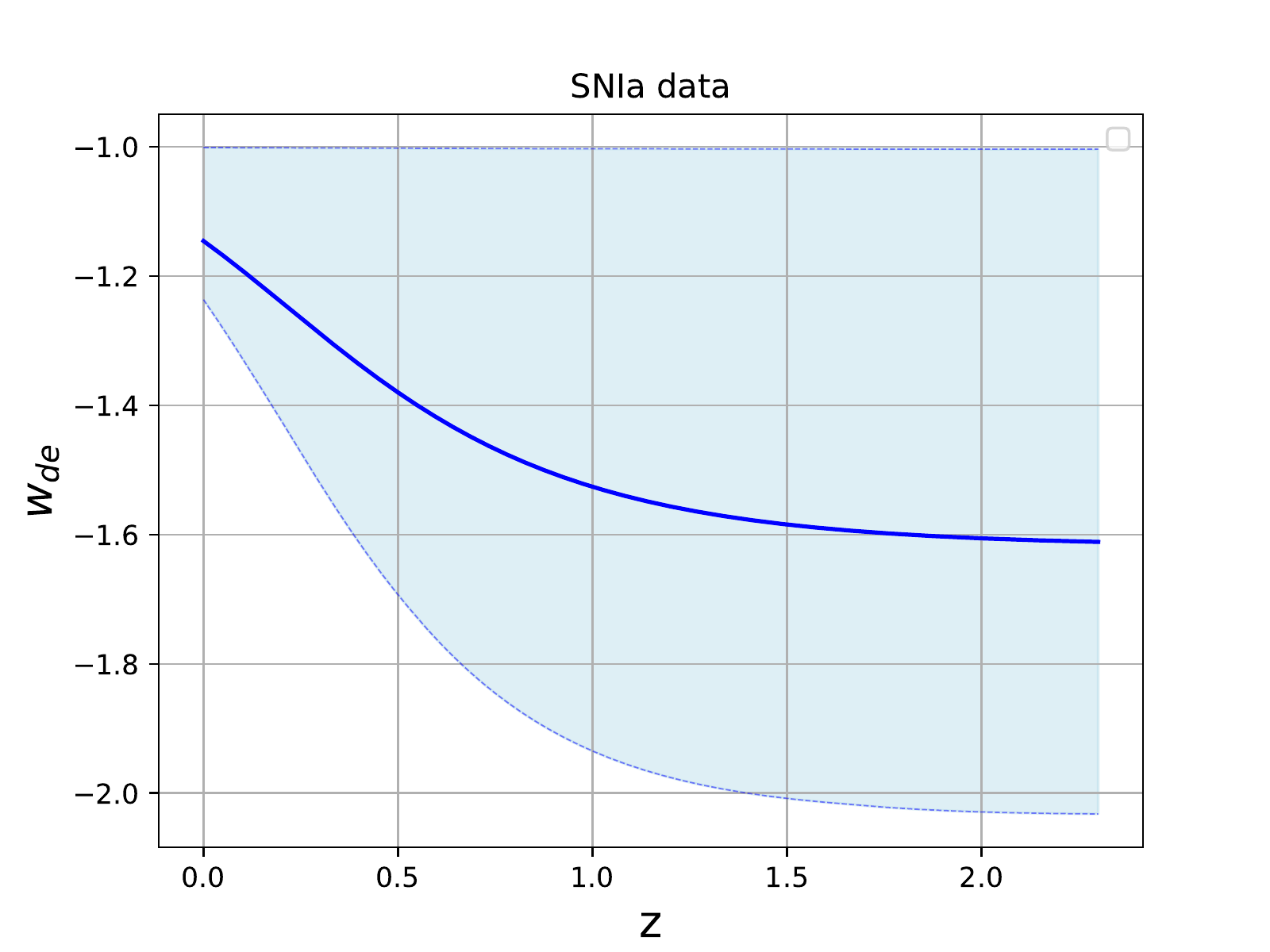}
	\includegraphics[width=8.5cm]{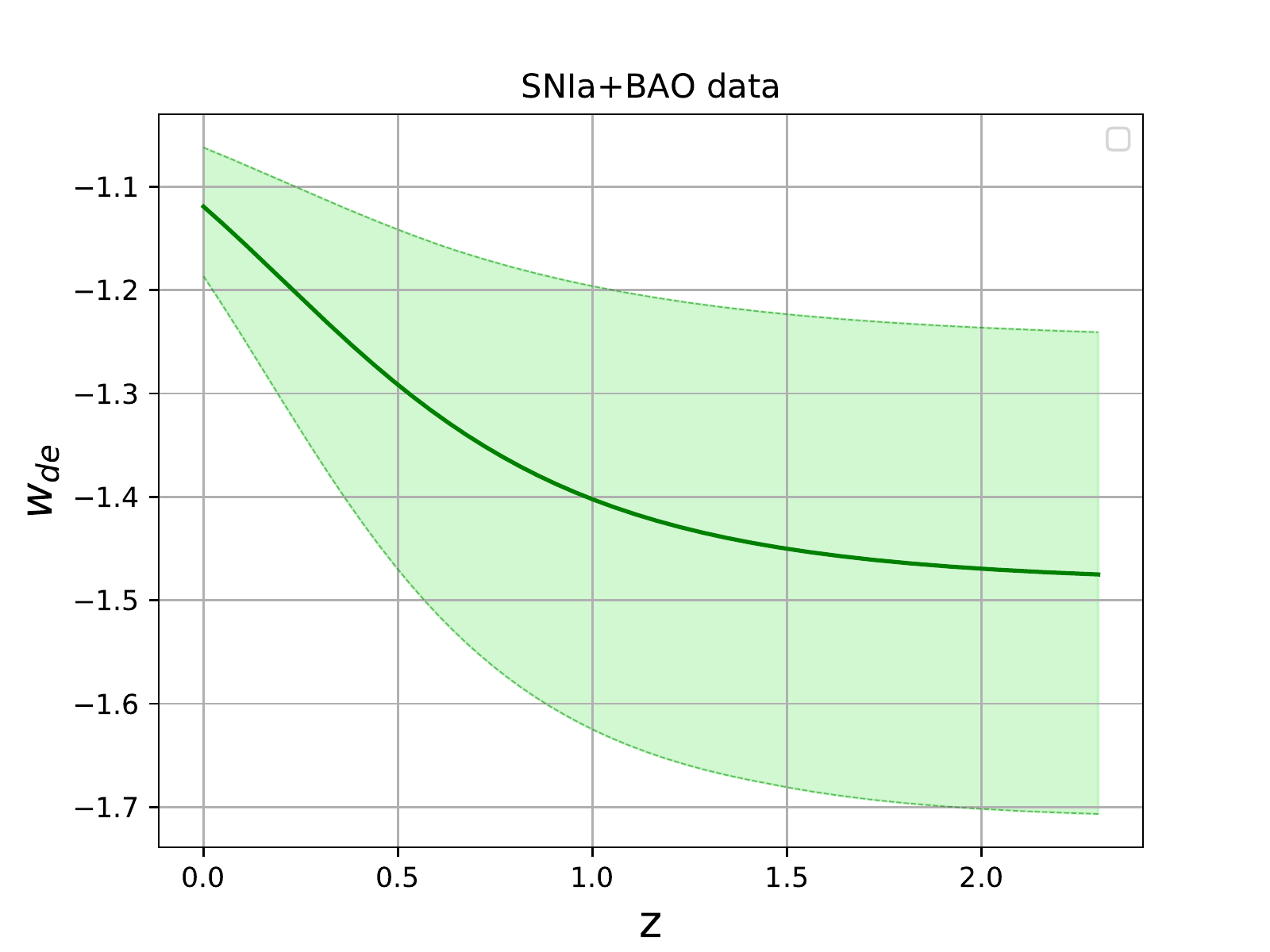}
	\includegraphics[width=8.5cm]{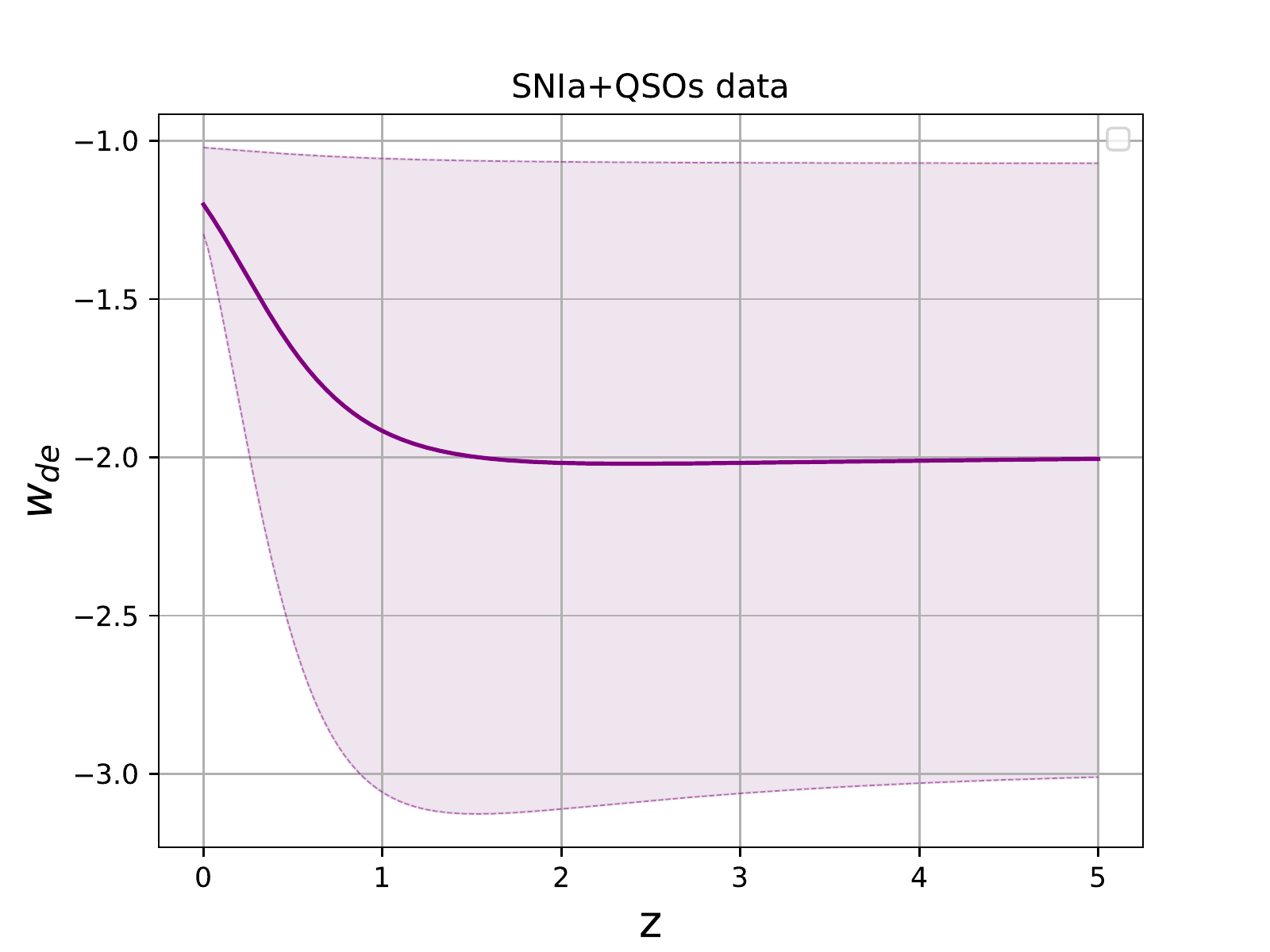}
	\includegraphics[width=8.5cm]{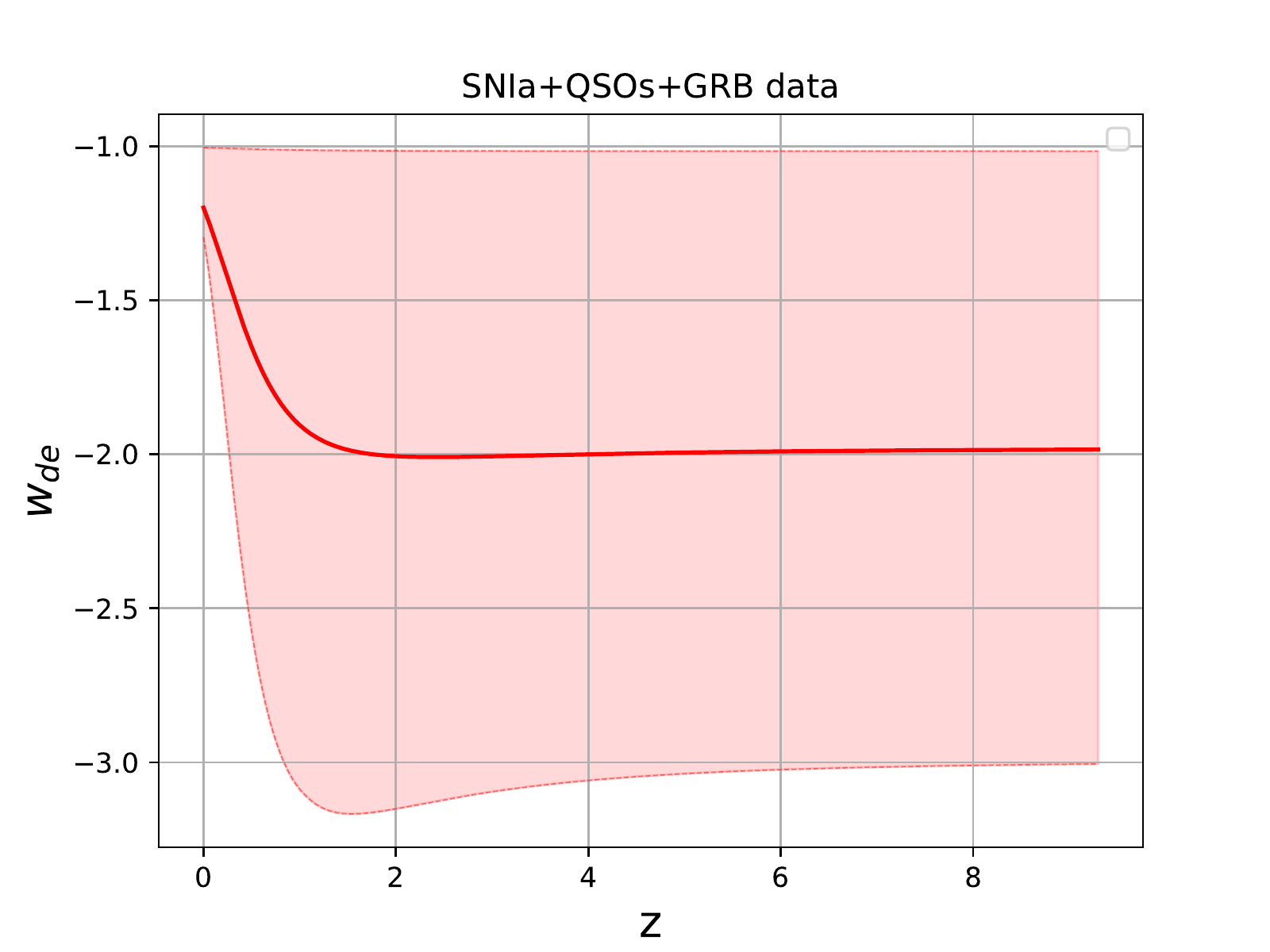}
	\caption{The upper-left panel: The redshift evolution of the constrained effective EoS parameter of the power-law $f(T)$ model together with upper and lower uncertainties band, obtained using the observational data for the Hubble diagram of SNIa. The upper-right panel: Same as upper-left panel, but by using the combination of Hubble diagram for SNIa + BAO observations. The lower-left panel: Same as the upper-left panel, but by using the combinations of the Hubble diagrams form SNIa and QSOs. The lowe-right panel: Same as the upper-left panel, but by using the combinations of the Hubble diagrams form SNIa, QSOs and GRBs.}
	\label{fig:fig9}
\end{figure*}
\begin{table}
	\centering
	\caption{Best fit with $1\sigma$, $2\sigma$ and $3\sigma$ confidence levels to the cosmological parameters of the power-law $f(T)$ model obtained from different samples considered in this work. Sample ({\it i}) includes Hubble diagram of SNIa, sample ({\it ii}) includes Hubble diagram of SNIa + BAO, sample ({\it iii}) includes Hubble diagrams of SNIa + QSOs and sample ({\it iv}) includes Hubble diagrams of SNIa+QSOs+GRBs.
	}
\fontsize{6pt}{6pt}
	\begin{tabular}{c  c  c c}
		\hline \hline
		  & $b$ & $\Omega_{m0}$&  \\
		\hline
		 sample {\it i} & $-0.61^{+0.20,+0.51,+0.66}_{-0.28,-0.43,-0.53}$ & $0.383^{+0.049,+0.075,+0.092}_{-0.035,-0.087,-0.11}$ &\\
		\hline
		 sample {\it ii} & $-0.48^{+0.12,+0.22,+0.22}_{-0.12,-0.20,-0.21}$ & $0.359^{+0.026,+0.043,+0.053}_{-0.023,-0.047,-0.058}$ &\\
		 \hline
		  sample {\it iii} & $-0.99^{+0.40,+0.64,+0.78}_{-0.35,-0.67,-0.85}$ & $0.448^{+0.057,+0.10,+0.12}_{-0.057,-0.10,-0.13}$&\\
		  \hline
		  sample {\it iv} & $-0.98^{+0.32,+0.61,+0.77}_{-0.32,-0.65,-0.87}$ & $0.446^{+0.051,+0.096,+0.12}_{-0.051,-0.10,-0.13}$\\ 
		\hline \hline
	\end{tabular}\label{tab:tab5}
\end{table}

\begin{table*}
	
	\centering
	\caption{Best-fit with $1\sigma$, $2\sigma$ and $3\sigma$ confidence intervals for the present values of the cosmographic parameters obtained from PADE cosmographic method and from the power-law $f(T)$ model (values inside the parenthesis), using the various samples considered in our analysis. Sample ({\it i}) includes Hubble diagram of SNIa. Sample ({\it ii}) includes Hubble diagram of SNIa + BAO measurements and sample ({\it iii}) includes Hubble diagrams of SNIa + QSOs. Sample ({\it iv}) includes Hubble diagrams of SNIa + QSOs + GRBs. }
	\fontsize{5pt}{5pt}
	\begin{tabular}{c c  c  c c c c }
		\hline \hline
		Model & $q_0$ & $j_0$ & $s_0$& $l_0$&$m_0$ \\
		\hline
		sample {\it i}& $-0.670^{+ 0.053,+0.10,+0.12}_{-0.053,-0.099,-0.12}$ & $ 1.50^{+0.28,+0.57,+0.72}_{-0.28,-0.57,-0.73}$ & $-0.47^{+0.59,+0.71,+0.93}_{-0.48,-0.86,-0.94}$ &  $3.25^{+0.94,+2.5,+2.9}_{-1.9,-2.1,-2.4}$ & $-6.5^{+7.1,+9.0,+9.8}_{-3.6,-11,-13}$
		\\
		&($-0.637^{+0.034,+0.068,+0.091}_{-0.034,-0.061,-0.077}$) &($ 1.25^{+0.11,+0.21,+0.25}_{-0.11,-0.22,-0.24}$) & ($-0.82^{+0.19,+0.45,+0.54}_{-0.25,-0.38,-0.44}$) &  ($4.92^{+0.76,+1.0,+1.1}_{-0.38,-1.4,-1.8}$) & ($-13.59^{+0.73,+2.2,+3.7}_{-1.1,-2.0,-2.2}$)
		\\
		\hline
		sample {\it ii}	&$-0.628^{+0.064,+0.13,+0.15}_{-0.064,-0.12,-0.14}$ & $1.27^{+0.36,+0.64,+0.71}_{-0.36,-0.68,-0.73}$ & $-0.43^{+0.68,+0.95,+1.3}_{-0.44,-1.2,-1.3}$ &  $4.1^{+1.1,+1.8,+2.2}_{-1.1,-1.9,-2.2}$ &
		$-15.3^{+3.4,+9.1,+11}_{-5.3,-7.4,-8.9}$\\
		&($-0.626^{+0.023,+0.045,+0.056}_{-0.023,-0.046,-0.053}$)& ($1.182^{+0.055,+0.095,+0.10}_{-0.055,-0.094,-0.10}$) & ($-0.70^{+0.13,+0.22,+0.30}_{-0.13,-0.22,-0.28}$) &  ($4.62^{+0.44,+0.70,+0.86}_{-0.36,-0.76,-0.96}$)&
		($-13.43^{+0.81,+1.5,+2.2}_{-0.81,-1.6,-2.0}$)\\
		\hline 
		sample {\it iii} &$-0.764^{+ 0.069,+0.14,+0.17}_{-0.069,-0.14,-0.15}$ & $2.10^{+ 0.43,+0.90,+1.1}_{-0.43,-0.92,-1.2}$ & $-0.07^{+1.0,+1.1,+1.1}_{-0.86,-1.8,-1.9}$ &  $3.3^{+1.2,+2.3,+2.8}_{-1.5,-2.2,-2.6}$ &
		$-9.1^{+3.8,+18,+19}_{-8.3,-11,-12}$\\
		&($-0.701^{+ 0.057,+0.11,+0.12}_{-0.057,-0.11,-0.12}$)& ($1.55^{+0.21,+0.47,+0.59}_{-0.32,-0.43,-0.49}$) & ($-1.20^{+0.16,+0.53,+0.68}_{-0.36,-0.41,-0.44}$) &  ($5.65^{+0.45,+0.52,+0.56}_{-0.082,-1.1,-1.6}$) &
		($-13.9^{+1.1,+2.1,+2.6}_{-0.92,-2.1,-3.7}$)\\
		\hline 
		sample {\it iv} & $-0.732^{+0.066,+0.15,+0.22}_{-0.075,-0.14,-0.18}$ &  $1.83^{+0.42,+0.88,+1.1}_{-0.42,-0.88,-1.1}$ & $-0.92^{+0.62,+1.7,+2.3}_{-0.95,-1.4,-1.7}$ &  $3.3^{+1.2,+2.3,+2.8}_{-1.5,-2.2,-2.6}$ &
		$-3.0^{+11.0,+12.0,+13.0}_{-14.0,-15.0,-16.0}$\\
		&($-0.697^{+ 0.053,+0.099,+0.12}_{-0.053,-0.11,-0.14}$)& ($1.53^{+0.20,+0.45,+0.61}_{-0.25,-0.41,-0.50}$)& ($-1.20^{+0.17,+0.51,+0.68}_{-0.31,-0.41,-0.46}$) &  ($5.67^{+0.43,+0.50,+0.55}_{-0.066,-1.0,-1.8}$) &
		($-13.8^{+1.0,+2.0,+2.6}_{-1.0,-2.1,-3.1}$)\\
		\hline \hline
		
	\end{tabular}\label{tab:tab6}
\end{table*}
\subsection{Constraints to $f(T)$ model using the Hubble diagrams and BAO observations}
In this subsection, we use the real observational data for Hubble diagrams of SNIa, QSOs and GRBs and place observational constraints on the cosmographic parameters for both the model-independent cosmographic method (defined based on the $P_{3,2}$ approximation) and the power-law $f(T)$ model. In addition, we extend our analysis by adding the BAO measurements as robust observations to the MCMC sample. If we find tensions between the cosmographic parameters of the power-law $f(T)$ model and those of the model-independent cosmographic method, this can be interpreted as a possible tool to reject the model. Here we use four different combinations of observational data, including sample ({\it i}) consisting of the Hubble diagrams of SNIa from the Pantheon catalogue \citep{Scolnic:2017caz}, sample ({\it ii}) consisting of the Hubble diagram of SNIa + BAO measurements, where we use the observational data for the BAO measurements from \citep{Mehrabi:2015hva}, sample ({\it iii}) consisting of the Hubble diagrams of SNIa + QSOs, where we use the binned data for QSOs from \citep{Lusso:2017hgz}, sample ({\it iv}) consisting of the Hubble diagrams of SNIa+QSOs+GRB, where the GRB data are composed of 141 data points in the redshift range $0.03<z<9.3$ collected in \citep{Escamilla-Rivera:2021vyw}. In Fi. (\ref{fig:fig8}), we plot the confidence regions of the cosmographic parameters $q_0$ and $j_0$ in the context of $P_{3,2}$ cosmographic method using different samples presented above. Also for each sample, we put the observational constrains on the cosmological parameters $b$ and $\Omega_{m0}$ (see Table \ref{tab:tab5}) and then obtain the best-fit values of the cosmographic parameters for the power-law $f(T)$ model from  Eq. (\ref{eq:eq20}). The best-fit values of the parameters $q_0$ and $j_0$ for the power-law $f(T)$ model are also shown in Fig. (\ref{fig:fig6}). For all samples used in our analysis, we find complete agreement within $1\sigma$ error between the cosmographic method and the power-law $f(T)$ model in the $q_0-j_0$ plane. Our numerical results for other cosmographic parameters are also shown in Table (\ref{tab:tab6}). For all samples, we see that the power-law $f(T)$ model is completely consistent with the model-independent cosmographic method. In particular, we obtain more closer consistency between the power-law $f(T)$ model and the cosmographic method when the BAO added to our MCMC sample (see the upper right panel of Fig. (\ref{fig:fig8}). Moreover, the power-law $f(T)$ model is consistent with the model-independent cosmographic method when we use the high redshift Hubble diagrams of QSOs and GRBs (see lower panels of Fig. \ref{fig:fig8}).
 Finally, substituting the constrained values of the cosmological parameters reported in Table (\ref{tab:tab5}) into Eq. (\ref{eq:eq5pp}), we show the redshift evolution of the effective EoS parameter of DE for the power-law $f(T)$ model in Fig. (\ref{fig:fig9}). On the upper-left panel, we have used the constrained values of $\Omega_{m0}$ and $b$, using sample ({\it i}). We see that the EoS parameter varies in the phantom regime, but the upper limit of the confidence band is fixed to critical value $w_{\Lambda}=-1.0$. This result indicates that our constraints on the effective EoS parameter $w_{de}$, using the observational data from the Hubble diagram of the SNIa, cannot discriminate the power-law $f(T)$ model from the standard $\Lambda$CDM cosmology. On the upper-right panel, we have used sample ({\it ii}) and showed that $w_{de}$ varies in the phantom regime. We observe that the confidence band of our constraints on the Eos parameter is more tighter compared to the previous case. This result is due to the including of the BAO observation as robust observational measurements in our analysis. In addition, we observe that the effective EoS parameter of the power-law $f(T)$ model completely deviates from the standard value $w_{\Lambda}=-1.0$. On the lower-left panel (lower-right panel), we have used the results of the sample ({\it iii}) (sample {\it iv}) presented in Table (\ref{tab:tab5}) and extended our constraints on the effective EoS parameter to the higher redshifts compared to the previous cases. Same as sample ({\it ii}), we observe the phantom-like behavior of the effective EoS parameter of DE but with larger uncertainties. In particular, the upper-band of our constraints is very close to $w_{\Lambda}=-1.0$, but cannot cross it. Moreover, the best-fit value of the EoS parameter reaches to the constant value $w_{de}=-2.0$ at redshifts higher than $z\simeq 1$. In overall, using the combination of Hubble diagrams from SNIa, QSOs and GRBs, we observed the phantom-like EoS parameter with upper-limit close to $w_{\Lambda}=-1$. While, the combination of SNIa Hubble diagram and BAO observation can put tighter constrains on the EoS parameter in phantom regime with a significant deviation from $w_{\Lambda}=-1.0$.  
\section{Conclusions} \label{sect:conlusion}
The cosmographic approach is a model-independent way to reconstruct the expansion rate of the Universe, without presupposing particular cosmological model. In this framework, we can expand the Hubble parameter around its present value using the linear Taylor series or other mathematical polynomials such as the rational PADE approximation. We note that the Taylor series and the rational PADE series are the mathematical approximations to reconstruct the Hubble parameter and consequently the distance modulus as a function of redshift. Therefor, we should be concerned about error propagation of such mathematical approximations in our calculations. In fact, we cannot use the cosmographic method based on a given mathematical approximation, if the error propagation is large enough in an unreal result. One possible way to examine the cosmographic method is to use mock data for the Hubble diagrams of SNIa (up to redshifts $\sim 2$), QSOs (\jt{up to $z\sim 5$}) \jt{and GRBs extended to $z\sim 9$}. By using mock data of SNIa, QSOs \jt{and GRBs} generated based on the cosmological model, we can constrain the cosmographic parameters in the context of cosmographic method. In parallel, using the same data, we obtain the best-fit values of the cosmographic parameters of the exact cosmological model. In order to accept the cosmographic method as a useful and powerful model-independent approach, we expect the best-fit cosmographic parameters of the model and the confidence regions obtained with the cosmographic method to match. We mention that the mock data are generated using the model. Hence, the model cannot automatically deviate from the mock data. In other words, any difference between the best-fit cosmographic parameters of the model and the confidence regions of cosmographic method can be interpreted as a defect of the cosmographic method due to the error truncation of the mathematical approximation. In this paper, we presented a new study on the power-law $f(T)$ model in the context of cosmographic method, using the fourth- and fifth- order Taylor series in addition to the rational PADE polynomials $(3,2)$ and $(2,2)$ . At the first step, using mock data for the Hubble diagram of SNIa, we found that all versions of the cosmographic methods, except the one based on the fourth-order Taylor series, can work well as the model-independent method for reconstructing the distance modulus up to redshift $z\sim 2.3$. Moreover, we found a strong deviation (larger than $3\sigma$) between the cosmographic parameter $l_0$ obtained with the cosmographic method based on the foruth-order Taylor series and that of the power-law $f(T)$ model. This deviation demonstrates the inefficiency of the fourth-order Taylor series compared to other mathematical approximations considered in this work. In the next step, we extended our study to the high redshifts $z\sim 5$ using mock data for the Hubble diagram of QSOs. In doing so, we found that the foruth-order Taylor series is not suitable for reconstructing the distance modulus at high redshifts.  In addition, we found that other mathematical approximations investigated in this work can work well at redshifts up to $z\sim 5$ and can therefore be considered suitable tools to study the power-law $f(T)$ model in the context of cosmographic method. \jt{ Finally we extend our analysis to extremely high redshifts $z\sim 9$ by using the mock Hubble diagrams of GRBs. We found that both fourth and fifth-order Taylor series are falsified to reconstruct the distance modulus, while both PADE approximations can work properly at very high redshifts where GRB explosions can be observed.} In particular, we concluded that the rational PADE $(3,2)$ approximation is the best approximation to reconstruct the distance modulus in the framework of cosmographic method. In this framework, we investigated the power-law $f(T)$ model using the real observational data from Hubble diagrams of SNIa, QSOs, GRBs and the BAO measurements. We found that the power-law $f(T)$ model is well consistent with the various samples of the combined observational data within the context of PADE cosmographic method. Moreover, this model agrees with the high redshifts QSOs and GRBs observations extended to $z\sim 5$ and $z\sim 9$, respectively. In addition, we observed that combining SNIa with the BAO observations leads to a better agreement between the best-fit values of the cosmographic parameters of the power-law $f(T)$ model and those of the model-independent cosmographic approach. Finally, we showed that the observational constraints on the EoS parameter of the effective DE, described in the power-law $f(T)$ model, represent the phantom like EoS parameter, especially when the BAO measurements are included in our analysis.\\   

\textbf{DATA AVAILABILITY}\\
The data underlying this article will be shared on reasonable request to the corresponding author.

\bibliographystyle{mnras}
\bibliography{ref}
\label{lastpage}

\end{document}